%% file: zvrk.tex
\PassOptionsToPackage{unicode}{hyperref}
\PassOptionsToPackage{hyphens}{url}
\PassOptionsToPackage{dvipsnames,svgnames,x11names}{xcolor}
\documentclass[
  astrosymb, twocolumn, twocolappendix, tighten]{aastex631}
\usepackage{amsmath,amssymb}
\usepackage{iftex}
\ifPDFTeX
  \usepackage[T1]{fontenc}
  \usepackage[utf8]{inputenc}
  \usepackage{textcomp} 
\else 
  \usepackage{unicode-math} 
  \defaultfontfeatures{Scale=MatchLowercase}
  \defaultfontfeatures[\rmfamily]{Ligatures=TeX,Scale=1}
\fi
\usepackage{lmodern}
\ifPDFTeX\else
\fi
\IfFileExists{upquote.sty}{\usepackage{upquote}}{}
\IfFileExists{microtype.sty}{
  \usepackage[]{microtype}
  \UseMicrotypeSet[protrusion]{basicmath} 
}{}
\makeatletter
\@ifundefined{KOMAClassName}{
  \IfFileExists{parskip.sty}{%
    \usepackage{parskip}
  }{
    \setlength{\parindent}{0pt}
    \setlength{\parskip}{6pt plus 2pt minus 1pt}}
}{
  \KOMAoptions{parskip=half}}
\makeatother
\usepackage{xcolor}
\usepackage{longtable,booktabs,array}
\usepackage{calc} 
\usepackage{etoolbox}
\makeatletter
\patchcmd\longtable{\par}{\if@noskipsec\mbox{}\fi\par}{}{}
\makeatother
\IfFileExists{footnotehyper.sty}{\usepackage{footnotehyper}}{\usepackage{footnote}}
\makesavenoteenv{longtable}
\usepackage{graphicx}
\makeatletter
\def\maxwidth{\ifdim\Gin@nat@width>\linewidth\linewidth\else\Gin@nat@width\fi}
\def\maxheight{\ifdim\Gin@nat@height>\textheight\textheight\else\Gin@nat@height\fi}
\makeatother
\setkeys{Gin}{width=\maxwidth,height=\maxheight,keepaspectratio}
\makeatletter
\def\fps@figure{htbp}
\makeatother
\providecommand{\tightlist}{%
  \setlength{\itemsep}{0pt}\setlength{\parskip}{0pt}}
\input{macros.tex}

\usepackage{mathptmx,txfonts,tikz,bm}
\ifLuaTeX
  \usepackage{selnolig}  
\fi
\usepackage[]{natbib}
\usepackage{bookmark}
\IfFileExists{xurl.sty}{\usepackage{xurl}}{} 
\hypersetup{
  colorlinks=true,
  linkcolor={xlinkcolor},
  filecolor={Maroon},
  citecolor={xlinkcolor},
  urlcolor={xlinkcolor},
  pdfcreator={LaTeX via pandoc}}

\begin{document}

\vspace*{-17.3ex}
\shorttitle{Gasing Pangkah I: Zvrk}
\title{The \textit{Gasing Pangkah}\footnote{\textit{Gasing pangkah} is a traditional Malay competitive game of fighting spinning tops, whose name we adopt as the name of our collaboration.} Collaboration: I. Asteroseismic Identification \\ and Characterisation of a Rapidly-Rotating Engulfment Candidate}
\input{preamble}
\begin{abstract}
We report the discovery and characterisation of TIC\ 350842552 (``Zvrk''), an apparently isolated, rapidly-rotating ($P_\text{rot} \sim 99\ \mathrm{d}$) red giant observed by TESS in its Southern Continuous Viewing Zone. The star's fast surface rotation is independently verified by the use of p-mode asteroseismology, strong periodicity in TESS and ASAS-SN photometry, and measurements of spectroscopic rotational broadening. A two-component fit to APOGEE spectra indicates a coverage fraction of its surface features consistent with the amplitude of the photometric rotational signal. Variations in the amplitude of its photometric modulations over time suggest the evolution of its surface morphology, and therefore enhanced magnetic activity. We further develop and deploy new asteroseismic techniques to characterise radial differential rotation, and find weak evidence for rotational shear within Zvrk's convective envelope. This feature, in combination with such a high surface rotation rate, is incompatible with models of angular-momentum transport in single-star evolution. Spectroscopic abundance estimates also indicate a high lithium abundance, among other chemical anomalies. Taken together, all of these suggest a planet-ingestion scenario for the formation of this rotational configuration, various models for which we examine in detail.
\keywords{Asteroseismology (73), Red giant stars (1372), Stellar rotation (1629), Star-planet interactions (2177)}
\end{abstract}

\section{Introduction}\label{introduction}

As stars evolve up the red giant branch (RGB), their radiative cores contract (and spin up) while their convective envelopes expand (and slow down in their rotation). While seismic rotational measurements with evolved stars indicate that subgiant convective envelopes rotate faster than strict angular momentum conservation would suggest \citep[e.g.][]{mosser_spin_2012, triana_internal_2017, deheuvels_seismic_2020}, envelope rotation rates nonetheless are predicted to decrease dramatically with further evolution, making them challenging to measure for more evolved red giants at higher luminosity. Indeed, many existing \edit1{seismic} techniques for rotation measurements in these evolved stars \edit1{require} the rotation of the stellar envelope to be ignored \citep[e.g.][]{gehan_core_2018, mosser_period_2018}. In the regime of (relatively) strong coupling between the interior g-mode cavity to the exterior p-mode cavity, this simplifying assumption has enabled the measurement of core rotation rates in red giants of intermediate luminosity in the \emph{Kepler} sample en masse, at the expense of sacrificing information about \edit1{their} surface rotation. \edit1{Since} the surfaces of red giants are not expected to rotate \edit1{much}, a loss of sensitivity \edit1{of this kind} is ordinarily considered acceptable.

\edit1{Nonetheless}, a small fraction of the \emph{Kepler} red giant sample does exhibit nontrivial photometric variability, attributable to surface features rotating into and out of view \citep[e.g.][]{tayar_rapid_2015, ceillier_surface_2017}. Rotation rates as high as these are not generally compatible with standard descriptions of angular momentum transport \edit1{for} less evolved stars \citep{aerts_angular_2019, fuller_slowing_2019}, and \edit1{may suggest} that these rapidly-rotating red giants (RRRGs) passed through evolutionary scenarios quite unlike the standard picture of single-star evolution \citep[e.g.][]{massarotti_rotational_2008, carlberg_frequency_2011}. Possible causes of this spin-up include binary interactions \citep[e.g.][]{tayar_rapid_2015, casey_tidal_2019} or engulfment events \citep{lau_hot_2022, oconnor_giant_2023}. The latter \edit1{could} be associated with overabundance in light elements, such as lithium, in the envelopes of first ascent red giants \citep[e.g.][]{aguileragomez_lithium_2016a, aguileragomez_lithium_2016b, soaresfurtado_lithium_2021}, and may illuminate the evolution of late-stage planetary systems \citep[e.g.][]{de_infrared_2023}.

\edit1{As with} main-sequence stars, we expect that \edit1{sensitivity to} surface rotation in evolved stars will be provided by their exterior p-mode cavity, \edit1{but this is} only indirectly \edit1{constrained} by the mixed-mode frequencies that are typically measured. While coupling between the two mode cavities would in most cases prevent direct measurement of these pure p-mode rotation rates, the strength of this coupling in first-ascent red giants decreases dramatically with evolution \citep[e.g.][]{farnir_eggmimosa_2021, jiang_evolution_2022, ong_rotation_2}. The frequencies of p-dominated mixed modes therefore become observationally indistinguishable from those of the pure p-modes underlying them in higher-luminosity RRRGs, which are in any case mostly convective by spatial extent. Although traditional computational methods for mixed-mode asteroseismology --- in particular relating mode frequencies to interior structure --- have been very expensive to apply to these most highly luminous red giants \citep{jcd_aarhus_2020}, recent theoretical developments now permit the numerical analysis of red-giant p-modes to be performed in a likewise decoupled fashion from the interior g-mode cavity, thereby circumventing these computational difficulties \citep{ball_surface_2018, ong_semianalytic_2020}.

RRRGs are frequently observed to have suppressed oscillations, making \edit1{their} asteroseismic characterisation challenging. This suppression is often also attributed to tidal synchronisation with a close companion \citep{gaulme_active_2020}. Conversely, the overwhelming majority of evolved RRRGs which do exhibit pulsations \edit1{lack} detectable orbital companions, rendering their rotation all the more difficult to explain. Owing to these observational and \edit1{methodological challenges}, pulsating RRRGs \edit1{do not easily yield} detailed asteroseismic constraints on their internal structure and rotation.

\edit1{In this work, we report such constraints,} arising from the contemporaneous detection and characterisation of both rotational and asteroseismic signals in TIC~350842552 (which we will refer to informally as ``Zvrk''\footnote{Serbo-Croat: ``spinning top''} in the remainder of this work, for brevity), a \edit1{bright ($m_V = 10.24$)}, highly evolved, and likely isolated RRRG. \edit1{We confirm a rapid asteroseismic rotation rate using both direct photometry --- from the NASA TESS mission, and separately with the ground-based ASAS-SN network --- and rotational broadening from different ground-based spectrographs. Moreover, we find Zvrk to exhibit unusual spectroscopic absorption signatures, which are difficult to explain with single-star evolutionary modelling. Given that Zvrk is a late-stage giant, --- far more} evolved than almost the entire \emph{Kepler} sample of RRRGs \citep{ceillier_surface_2017} \edit1{--- all of its observed modes may be well approximated as being pure p-modes, permitting new analytic developments in their analysis and interpretation to be applied.} Rather than characterising a two-zone core-vs-envelope rotational contrast as is typically examined in less evolved red giants, we find instead weak evidence for radial differential rotation within Zvrk's convective envelope.

Zvrk's high rotation rate, possible rotational shear, and anomalous chemical abundances are highly suggestive of an ingestion scenario for its formation, on which we place preliminary constraints. We conclude with some discussion about the astrophysical significance of this finding, methodological implications of the techniques that were developed for this work, and suggestions for future work.

\section{Observational Characterisation}\label{observational-characterisation}

Zvrk was \edit1{first} flagged for asteroseismic analysis as part of an ongoing large-scale search for unusual oscillation signatures \edit1{among stars with} high lithium abundance. For this purpose, an overlapping sample of stars with both preliminary GALAH DR4 lithium abundance estimates, and TESS coverage in the Southern continuous viewing zone (CVZ), was constructed (Montet et al., in prep). Targets in this list were subjected to preliminary asteroseismic analysis, to constrain the global p-mode asteroseismic parameters \(\Dnu\) and \(\numax\). This was done using the 2D autocorrelation function technique \citep[2DACF:][]{viani_dnu_2019}, applied to the SPOC light curves from the first two Southern CVZ campaigns in TESS Cycles 1 and 3. For Zvrk in particular, this procedure yielded \(\Dnu = (1.23 \pm 0.01)\ \mathrm{\mu Hz}; \numax = (7.5 \pm 0.3)\ \mathrm{\mu Hz}\).

Among this sample, Zvrk stood out because of an unusually large surface lithium abundance (\(\mathrm{A(Li)} \sim 3.2\ \mathrm{dex}\); cf.~\autoref{sec:abundance}). Also unusually for a red giant, it was observed by TESS at two-minute cadence: in Cycle 1 (Sectors 1-13), Cycle 3 (Sectors 27-39), and \edit1{Cycle 5 (ongoing; not used in this study). As such, we were able to rely on} the publicly-available presearch data conditioning simple aperture photometry (PDCSAP) lightcurves instead \edit1{of performing our own reduction. \edit2{The resulting time series spans 3 years --- 2 yearlong spans with a 1-year gap.} Zvrk's PDCSAP} power spectrum exhibits unusual features, whose structure departs significantly from that seen in similarly evolved red giants \citep[e.g.][]{stello_nonradial_2014, yu_luminous_2020}, and which defeated preliminary attempts at p-mode identification.

\begin{figure*}[htbp]
    \centering
    \includegraphics[width=\textwidth]{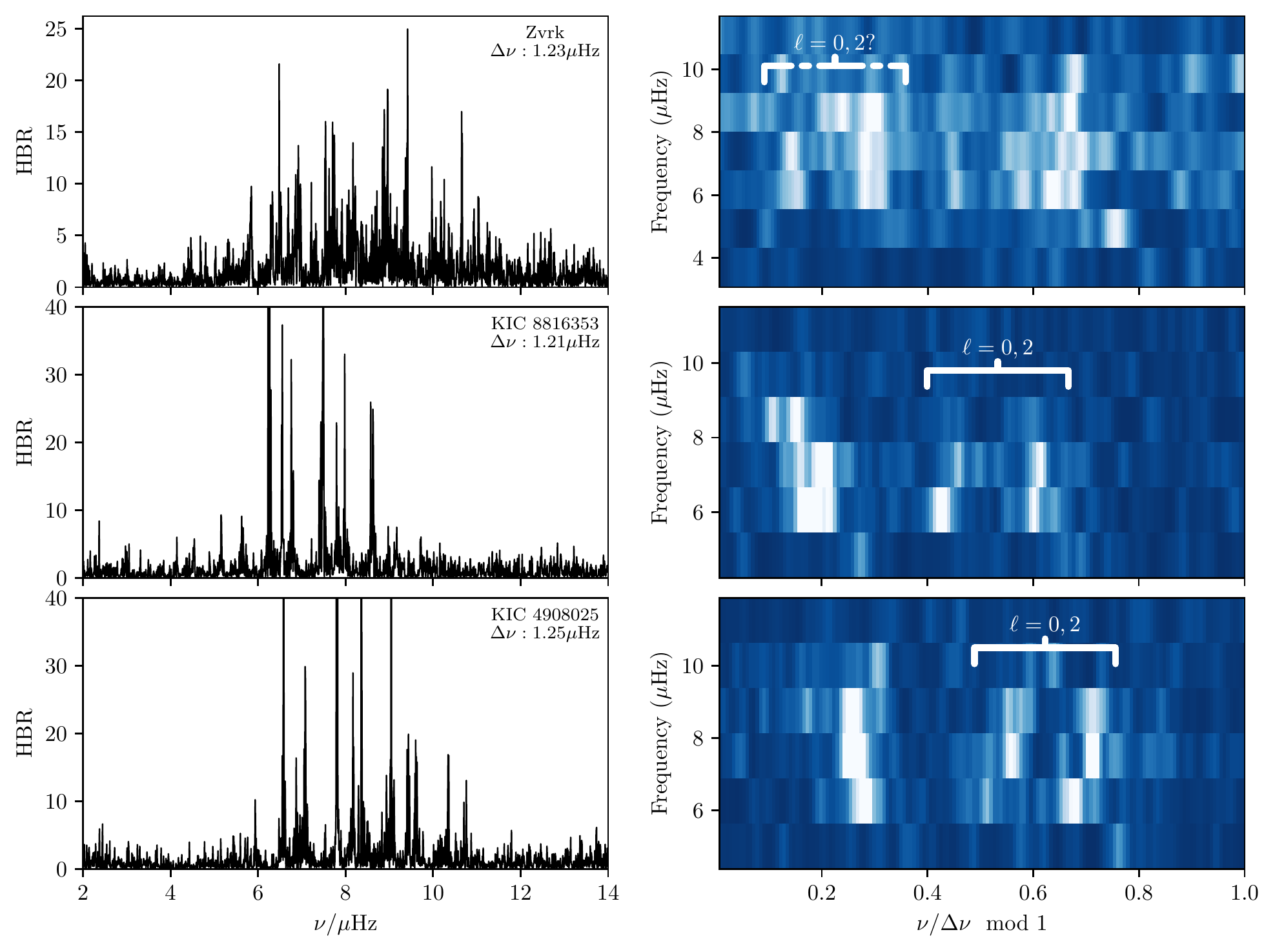}
    \caption{Comparison between Zvrk's oscillation power spectra with those from \textit{Kepler} red giants KIC 8816353 and KIC 4908025\edit1{, which have similar values of \Dnu}. We show background-divided periodograms \edit2{(yielding height-to-background ratios)} for all three stars in the left column of panels; in the right column of panels, we show the corresponding echelle power diagrams, smoothed by Gaussian kernels with widths matching the resolution of the power spectra, to more clearly show structure. \edit2{In all cases, the granulation background was estimated using the running-log-window median filter implemented in \texttt{lightkurve} \citep{lightkurve}}. The oscillation spectra of the \textit{Kepler} red giants show the expected mode pattern for evolved red giants: three distinct ridges in the \'echelle diagram, with $\ell=2,0$ ridges centered at $\epsilon_p = \nu/\Delta\nu\mod 1 \sim0.6$. By contrast, the power spectrum for Zvrk contains many additional peaks that does not conform to the expected mode pattern, with what was initially identified as the $\ell=2,0$ ridges located far from the typical value. }
    \label{fig:echelle-comparison}
\end{figure*}

\subsection{Mode Frequencies and Rotational Splittings}\label{mode-frequencies-and-rotational-splittings}

\label{sec:nested}

Further asteroseismic analysis requires integer indices \(n_p, \ell, m\) to be assigned to any mode frequencies identified in the observed power spectrum. The shape of the double ridges seen on Zvrk's frequency échelle diagram (top right panel of \cref{fig:echelle-comparison}) evokes that of other p-mode oscillators, as previously observed en masse with \emph{Kepler} and TESS (right column of \cref{fig:echelle-comparison}), and suggest identification as being modes of low, even degree (\(\ell = 0, 2\)). In p-mode oscillators, these are known to form double ridges of this kind as a result of the p-mode asymptotic eigenvalue equation
\begin{equation}
\nu_{n_p\ell} \sim \Dnu\left(n_p + {\ell \over 2} + \epsilon_p\right) + {\ell(\ell+1)}{\delta\nu_{02}\over 6} + \mathcal{O}(1/\nu),\label{eq:asymptotic}
\end{equation} in the absence of rotation. Here \(\Delta\nu\) and \(\delta\nu_{02}\) are the large and small frequency separations from the standard phenomenological description of p-mode asteroseismology, and \(\epsilon_p\) is a slowly-varying phase function. Under this putative identification, \edit1{the remaining peaks would} be attributed to oscillations of dipole modes, which are known in other red giants to be disrupted by mode mixing with interior g-modes, producing a complex forest of gravitoacoustic mixed modes. Visually, the morphology of this part of Zvrk's echelle diagram strongly resembles that seen in core-helium-burning stars, where the coupling between the interior g-mode and exterior p-mode cavities is known to be strong. Indeed, this is the mode identification also arrived at by the use of existing ``data-driven'' automated methods, trained on the \emph{Kepler} sample of oscillating red giants \citep[e.g.~PBJam:][]{pbjam}.

\begin{figure*}
\centering
\annotate{\includegraphics[width=.475\textwidth]{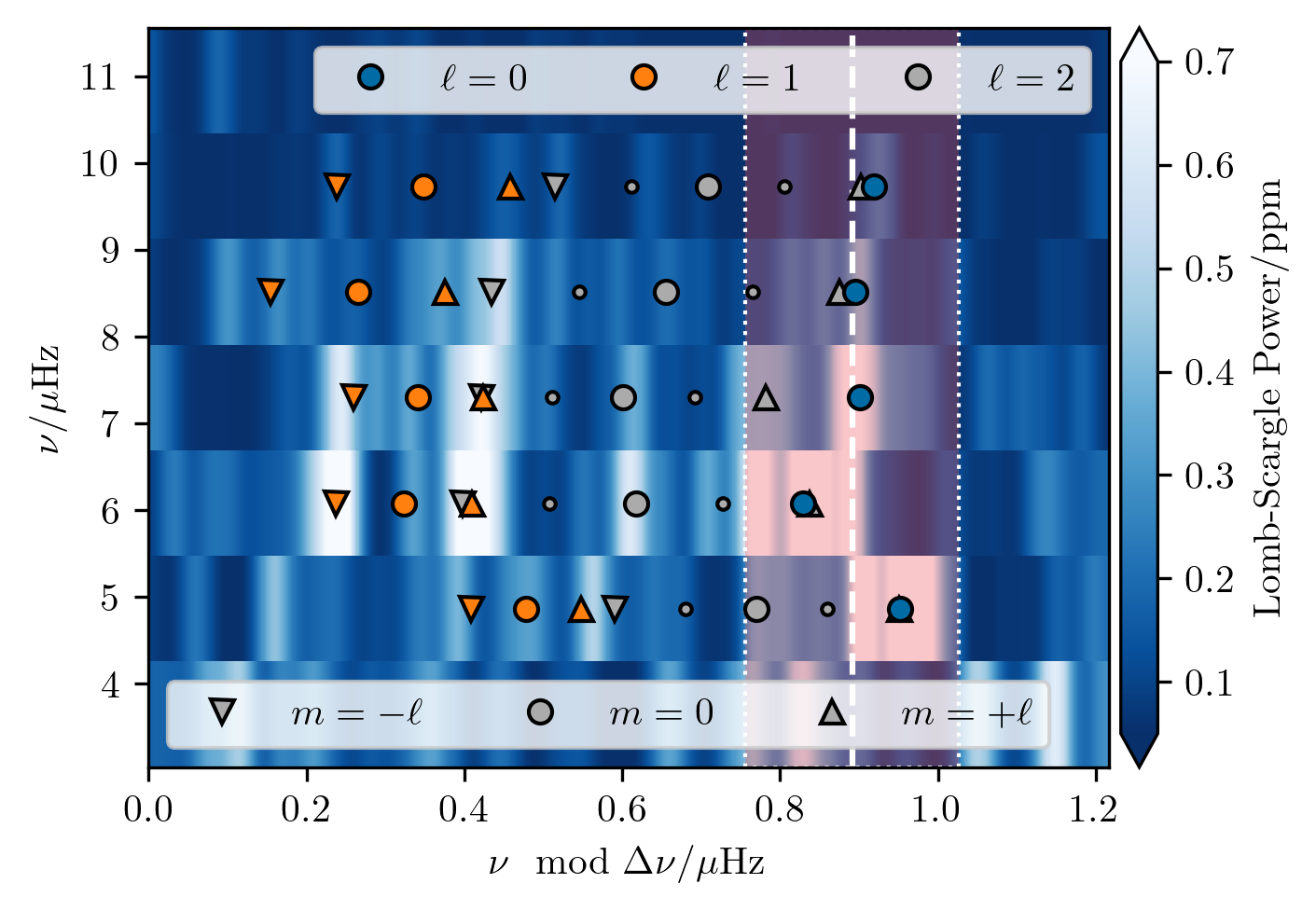}}{\node[white] at (.15, .9){\textbf{(a)}};}
\annotate{\includegraphics[width=.475\textwidth]{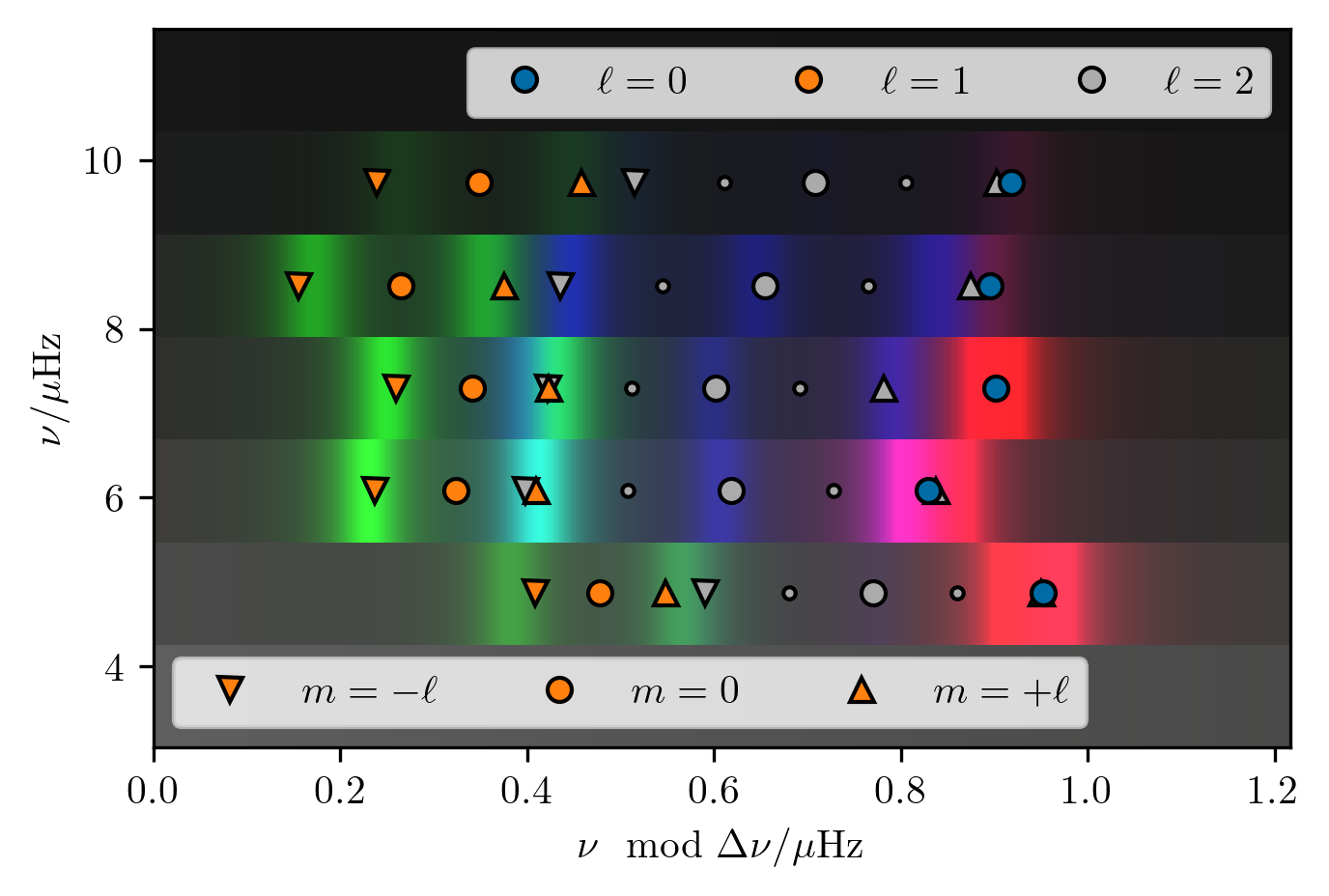}}{\node[white] at (.15, .9){\textbf{(c)}};}
\annotate{\includegraphics[width=.95\textwidth]{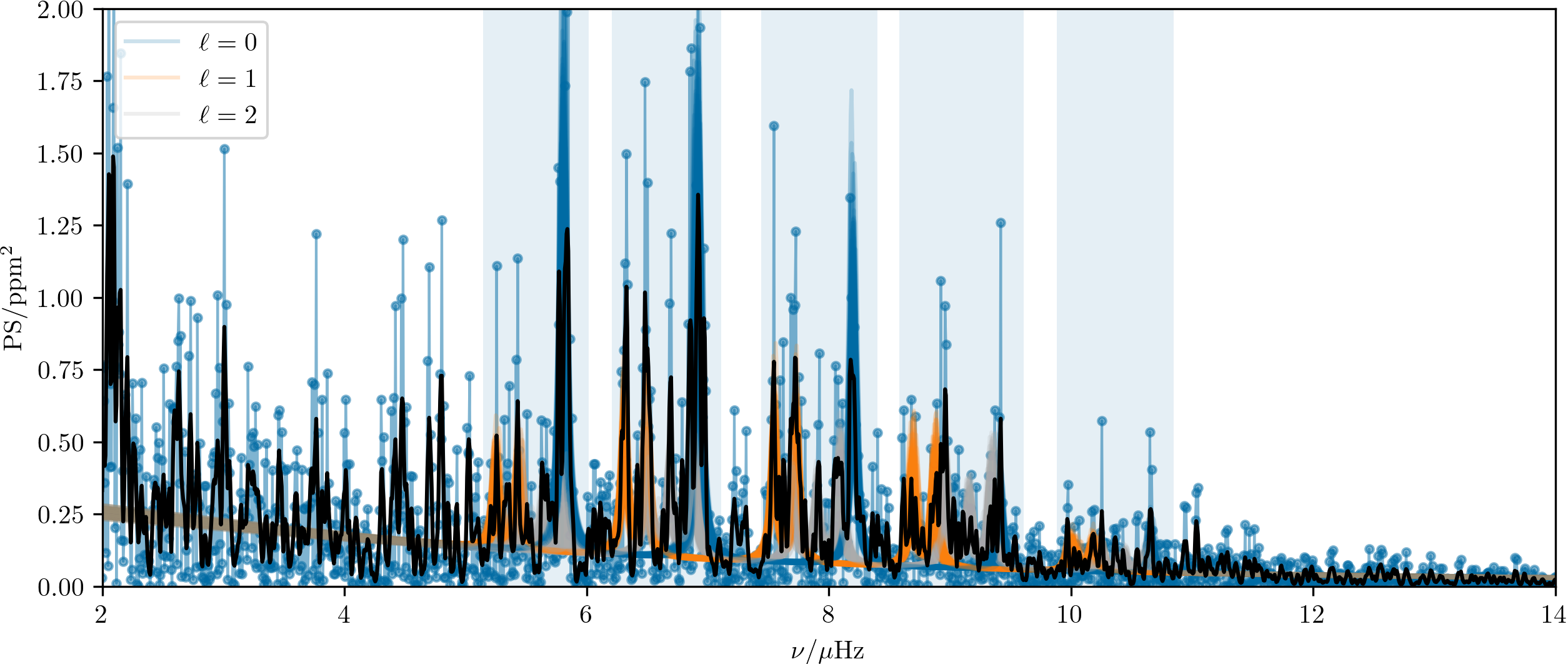}}{\node at (.95, .9){\textbf{(b)}};}
\caption{Asteroseismic characterisation of Zvrk. \textbf{(a)}: \'Echelle power diagram showing putative mode identification, manually constructed from naive peakfinding. The red shaded region indicates the allowable region for radial modes, \edit1{$\epsilon_p \sim 0.73\pm0.11$, as implied by} measurements from the \textit{Kepler} field. \textbf{(b)}: Samples from the joint posterior distribution for all parameters describing our model of the power spectrum (semitransparent curves, \cref{eq:model}; different colors show portions of the power spectrum attributed to modes of different degrees), overplotted against the raw power spectrum (filled circles joined with lines), and compared with a smoothed power spectrum (black line, \edit1{using} a Gaussian kernel of width $0.008\ \mu$Hz). \textbf{(c)}: Colour-channel \'echelle power diagram showing same samples from posterior distribution. The different colour channels show power attributed to modes of different degree: red for $\ell = 0$, green for $\ell = 1$, and blue for $\ell = 2$. Our original manual (not fitted) mode identification is overplotted for comparison. An animated version of this panel, to accommodate colourblind readers, is available in the online edition of this paper. \edit1{This animation shows each of the colour channels separately for 3 seconds each, to illustrate the relative significance of modes of each degree, and then shows all colour channels coadded for 3 seconds, as in the static version of the figure.}\label{fig:asteroseismology}}
\end{figure*}

However, such an identification would be in significant tension with other known properties of stochastically-excited red giant oscillations. In particular:

\begin{itemize}
\tightlist
\item
  Red clump stars are known to yield much higher values of both \Dnu~and \numax~than we have measured from Zvrk. The measured values of these quantities instead strongly favour classification of Zvrk as either a first-ascent red giant, or a shell-helium-burning asymptotic giant branch (AGB) star.
\item
  Both single-star stellar modelling \citep[e.g.][]{deheuvels_seismic_2022} and observational measurements \citep[e.g.][]{stello_asteroseismic_2013, mosser_mixed_2014} of gravitoacoustic mixed modes in the \emph{Kepler} sample indicate that \(\Delta\Pi_1\), the dipole-mode period spacing associated with the interior g-mode cavity, lies in a narrow band of allowed values with \Dnu~for both first-ascent RGB stars, and AGB stars. At this value of \Dnu, \(\Delta\Pi_1\) \edit1{is} far too small, and the mixed-mode coupling strength too weak, to cause significant departures in the frequencies of the observed dipole modes from those of simple p-modes. This can be seen to be the case for the two \emph{Kepler} giants shown in \cref{fig:echelle-comparison}, where the dipole modes can be seen to form a single ridge, indicating the observational unavailability of g-mode mixing. This being the case, we ought not to expect it to be available from Zvrk either. Thus, the unusual structure of the peaks lying away from the double ridge cannot be identified as dipole mixed modes.
\item
  This mode identification would also imply a radial p-mode phase offset of \(\epsilon_p \sim 0.25\). However, the value of \(\epsilon_p\) for radial modes in first-ascent red giants follows an extremely tight and robust relation with \Dnu, both in theoretical studies of stellar models \citep[e.g.][]{white_calculating_2011, ong_structural_2019}, as well as in existing large-scale characterisation of the \emph{Kepler} sample (e.g.~\citealt{mosser_universal_2011};\citealt{yu_luminous_2020}; cf.~Figs. 10b and c of the latter). Thus, such a value of \(\epsilon_p\) would be in significant tension with the value of \(0.7\) implied by the \emph{Kepler} sample for red giants close to our nominal value of \(\Dnu = 1.23\ \mathrm{\mu Hz}\). We illustrate this in \cref{fig:echelle-comparison}, where the position of the double-ridge feature in Zvrk's echelle power diagram can be seen to be offset from that shown for two other \emph{Kepler} stars, both also first-ascent red giants. \edit1{We find that only the first-ascent RGB value of $\epsilon_p$ explains the peaks in the power spectrum, as a similar sequence constructed using only AGB stars} \citep[e.g.][]{kallinger_evolutionary_2012} would require lower values of \(\epsilon_p\) than shown for Zvrk in \cref{fig:echelle-comparison}.
\end{itemize}

We show instead in \cref{fig:asteroseismology}a our proposed alternative mode identification, taking all of the above constraints from the \emph{Kepler} field into account. By necessity, our radial modes are anchored by \(\epsilon_p\) measurements from the \emph{Kepler} field. To illustrate this, we show the average value of \(\epsilon_p\) for all stars within \(0.5\ \mu\)Hz of \(\Dnu\) to Zvrk in the catalogue of \citet{yu_luminous_2020}, and their standard deviation, using the white dashed line and red shaded interval, respectively.

In the presence of rotation, each mode at degree \(\ell\) and radial order \(n_p\) splits into a \((2\ell+1)\)-tuplet of peaks in the power spectrum, with the distribution of observed power between peaks being determined by the inclination of the stellar rotational axis \citep{pesnell_observable_1985, gizon_inclination_2003}. Rather than being modes of even degree, we instead identify the double ridge as being rotationally-split doublets of dipole (\(\ell = 1\)) modes, viewed close to equator-on.

\edit1{Rotation \cite[e.g.][]{gaulme_active_2020} and magnetism \cite[e.g.][]{fuller_fields_2015} are known to suppress oscillations in general, so a mode identification including rotation would be consistent with the apparently lower \edit2{height-to-background ratio} shown for Zvrk in \cref{fig:echelle-comparison} than for the \textit{Kepler} stars. Moreover, this suppression is known to act on dipole modes preferentially, so our proposal of clear dipole modes also requires a consistent set of quadrupole modes to be identified.} In such an equator-on configuration, \edit1{only $\ell+1$ components are visible in each multiplet}; our identification of the quadrupole modes must then explain the remaining peaks using rotationally-split triplets. While our putative identification of the quadrupole modes is unfortunate from the perspective of fitting the rotational splittings (as the quadrupole-mode triplets straddle both the dipole and radial modes \edit2{in a degenerate fashion}), \edit2{the structure of the observational window function (which we discuss in more detail in \autoref{sec:window}) is such that it is unlikely that we have assigned identifications to spurious artifacts}. Our proposed position for \edit2{the quadrupole modes} is \edit1{also} required for general consistency with the values of \(r_{02}\) typically returned from stellar models with comparable \(\Dnu\) \citep[e.g.][]{white_calculating_2011}.

In keeping with the star's evolved state, all of these modes are proposed to be essentially decoupled from the interior g-mode cavity. Since g-mode mixing is not an observational concern, we may directly repurpose existing techniques for the derivation of p-mode frequencies from power spectra --- peakbagging --- to this star. For this purpose, we fit an ansatz model in the form
\begin{equation}
    f(\nu) = \sum_i \sum_{m=-{\ell_i}}^{\ell_i}{H_i r(m, \ell_i, i_\star) \over 1 + (\nu_i + m \delta\nu_i - \nu)^2/\Gamma_i^2} + \mathrm{B}(\nu)\label{eq:model}
\end{equation}
directly to the power spectrum. The asteroseismic component of this power spectral model can be seen to be a sum of Lorentzians parameterised by the nonrotating frequencies \(\nu_i\), the mode heights \(H_i\), the mode linewidths \(\Gamma_i\) (which describe the damping rates of the modes), the rotational multiplet splittings \(\delta\nu_i\), as well as the inclination \(i_\star\) of the rotational axis (through the visibility ratios \(r\) of each multiplet component). \edit2{These ratios \cite[e.g.][]{gizon_inclination_2003} are such that peaks of the same $\ell$ and $|m|$ are assigned the same height}. In addition to this, we choose to describe our background model (B) with a combination of a single Harvey profile and a white-noise term.

\edit1{To infer Bayesian posterior distributions of all of these parameters, we must also specify their prior distributions.} We place flat priors on each \(\nu_i\) in windows \(0.2\ \mu\)Hz wide, centered on each of the manually identified values shown in \cref{fig:asteroseismology}a; a flat prior on \(\mu = \cos i_\star\) for isotropy; flat priors on the logarithms of the mode lifetimes, heights, and all parameters of our noise model; and flat priors on the widths of the rotational splittings. For our main analysis in this section we moreover pool the rotational splittings of the dipole and quadrupole modes separately, assigning either \(\delta\nu_{\text{rot}, \ell=1}\) or \(\delta\nu_{\text{rot}, \ell=2}\) to each nonradial mode depending on its degree \(\ell_i\). Additionally, we performed this exercise with and without pooling of the linewidths \(\Gamma_i\). However, we found that, when linewidths were fitted on a per-mode basis, the posterior distributions of the linewidths for \edit3{specifically the modes with the lowest signal-to-noise ratios} were very poorly constrained, and therefore permitted to take on unphysically large values at low amplitudes, given our uninformative prior. \edit2{It is for this reason that, in other mode-fitting pipelines, the linewidths are often instead constrained to vary slowly as a function of frequency \citep[e.g.][]{appourchaux_oscillation_2016,kuszlewicz_bayesian_2019,hall_weakened_2021}, rather than permitted to each vary freely. \cite{kamiaka_reliability_2018} also find systematic biases in the mode heights and rotational widths when the linewidths are permitted to each vary freely, if peaks in the power spectrum are close to overlapping.} Since \edit3{we do not derive any physical interpretation from these linewidths}, and in any case their dependence on the mode frequency \edit2{being} weak at the low frequencies that we consider here \citep[e.g.][]{vrard_amplitude_2018}, we restrict our attention to results derived with a single linewidth \(\Gamma\) being fitted against all modes, to simplify our analysis.

With this parameterisation, these priors, and the standard likelihood function for describing values of the power spectrum \(P_j\) by the \(\chi^2\) distribution with 2 degrees of freedom \citep{anderson_modeling_1990},
\begin{equation}
    \mathcal{L}(\theta) \propto \exp \left[-\sum_j \left(\log f(\nu_j, \theta) + {P_j \over f(\nu_j, \theta)}\right)\right],\label{eq:lnlike2dof}
\end{equation}
in hand, we generate draws from the posterior distribution, constrained by TESS, using the nested sampling algorithm, as implemented in the \texttt{dynesty} Python package. In \cref{fig:asteroseismology}b, we show 100 draws from the posterior distribution overplotted on the power spectrum, with contributions to \cref{eq:model} at different degree \(\ell\) indicated with different colours of curves. The averages of these samples are shown with a colour-separated echelle power diagram in \cref{fig:asteroseismology}c, for easier graphical comparison with our mode identification by hand.

More details about the results of this procedure can be found in the Appendix, where we show the joint posterior distributions for all pooled seismic quantities, as well as the parameters of our combined red- and white-noise background model (\cref{fig:shared}). There, we also report the medians and the widths of the \(1\sigma\) credible regions of the posterior distributions for the fitted nonrotating (\(m=0\)) mode frequencies in \cref{tab:peakbag}, \edit1{and evaluate the statistical significance of our asteroseismic detection}. From our radial mode frequencies alone, we re-fit \(\Delta\nu\) to yield \(1.21 \pm 0.01\ \mu\)Hz.

\begin{figure}
\centering
\includegraphics{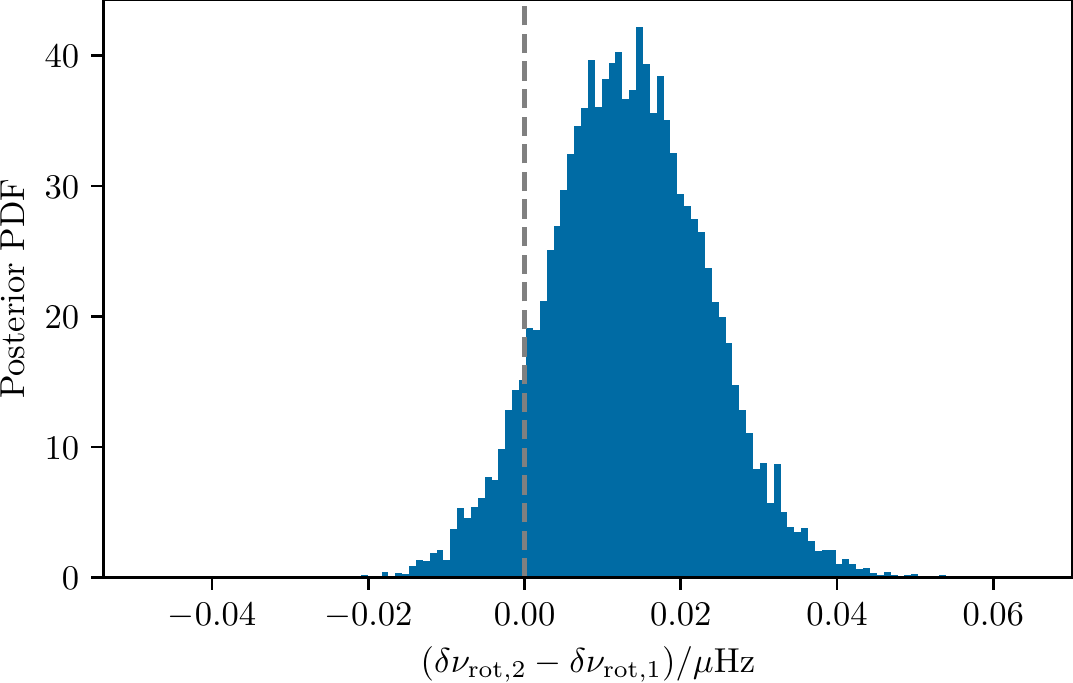}
\caption{Rough estimate of rotational shear, characterised through the marginal posterior distribution for the difference \(\delta\nu_{\text{rot},2} - \delta\nu_{\text{rot},1}\) in the widths of the multiplet splittings for modes of different degree. The distribution can be seen to be a little more than 1\(\sigma\) discrepant with uniform rotation.\label{fig:shear-obs}}
\end{figure}

For rotation in particular, we find that the averaged \(\ell = 1\) and \(\ell = 2\) rotational splittings to be \(\delta\nu_{\text{rot}, 1} = 0.091 \pm 0.007\ \mu\text{Hz}\) and \(\delta\nu_{\text{rot}, 2} = 0.104\pm0.007\ \mu\text{Hz}\), with a corresponding pooled inclination of \(i_\star = {84 ^{+4}_{-5}} ^\circ\). Both rotational splittings are larger than the pooled linewidth of \(\Gamma = 0.07\pm0.01\ \mu\text{Hz}\). Moreover, we note that the marginal posterior distribution in the difference between the dipole and quadrupole rotational splittings, which we show in \cref{fig:shear-obs}, suggests that they are discrepant from each other, which in turn would indicate spatial variations in the rotation rate. We examine this in more detail in \autoref{sec:inversion}.

\edit3{We must caution that our reported seismic quantities are contingent on the mode identification that we have assigned to the peaks in the power spectrum. While we have described how some alternative mode identifications (e.g. at different evolutionary stages) may be unphysical, the numerical machinery required to exhaustively denumerate and account for all plausible alternative choices of mode identification does not yet exist. For this reason, all of our reported quantities must be interpreted --- as is customary in the analysis of solar-like oscillators --- as being derived from a conditional distribution on the seismic parameters (i.e. conditioned on our choice of mode identification).}

\begin{figure*}
\centering
\annotate{\includegraphics[width=.95\textwidth]{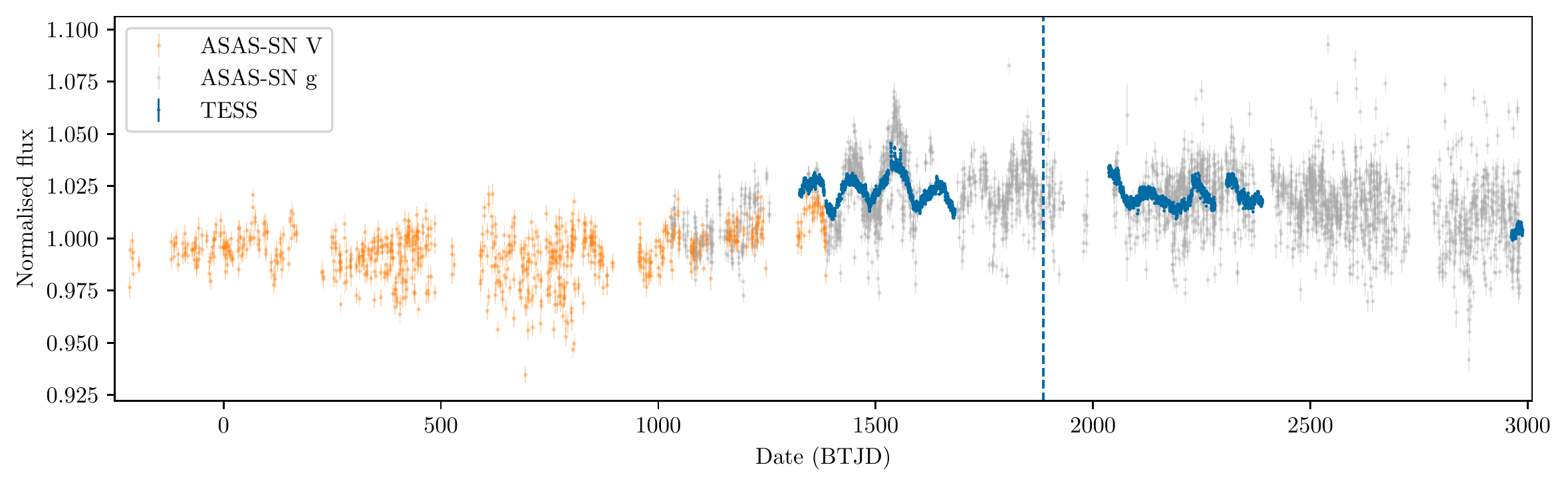}}{\node at (.95, .9){\textbf{(a)}};}
\annotate{\includegraphics[width=.95\textwidth]{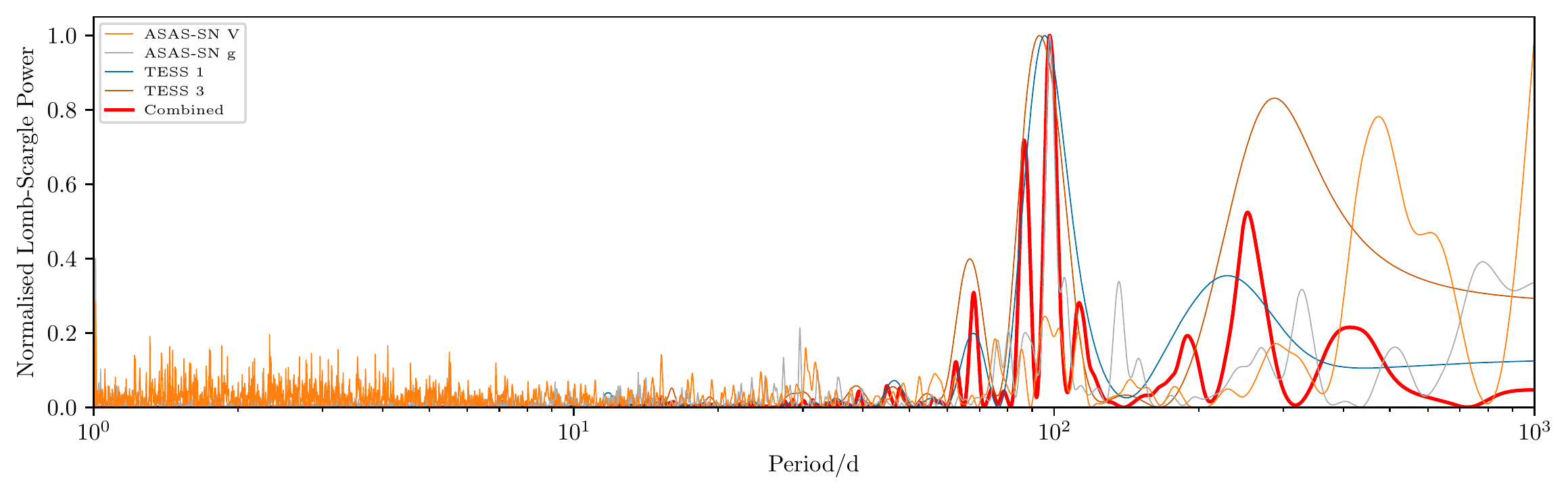}}{\node at (.95, .9){\textbf{(b)}};}
\annotate{\includegraphics[width=.95\textwidth]{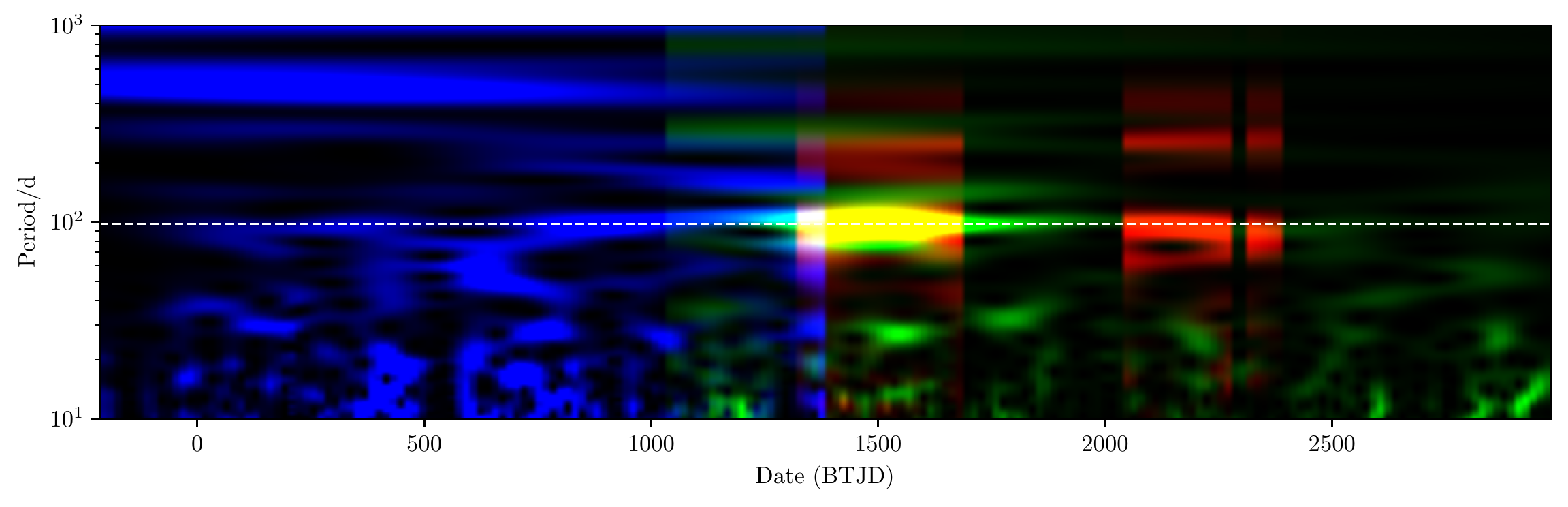}}{
\node at (.95, 1){\textbf{(c)}};
\draw[blue, <->, inner sep=1](.0615, 1) -- (.525, 1) node[midway, label=below:{\tiny ASAS-SN V}]{};
\draw[red, <->, inner sep=1](.507, 1.05) -- (.613, 1.05) node[midway, label=below:{\tiny TESS Cycle 1}]{};
\draw[red, <->, inner sep=1](.71, 1.05) -- (.816, 1.05) node[midway, label=below:{\tiny TESS Cycle 3}]{};
\draw[ForestGreen, <->, inner sep=1](.42, 1.1) -- (.988, 1.1) node[midway, label=below:{\tiny ASAS-SN g}]{};
}
\caption{Photometric characterisation of Zvrk. \textbf{(a)} Stitched TESS aperture photometry (blue), shown over data from ASAS-SN in $V$-band (orange) and $g$-band (gray). The date of the APOGEE visit is shown with the vertical dashed line. \textbf{(b)} Lomb-Scargle power spectral densities from different TESS CVZ cycles and ASAS-SN bandpasses, normalised to give unity at their maximum values. A combined power spectral density incorporating all data sets (having applied offsets shown in panel a) is also shown with the red curve. From the most prominent peak of this combined power spectrum we infer a characteristic period of 99 days. \textbf{(c)} Color-separated frequency-time power diagram, with the intensity at each pixel showing the Lomb-Scargle-wavelet-transform power associated with a given time and oscillation period. Different colour channels correspond to different instruments, with TESS shown in red, ASAS-SN g in green, and ASAS-SN V in blue. A nominal oscillation period of 99 days, from our global Lomb-Scargle analysis, is marked out with the horizontal dashed line. An animated version of this panel, to accommodate colourblind readers, is available in the online edition of this paper. \edit1{The animation shows the wavelet power diagram for ASAS-SN V-band, then ASAS-SN g-band, and then TESS, separately normalised with the Viridis colourmap but on the same frequency-time axes, for 3 seconds each.} \label{fig:photometry}}
\end{figure*}
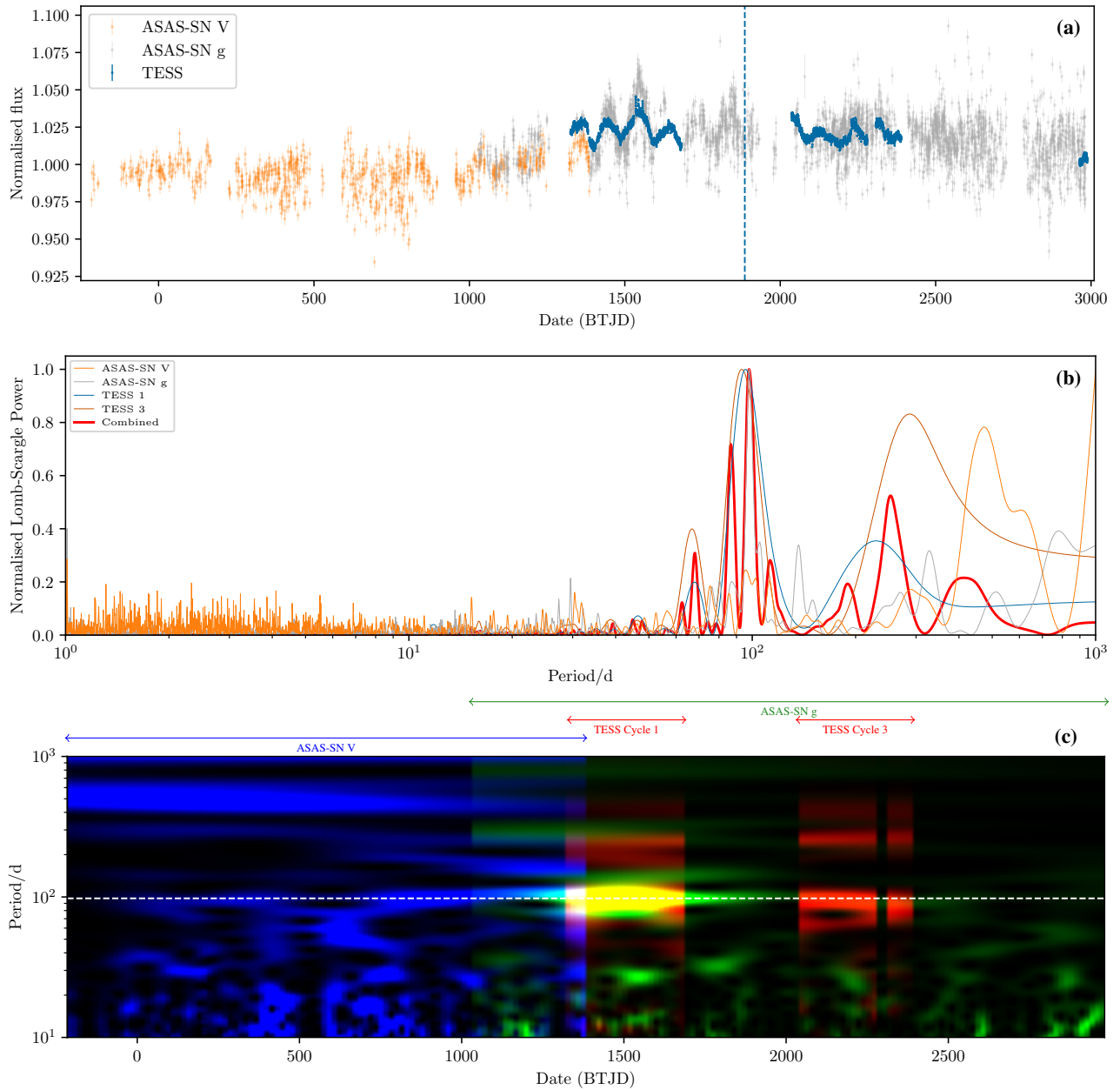

\subsection{TESS and ASAS-SN Photometry}\label{tess-and-asas-sn-photometry}

\label{sec:photometry}

In addition to asteroseismology, \emph{Kepler} data suggest that we should in principle also be able to derive photometric surface rotation rates for rotating red giants. The rotational frequencies obtained from our seismic analysis correspond to rotational periods of potentially up to 126 days (from our slower dipole-mode rotational frequency), spanning multiple 27-day sectors. However, the short-cadence PDC-SAP lightcurves from which our asteroseismic analysis above is derived have been aggressively detrended to eliminate long-term temporal variability. As a secondary result of this detrending, discontinuities are also introduced between sectors. In consequence of these, rotational periods longer than 27 days cannot be observed directly from this data product \citep[cf.][]{avallone_rotation_2022}. Alternative techniques \citep[e.g.][]{lu_astraea_2020, claytor_recovery_2022} rely on pattern-matching against single sectors and spectroscopic measurements, rather than attempting direct measurement. Given that Zvrk is both rotationally and spectroscopically unusual, and therefore unlikely to be well-described by the training sets of machine-learning pattern matching, we opt to undertake direct measurement instead.

We thus construct our own custom aperture-photometry light curves, correcting for TESS instrumental systematics by applying the pixel-level decorrelation technique \citep[PLD:][]{deming_spitzer_2015}, as implemented in the \texttt{lightkurve} Python package, to the TESS target pixel files. While this suffices to recover slow variability within each TESS sector, the long rotational periods under consideration also require us to stitch TESS sectors together in a principled fashion. Again, since our asteroseismic analysis suggests rotational periods which are far longer than the duration of a single sector, we opt to do this stitching by naive linear extrapolation. Specifically, we fit a linear function to the last 5 days of each sector, and extrapolate the fitted function to the first 5 days of the following sector. The observed flux in the following sector is normalised such that its median flux matches the median value inferred based on the extrapolatation. This procedure is repeated for each consecutive sector. We show the results of this procedure with the blue points in \cref{fig:photometry}. Note that this procedure relies on there not being gaps between sectors, and so the missing sector 36 in Cycle 3 results in sectors 37-39 not being stitched to the preceding sector 35. Nonetheless, a clear periodic variation can be seen to emerge in Cycle 1, all of whose 13 consecutive sectors are successfully treated with this procedure, with a period of roughly 100 days. Likewise, similar periodicity can be seen in the first 9 consecutive sectors of Cycle 3, albeit at reduced amplitude.

This periodicity persists across variations in our stitching procedure, e.g.~omitting the first day of each sector in the linear fit and renormalisation, or using Gaussian process regression for sector stitching. Nonetheless, to rule out these variations having been caused by by latent, unaccounted TESS instrumental systematics, or artificially and inadvertently introduced by these choices of detrending or stitching procedures, we also examined Zvrk through independent ground-based photometry. We do so by way of the ASAS-SN network of ground-based telescopes \citep{shappee_asassn_2014, kochanek_asassn_2017, hart_asassn_2023}. Since Zvrk is fairly bright (\(m_V = 10.24\)), we used aperture photometry with a wider aperture than standard: we construct lightcurves over a 3-pixel-wide aperture, and treat each dither exposure as an independent point in the time series. The residual instrumental systematic perturbations which affect each of the ASAS-SN detectors (aside from lunation and annual variations) remain temporally uncorrelated from those afflicting TESS. Thus, over longer timescales we may still rely on ASAS-SN to anchor our absolute photometric normalisation. We compare in \cref{fig:photometry}a the normalised ASAS-SN \(V\)-band and \(g\)-band photometry with our TESS lightcurves, choosing to display each consecutive run of TESS sectors in such a way as to match the median normalisation of the contemporaneous ASAS-SN \(g\)-band data points, which as a whole are themselves shown median-normalised. The ASAS-SN \(V\)-band data have also been scaled to yield equivalent median normalisation over the overlap period where contemporaneous observations exist with \(g\)-band.

The ASAS-SN data exhibit periodicity contemporaneous with that shown in TESS Cycle 1, across multiple cameras, confirming the photometric variations first seen from the TESS data. In the small period of overlap between TESS Cycle 1 and ASAS-SN \(V\)-band coverage, the temporal variations in the ASAS-SN data can also be seen to be apparently in phase with the TESS modulation signal. Furthermore, this periodicity can be seen to continue in the ASAS-SN data for at least two more rotational periods after the end of TESS Cycle 1 (i.e.~after a BTJD of about 1690), suggesting the persistence of any surface morphology causing the rotational signal.

Many techniques exist for the derivation of the period of this kind of quasiperiodic variability. For our analysis, we rely on the standard Lomb-Scargle technique \citep[e.g.][ and references therein]{lomb_ls_1976, scargle_studies_1982, vanderplas_understanding_2018}, since it allows us to combine multiple data sets with different temporal sampling characteristics, and has already been used extensively in the literature. We show in \cref{fig:photometry}b the normalised Lomb-Scargle periodograms for different subsets of the data shown in panel (a), and additionally a periodogram generated from the combined data set, shown with the thick red line. We can see that all of the data sets exhibit a large peak at a period of between 90 to 100 days. Interpreting the combined periodogram as being a log-likelihood function, we use the inverse curvature of this peak (or, equivalently, its half-width) as an estimate of the uncertainty on its location \citep{ivezic_statistics_2014, vanderplas_understanding_2018}, and in this fashion find it to be at \(99 \pm 3\) days. The scatter in the location of this peak in the other periodograms shown is 2 days, which we adopt as an estimate of variability between our data sets. To estimate the systematic uncertainty owing to our choice of methodology, we also fit a Gaussian process using a quasiperiodic kernel against the ASAS-SN \(g\)-band data set, obtaining a periodicity of \(104 \pm 7\) days. We adopt the discrepancy between the two techniques as an estimate of our systematic error. Thus, we report a rotation rate of \(99 \pm 3 \text{(stat)} \pm 5 \text{(sys)}\) days.

The peak-to-peak amplitude of this variability varies between 2-4\% depending on the instrument measuring it. The larger amplitude of the modulation as observed by ASAS-SN compared to TESS reinforces an interpretation of this signal as originating from a rotational modulation. A fixed temperature contrast between cool surface features, such as spots or large-scale convective patterns \citep[e.g.][]{paladini_large_2018}, and the surrounding stellar photosphere, will yield a greater reduction in intensity compared to the rest of the disk when observed in a bluer bandpass than in a redder one, and ASAS-SN's \(V\)-band (\(\lambda \sim 540\ \mathrm{nm}\)) and \(g\)-band (\(\lambda \sim 477\ \mathrm{nm}\)) observations are indeed bluer than TESS (\(\lambda \sim 787\ \mathrm{nm}\)). If these features are indeed convective in origin, then we should expect them to evolve over the course of the convective turnover timescale. Alternatively, if they arise from near-surface magnetism like that seen in active main-sequence dwarfs, we should also expect their shapes and positions to change over many rotational periods. Either way, this will affect the properties of the observed photometric modulations. The shapes and sizes of such features will affect its amplitude, and their positions will, in the presence of latitudinal differential rotation, modify its apparent period.

The prominence of this photometric rotational signal and, to some extent, its period, do indeed appear to vary over time. The peak of the Lomb-Scargle periodogram for TESS Cycle 1 is located at a period of 96 days, and that of Cycle 3 at 93 days, suggesting some temporal variations in the putative rotational frequency. We investigate this with more detail in \cref{fig:photometry}c, which shows a frequency-time power diagram generated using the Lomb-Scargle wavelet power spectrum utility function of the ASAS-SN Sky Patrol client \citep{hart_asassn_2023}, over which we mark out a nominal period of 99 days with a horizontal line. The wavelet-transform power in different bandpasses is shown on this diagram in different colour channels, with display normalisation chosen to saturate each colour channel by a factor of 1.5, so as to enhance the visibility of lower-amplitude features. The peaks of 100 days in our periodograms in panel (b) correspond to a horizontal band of power on this frequency-time power diagram. Heuristically, the Lomb-Scargle periodograms (up to convolution against a kernel determined by the temporal resolution of this diagram) may be recovered by simply integrating over time. However, for the horizontal band at 99 days, such an integral can be seen to be largely dominated by an episode of large photometric amplitude centred at around BTJD 1500, which, fortuitously, was covered contemporaneously by ASAS-SN \(g\)-band, TESS, and (at least at its beginning) ASAS-SN \(V\)-band. Moreover, while this rotational peak is substantially less prominent in the ASAS-SN \(V\)-band periodogram compared to the others, our wavelet analysis reveals that this is because of time evolution: oscillatory variability at this period can be seen to emerge only after BTJD 700 or so, more than halfway through our temporal coverage in \(V\)-band.

\subsection{Spectroscopy}\label{spectroscopy}

\label{sec:spec}

To further test the interpretation of our quasiperiodic signal as being of rotational origin, we subject Zvrk to the LEOPARD two-component fitting procedure described in \citet{cao_starspots_2022}, applied to an APOGEE spectrum taken during a visit to Zvrk made at MJD 58886 = BTJD 1886, at a time where contemporaneous ASAS-SN \(g\)-band coverage indicates the persistence of the photometric modulation signal (vertical line in \cref{fig:photometry}a). The LEOPARD procedure returns a disk area coverage fraction of \(f_\text{spot} = 0.02\), and a temperature contrast of \(x_\text{spot} = T_\text{spot}/T_\text{surf} = 0.8\). Assuming that secondary component corresponds to a single large, morphologically concentrated, surface feature, its coverage fraction and intensity contrast would produce a roughly \(2\%\) photometric variability amplitude, which is consistent with that observed in \autoref{sec:photometry}. However, the presence of only a single APOGEE visit prevents us from assessing properties of the evolution of these putative surface features.

To supplement our asteroseismic measurement, spectroscopic measurements of e.g.~the stellar metallicity and effective temperature are also necessary as inputs into stellar modelling for more precise constraints on stellar properties. Such measurements for Zvrk already exist in the APOGEE DR17 catalogue \citep[wherein it is found as 2MASS ID 05592585-5911542]{apogeedr17}, derived from fitting a spectral template (given values of \(\teff\), \(\log g\), and abundances) against the spectrum taken during its single APOGEE visit. Zvrk's fitted values from DR17 are \(\teff = 4320 \pm 80\ \mathrm{K}\), \(\log g_\text{spec} = 2.11 \pm 0.08\ \mathrm{dex}\)\edit1{, $\mathrm{[Fe/H]} = -0.24 \pm 0.08$ dex}, and \(\mathrm{[M/H]} = -0.21 \pm 0.08\) dex.

This value of the effective temperature, in conjunction with the solar-calibrated scaling relation \(\numax \sim g / \teff\), in turn yields \(\log g_\text{seis} = 1.76 \pm 0.03\), which is significantly lower than the APOGEE DR17 value. This is likely a result of rotation: \(\log g\) determines the amount of pressure broadening applied to the template spectrum, which is highly degenerate with rotational broadening \citep[cf.][]{thompson_noninteracting_2019}; \edit1{the APOGEE pipeline in turn does not fit} rotational broadening simultaneously with \(\log g\) \edit1{for giants}.

To correct for this bias, as is often done in other works combining asteroseismology with spectroscopy, we may in principle employ our asteroseismic measurement of \(\log g\) as a direct constraint on the template-fitting procedure, and iterate between fitting for the remaining quantities with it held fixed, and refining asteroseismic \(\log g\) using the refitted \teff, until the fitted values converge. However, this iterative procedure is typically performed with far more precise constraints on \(\numax\) than we have available here. Conversely, using our asteroseismic surface gravity here \edit1{only slightly changes the} fitted temperature and metallicity\footnote{\edit1{The APOGEE data quality flags \texttt{ascap\_flag}, \texttt{m\_h\_flag}, and \texttt{teff\_flag} are all 0 (i.e. nominal): only the \texttt{logg\_flag} indicates an issue in deriving the final reported value, as a consequence of excess line broadening.}}.
We therefore adopt the nominal APOGEE values for \(\teff\) and metallicity, and their uncertainties, in our subsequent analysis.

We may then estimate a spectroscopic \(V \sin i\) through the differential Doppler-broadening technique also used in \citet{tayar_spinning_2022}, which estimates the optimal amount of additional broadening required to reconcile the fitted spectral template with the observed APOGEE spectrum. Holding \(\log g\) fixed at its seismic value for this purpose, we obtain that \(V \sin i = 10.1 \pm 0.7\ \mathrm{km s^{-1}}\).

\begin{figure*}[htbp]
    \centering
    \annotate{\includegraphics[height=3.2 in]{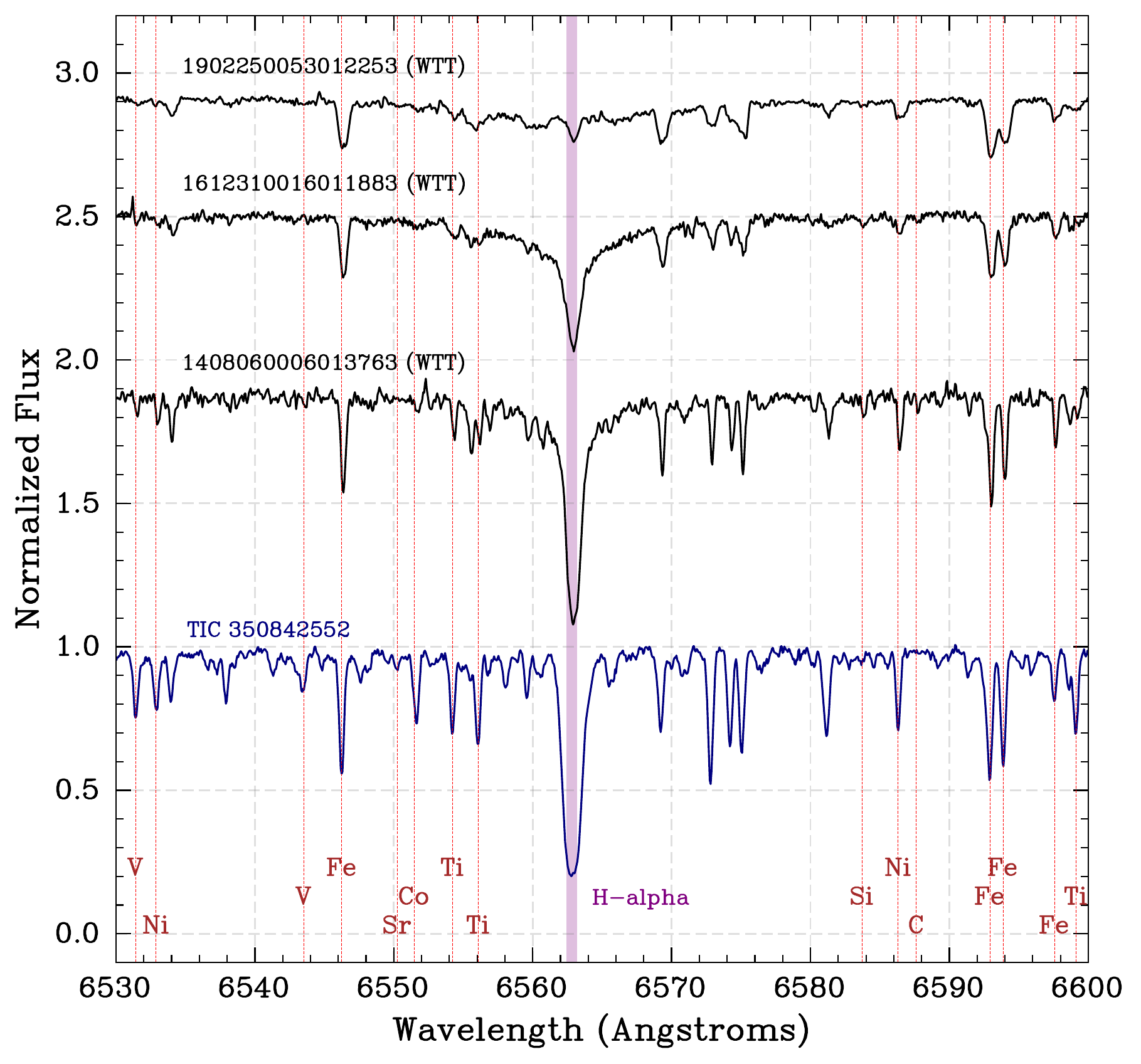}}{\node at (.9, .95){\textbf{(a)}};}
    \annotate{\includegraphics[height=3.2 in]{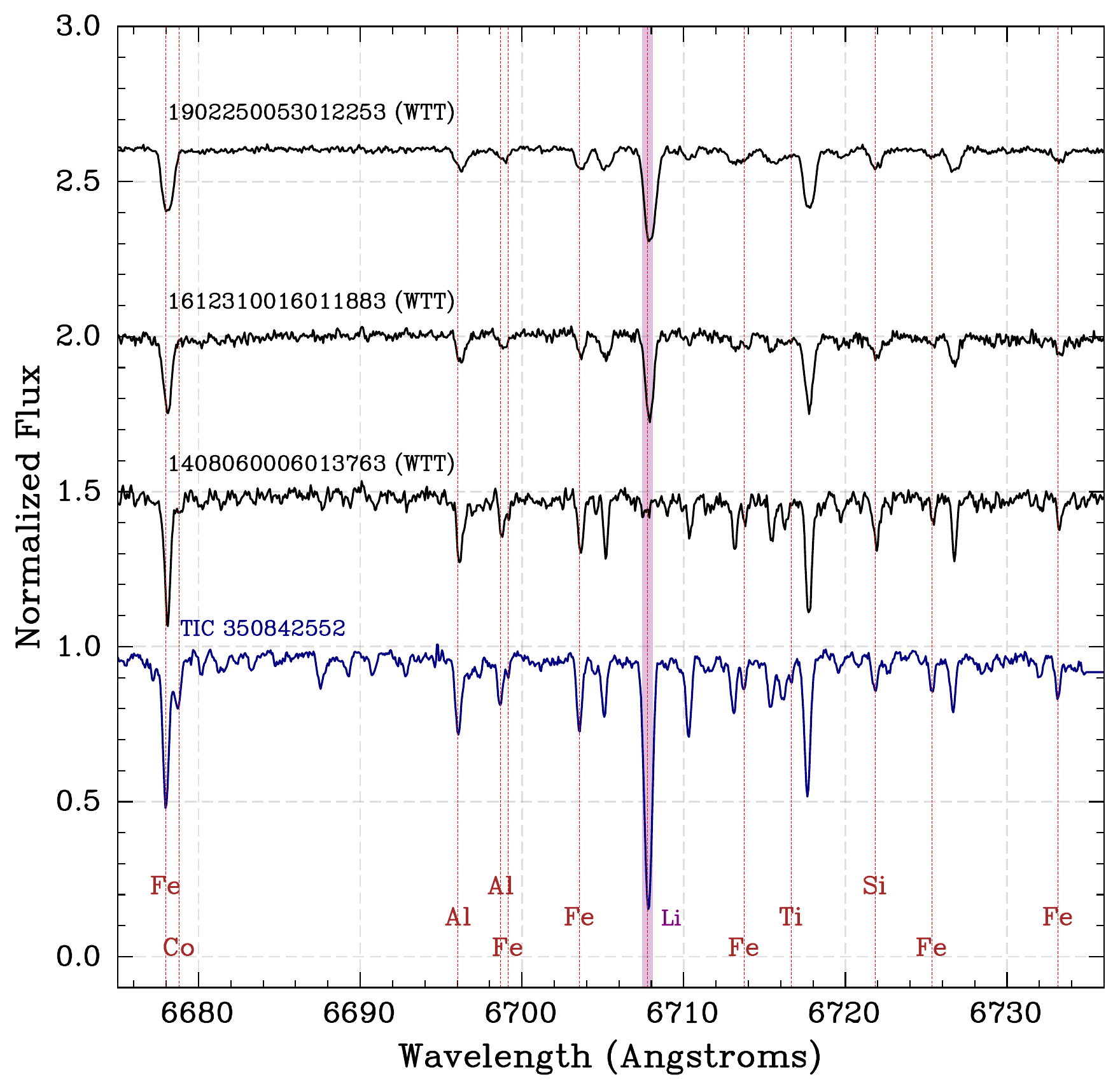}}{\node at (.9, .95){\textbf{(b)}};}
    \caption{Excerpts of high-resolution GALAH spectra, showing \textbf{(a)} H$\alpha$, and \textbf{(b)} lithium absorption features. Zvrk's spectrum is shown with the blue solid curve, while comparisons with weak-lined T Tauri stars from \cite{xing_lithium_2010} are shown in black.}
    \label{fig:abund}
\end{figure*}

Several spectral features which are also important to our analysis --- specifically, pertaining to Zvrk's H\(\alpha\) (\cref{fig:abund}a) and lithium (\cref{fig:abund}b) absorption features --- lie at optical \edit1{wavelengths}, outside of APOGEE's near-IR coverage. Thus, in addition to APOGEE spectra, we also obtained a high-resolution GALAH spectrum \edit1{(in which catalogue it is found by its Gaia DR3 ID 4764985981182798976)} covering \edit1{those optical wavelengths}. \edit1{The stellar parameters returned by the GALAH analysis pipeline \citep{buder_galah_2021} are} \(\log g = 1.7\pm 0.2\ \mathrm{dex}\), \(\teff = 4250 \pm 100\ \mathrm{K}\), \(V\sin i = 9.5 \pm 1.0\ \mathrm{km s^{-1}}\), and \(\mathrm{[Fe/H]} = -0.28 \pm 0.05\ \mathrm{dex}\). We adopt the APOGEE values in our asteroseismic analysis so as to minimise systematic discrepancies between our derived seismic quantities and, e.g., those previously determined for members of the APOKASC sample \citep{serenelli_apokasc_2017, pinsonneault_apokasc_2018, pinsonneault_apokasc_2023}. Nonetheless, the GALAH spectrum will be critical for our subsequent abundance analysis.

\section{Derived Stellar Properties}\label{derived-stellar-properties}

\subsection{Evolutionary Identification}\label{evolutionary-identification}

\label{sec:ttauri}

Our preceding asteroseismic analysis favours an interpretation of Zvrk as being a first-ascent red giant. However, \edit1{if we were to set this asteroseismology aside, or assume it to be spurious, Zvrk's} combination of high surface rotation, strong lithium absorption line, and low mean density may also be compatible with it being a pre-main-sequence T Tauri star. Were Zvrk actually to be a T Tauri star, we should also expect to see strong H\(\alpha\) line emission in its spectrum originating from an extensive circumstellar disk, with substantial amounts of ionised hydrogen present. Contrary to this, Zvrk's H\(\alpha\) line, measured with GALAH, can be seen to be a very deep absorption feature (\cref{fig:abund}a), comparable to other red giants (both first-ascent and core-helium-burning), rather than the emission expected of T Tauri stars. Morever, occultation by circumstellar material tends to produce irregular photometric variability with a characteristic ``dipping'' pattern, as opposed to the quasiperiodic variability shown in \autoref{sec:photometry}. In any case, the modulation amplitude of a few percent that we measure in \autoref{sec:photometry}, using multiple instruments, is also an order of magnitude smaller than that typically produced by such occultation.

H\(\alpha\) absorption and low photometric amplitudes do not entirely rule out Zvrk being a T Tauri star, as certain ``weak-lined'' T Tauri stars are known to exist, which lack significant disks, and, therefore, H\(\alpha\) emission. However, were Zvrk actually to be a weak-lined T Tauri star, its dynamics would also be extremely anomalous. While rapid for a red giant, Zvrk rotates an order of magnitude more slowly than essentially all rotational measurements of weak-lined T Tauri stars in the catalogue of \citet{bouvier_gaia_2016}. A comparably slowly rotating T Tauri star would therefore have to be much younger than their sample, drawn from NGC 2264 at 3 Myr old. This is also suggested by Zvrk's surface gravity, which is much lower than typical for a T Tauri star. However, Zvrk's lithium equivalent width of \(100\ \mathrm{m\AA}\) is lower than that of the NGC 2264 sample, rather than higher as we would expect from a younger object. The only candidate weak-lined T Tauri star of comparable rotational period and lithium abundance that we were able to find in the literature, V501 Aur, was later shown to be actually a misclassified red giant \citep{vanko_nature_2017}.

Since T Tauri stars are also known to be preferentially associated with natal comoving groups, such a young object would be likely found close by to other comoving objects. However, using the \texttt{Comove} software package \citep{kraus_comove_2022}, we are unable to find any co-moving sources (i.e.~with systemic velocities within 5 \(\mathrm{km\ s^{-1}}\) of Zvrk's) within a 25 pc search radius centred on Zvrk in the Gaia DR3 catalogue \citep{gaiadr3}. Thus, in addition to being weak-lined, and both rotationally and chemically unusual, Zvrk would also have to be kinematically anomalous for consistency with being a T Tauri star.

As such, although its chemical and rotational properties would make Zvrk an unusual red giant, our above discussion suggests it would be an even more unusual T Tauri star. Accordingly, we will proceed in our following analysis on the assumption that Zvrk is indeed a first-ascent red giant\edit1{, consistently with our asteroseismic analysis}.

\subsection{Detailed Asteroseismic Modelling}\label{detailed-asteroseismic-modelling}

\label{sec:opt}

Given the global asteroseismic properties \(\Dnu\) and \(\numax\), and our spectroscopic effective temperature, the so-called ``direct method'' inverts their usual scaling relations to produce mass and radius estimates. We do so in a similar fashion to the APOKASC3 catalogue of solar-like oscillators \citep{pinsonneault_apokasc_2023} in the APOGEE DR17 sample, including incorporating a structural correction factor \(f_{\Dnu} = 0.97\) via the prescription of \citet{sharma_stellar_2016}, and obtain for Zvrk that \(M_\star = (1.181 \pm 0.15) M_\odot\), \(R_\star = (23.7 \pm 1.1) R_\odot\).

The uncertainties on these estimates are large, as \(\numax\) is difficult to measure for evolved stars with comparatively few visible modes; the scaling relations themselves may also not be reliable at advanced stages of evolution (e.g.~\citealt{zinn_testing_2019}, whence the need for ad-hoc corrections like \(f_{\Dnu}\) in the first place). We turn to detailed modelling of stellar structure, as constrained by our measurements of individual mode frequencies, for more refined estimates of these properties. \edit1{As this modelling is done under standard assumptions about single-star evolution and chemical mixtures, it is unlikely to be representative of Zvrk's internal physics and absolute evolutionary age, given its rapid rotation and chemical anomalies. Nonetheless, such modelling will still meaningfully inform us about its present properties, evolutionary timescales, and internal structure.}

\begin{longtable}[]{@{}rcc@{}}
\caption{Global properties adopted in stellar modelling.\label{tab:t1}}\tabularnewline
\toprule\noalign{}
Constraint & Value & Reference \\
\midrule\noalign{}
\endfirsthead
\toprule\noalign{}
Constraint & Value & Reference \\
\midrule\noalign{}
\endhead
\bottomrule\noalign{}
\endlastfoot
\(\teff\) / K & \(4320 \pm 80\) & APOGEE DR17 \\
\(\mathrm{[M/H]}\) & \(-0.21 \pm 0.08\) & APOGEE DR17 \\
\(L / L_\odot\) & \(174 \pm 10\) & \citet{gaiadr2} \\
\(\Dnu/\mu\mathrm{Hz}\) & \(1.21 \pm 0.01\) & This work \\
\(\numax/\mu\mathrm{Hz}\) & \(7.5 \pm 0.3\) & This work \\
\end{longtable}

We use the implied \(m = 0\) mode frequencies from the \edit1{nested-sampling} procedure that we described in \autoref{sec:nested}, which we list in full in \cref{tab:peakbag} in the appendix, as the seismic inputs in our modelling. Supplementing these mode frequencies are constraints on some global stellar properties, which we list in \cref{tab:t1}. Using these, we search for stellar models which match the global properties and individual mode frequencies by minimising the sum-of-squares discrepancy function, where for a given stellar model described by parameters \(\theta\) we have
\begin{equation}
\begin{aligned}
    \chi^2_\text{tot}(\theta) &= \left(\teff(\theta) - \teff_\text{obs} \over \sigma_\teff\right)^2  +\left({\mathrm{[M/H]}}(\theta) - {\mathrm{[M/H]}}_\text{obs} \over \sigma_{\mathrm{[M/H]}}\right)^2 \\
    &+\left(L(\theta) - L_\text{obs} \over \sigma_L\right)^2 + \chi^2_\text{seis}(\theta),\label{eq:chi2}
\end{aligned}
\end{equation}
where \(\chi^2_\text{seis}\) is a cost function incorporating constraints from the individual mode frequencies, whose precise form depends on how we choose to correct for systematic errors in the near-surface physics of our stellar models (the asteroseismic ``surface term''). We tested both the parametric surface-term correction of \citet{ball_correction_2014}, and the nonparametric surface-insensitive penalty function of \citet{roxburgh_asteroseismic_2016}. As our results do not change significantly between the two, we report results obtained with the latter, which provides slightly larger uncertainties \citep{ong_differential_2021}.

Importantly, neither \(\Dnu\) nor \(\numax\) enter directly into the constraints derived in this fashion: for this exercise, they affect only ancillary properties of our search for best-fitting parameters \(\theta = (M_\star, Y_0, Z_0, \amlt, t)\). We search for stellar models using the differential-evolution optimisation scheme, as implemented in \citet{mier_yabox_2017}, which requires us to supply bounds on our search space. Based on our scaling-relation estimate for the seismic mass, we restrict our attention to \(M_\star/M_\odot \in [1.05, 1.3]\); we also restrict our attention to stellar models within \(\pm 25\%\) of our adopted value of \numax (more than \(\pm6\sigma\)), as a proxy for evolution up the RGB. For the remainder of our parameters, we adopt large bounds of \(Y_0 \in [0.248, 0.32]\), \(Z_0 \in [0.008, 0.02]\), and \(\amlt \in [1.7, 2.1]\).

For each trial \(\theta\), we generate an evolutionary track of stellar models using \mesa~r12778, using the chemical mixture of \citet{grevesse_standard_1998}. Stellar models were generated without overshooting or mass loss, and with an Eddington-gray model atmosphere. While we include the effects of the diffusion and settling of helium and heavy elements, we vary the overall mixing coefficient as a function of stellar mass according to the prescription of \citet{viani_investigating_2018}, in order to smoothly disable it at masses substantially higher than 1.25 \(M_\odot\). For models on each evolutionary track that lie within our permitted range of \(\numax\) values, we compute the radial p-mode frequencies, and those of the nonradial \(\pi\) modes as isolated using the prescription of \citet{ong_semianalytic_2020}, with the \gyre~stellar oscillation solver. We then run 400 iterations of the differential evolution optimisation algorithm, retaining all stellar models generated along the optimisation trajectory.

From optimisation, our best-fitting point estimate of \(\hat{\theta}_\text{MLE} = (\hat{M}, \hat{Y_0}, \hat{Z_0}, \hat{\amlt}, \hat{t})\) is \((1.18 M_\odot, 0.289, 0.0099, 1.863, 4.7\ \mathrm{Gyr})\). We will use this best-fitting model, and its associated evolutionary track, in our subsequent analysis requiring knowledge of Zvrk's temporal evolution or interior structure. More generally, to report uncertainties, we treat all of the models generated over the course of performing such optimisation as a set of nonuniform samples over our parameter space, \(\Theta = \{\theta_i\} = \{(M_i, Y_{0,i}, Z_{0,i}, \amlt_i, t_i)\}\). Associated with this set of samples are the values of a likelihood function defined through \cref{eq:chi2} as \(\mathcal{L}(\theta_i) = \exp[-\chi_\text{tot}^2(\theta_i)/2]\), evaluated for all \(\theta_i \in \Theta\). Were we to compute likelihood-weighted averages and quantiles of stellar properties with respect to \(\Theta\), as is often done with grid-based modelling \citep[e.g.][]{cunha_plato_2021}, we would effectively provide Bayesian posterior estimates of these properties and their uncertainties, with the prior distribution determined by the sampling function over \(\Theta\).

Such a procedure would yield estimates with respect to a uniform prior when used in grid-based modelling, where \(\Theta\) is sampled uniformly (typically using a discretised coordinate mesh or Sobol sequence). In our case, however, the sampling over \(\Theta\) is determined by our optimisation trajectory, which becomes increasingly densely sampled in the neighbourhood of the best-fitting point. Thus, a naive likelihood-weighted average would be prior-dominated in such a fashion as to both artificially select for the best-fitting result, and also systematically reduce the estimated uncertainties around it. Instead, it would be preferable for us to also return posterior estimates with respect to an uninformative, uniform prior over our input parameters, and to do this we must divide out the sampling function over \(\Theta\). We estimate the sampling function \(s(\theta)\) over \(\Theta\) using a kernel density estimator, and compute weighted quantiles with respect to modified likelihood weights \(w_i \propto \mathcal L(\theta_i) / s(\theta_i)\), with normalisation chosen to set \(\sum_i w_i = 1\). We report the resulting posterior medians and \(\pm 1\sigma\) quantiles for our input parameters, and Zvrk's radius, in \cref{tab:t2}. In addition, we provide values from a solar-calibrated model for comparison, where applicable.

As we might expect, the stellar properties that emerge from our detailed modelling are far more precise than, although still consistent with, the rough estimates obtained from scaling relations. Given this constraint on the stellar radius, and using our posterior median estimate of the stellar inclination, our spectroscopic rotational broadening implies a surface rotation rate of \(\Omega_\text{surf} / 2\pi = 0.099 \pm 0.007 \mu\text{Hz}\), which lies between the values that we obtain from the dipole and the quadrupole rotational splittings.

\begin{longtable}[]{@{}rcc@{}}
\caption{Global properties \edit1{and statistical errors} returned from stellar modelling with mode frequencies. \label{tab:t2}}\tabularnewline
\toprule\noalign{}
Quantity & Inferred Value & Remarks \\
\midrule\noalign{}
\endfirsthead
\toprule\noalign{}
Quantity & Inferred Value & Remarks \\
\midrule\noalign{}
\endhead
\bottomrule\noalign{}
\endlastfoot
\(M_\star/M_\odot\) & \(1.14^{+0.05}_{-0.03}\) & --- \\
\(R_\star/R_\odot\) & \(23.5^{+0.4}_{-0.2}\) & --- \\
\(\amlt\) & \(1.89^{+0.12}_{-0.08}\) & \(\alpha_\odot = 1.824\) \\
\(Y_0\) & \(0.273^{+0.03}_{-0.02}\) & \(Y_\odot = 0.268\) \\
\(Z_0\) & \(0.010 \pm 0.001\) & \(Z_\odot = 0.018\) \\
Model age/Gyr & \(5.4^{+1.0}_{-0.8}\) & Single-star Scenario \\
\end{longtable}

Finally, we consider systematic errors induced by variations on the techniques used for detailed modelling, which have been shown to be nontrivial for grid-based modelling of red giants \citep[e.g.][]{campante_detailed_2023}. We conducted a separate optimisation-based modelling exercise, with stellar models generated using \mesa~r15140, without diffusion and settling of helium and heavy elements, using a constant opacity in the atmosphere rather than the Eddington \(T-\tau\) relation, and with the chemical mixture of \citet{asplund_chemical_2009}. The mass-loss prescription of \citet{reimers_circumstellar_1975} was also used, with an efficiency parameter of \(\eta = 0.1\). For this exercise, only radial mode frequencies were incorporated into the asteroseismic constraint; these were computed from stellar models using the \texttt{ADIPLS} oscillation solver, and included in the penalty function \cref{eq:chi2} through the surface-term correction of \citet{ball_correction_2014}. A search for the best-fitting parameters was conducted via the \texttt{astero} module of \mesa, using the Nelder-Meade optimisation scheme, and with uncertainties estimated using a local Jacobian matrix evaluated by finite differencing on a bounding simplex around the best-fitting sampled value. This second optimisation exercise yields an estimate for the initial mass of \(1.142 \pm 0.050 M_\odot\) and a present mass of \(1.138 \pm 0.050 M_\odot\), as well as an estimated radius of \(23.44 \pm 0.11 R_\odot\); these are consistent with the values that we have reported in \cref{tab:t2}. The systematic differences between the initial and final mass, as well as between these results and \edit1{the masses and radii reported} in \cref{tab:t2}, appear to be dominated by the much larger statistical uncertainties. Since the values in \cref{tab:t2} include constraints from nonradial modes, we adopt them as our preferred values for subsequent analysis. We also adopt the best-fitting \mesa~model from this optimisation exercise as a point estimate of Zvrk's interior structure, which we will use in subsequent analysis.

\subsection{Abundance Analysis}\label{abundance-analysis}

\label{sec:abundance}

First-ascent red giants of comparable metallicity exhibit a tight correlation between stellar mass and the \([\mathrm{C/N}]\) \edit1{abundance} ratio\edit1{, as} increasing the stellar mass results in a more massive convective core, in which main-sequence nuclear processing occurs through the CNO cycle. At core hydrogen exhaustion, more massive stars will therefore have larger residual quantities of unreacted \(^{14}\mathrm{N}\) present, which is then redistributed into the envelope during first dredge-up.

\begin{figure}
\centering
\includegraphics{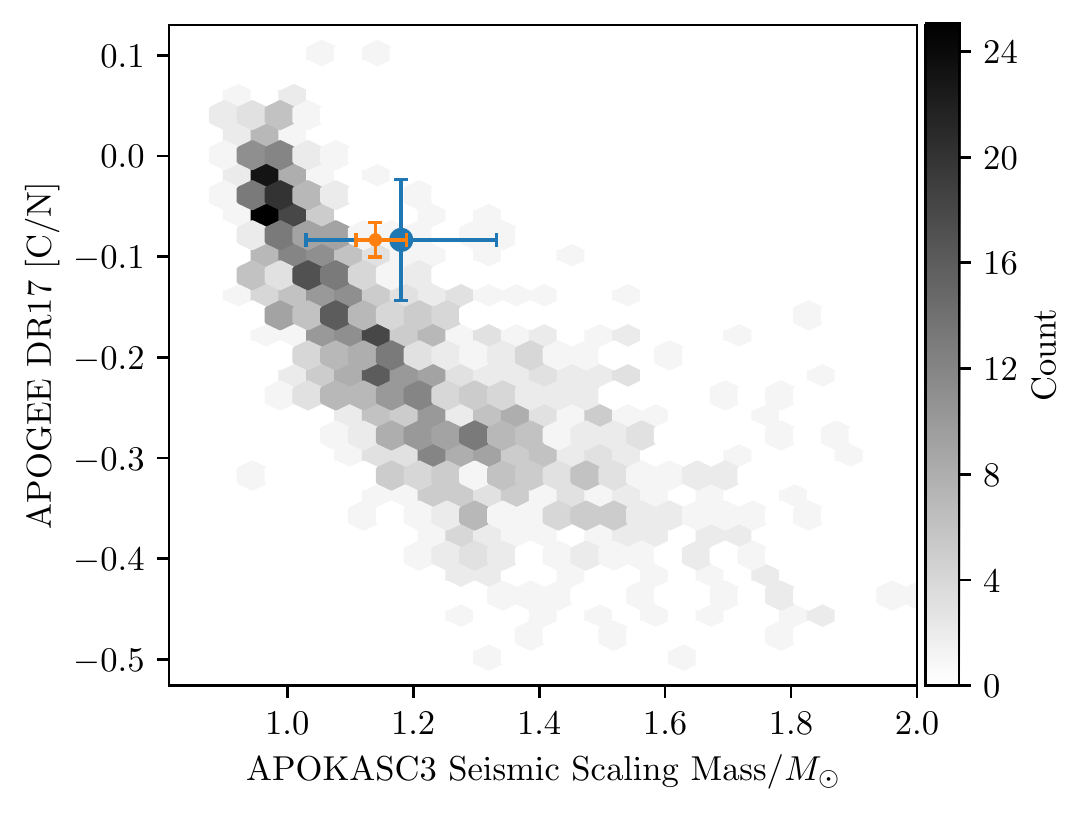}
\caption{Comparison of Zvrk to the seismic mass-\([\mathrm{C/N}]\) sequence of the APOKASC3 observational sample, with a metallicity cut of 0.1 dex around the nominal value for Zvrk, and restricting attention to only first-ascent red giant stars (per identification supplied by APOKASC3). Zvrk's seismic mass and relative abundances can be seen to lie above this sequence. We plot the location of Zvrk on this diagram using both its seismic mass constrained by detailed modelling against individual mode frequencies, and statistical error in {[}C/N{]} (orange), to indicate our actual precision of relative inference, but also show its scaling-relation mass (blue) and total uncertainty in {[}C/N{]}, for an apples-to-apples comparison against the scaling-relation masses in the APOKASC3 catalogue. \label{fig:cn}}
\end{figure}

We show this \edit1{mass-vs-[C/N]} sequence with APOKASC3 seismic masses \citep{pinsonneault_apokasc_2023} and APOGEE DR17 {[}C/N{]} ratios in \cref{fig:cn}, restricting our attention to only first-ascent red giants with \edit1{[M/H]} within 0.1 dex of Zvrk's, using the greyscale histogram bins. We indicate Zvrk's position on this diagram using its seismic mass from individual mode frequencies (\autoref{sec:opt}), and \([\mathrm{C/N}]= -0.08 \pm 0.017 \text{(stat)} \pm 0.057 \text{(sys)}\) from APOGEE DR17, with the orange data point, under the optimistic assumption that an intra-sample comparison of this kind --- with both instrumental and methodological homogeneity --- may be assessed by way of the statistical uncertainties on {[}C/N{]}. This combination of constraints can be seen to lie above the APOKASC3 mass-\edit1{[C/N]} sequence, in turn suggesting \edit1{either that Zvrk formed with an unusually high initial carbon enrichment, or} that the material in Zvrk's convective envelope is less nuclear-processed than would be typical, for an ordinary red giant of comparable mass, radius, and metallicity.

However, this would not be a comparison \edit1{of like with like, as the APOKASC3 seismic masses} were obtained by inverting scaling relations, rather than detailed modelling as we have done. Our only avenue for \edit1{such a} comparison is through the less precise scaling-relation mass, since only these and not detailed-modelling seismic masses are available (and for that matter, feasible) for the APOKASC3 catalogue. Moreover, while the intrasample systematic error on {[}C/N{]} will be smaller than the empirical absolutely-calibrated value of \citet{hayes_bacchus_2022} that we have quoted above, it will nonetheless be larger than the statistical error alone. We illustrate the most pessimistic combination of constraints on Zvrk --- scaling-relation mass and full systematic uncertainties --- with the blue data point in \cref{fig:cn}. While this lies further away from the \edit1{mass-[C/N]} sequence, the larger uncertainties render this only weakly suggestive.

\begin{figure}
\centering
\includegraphics{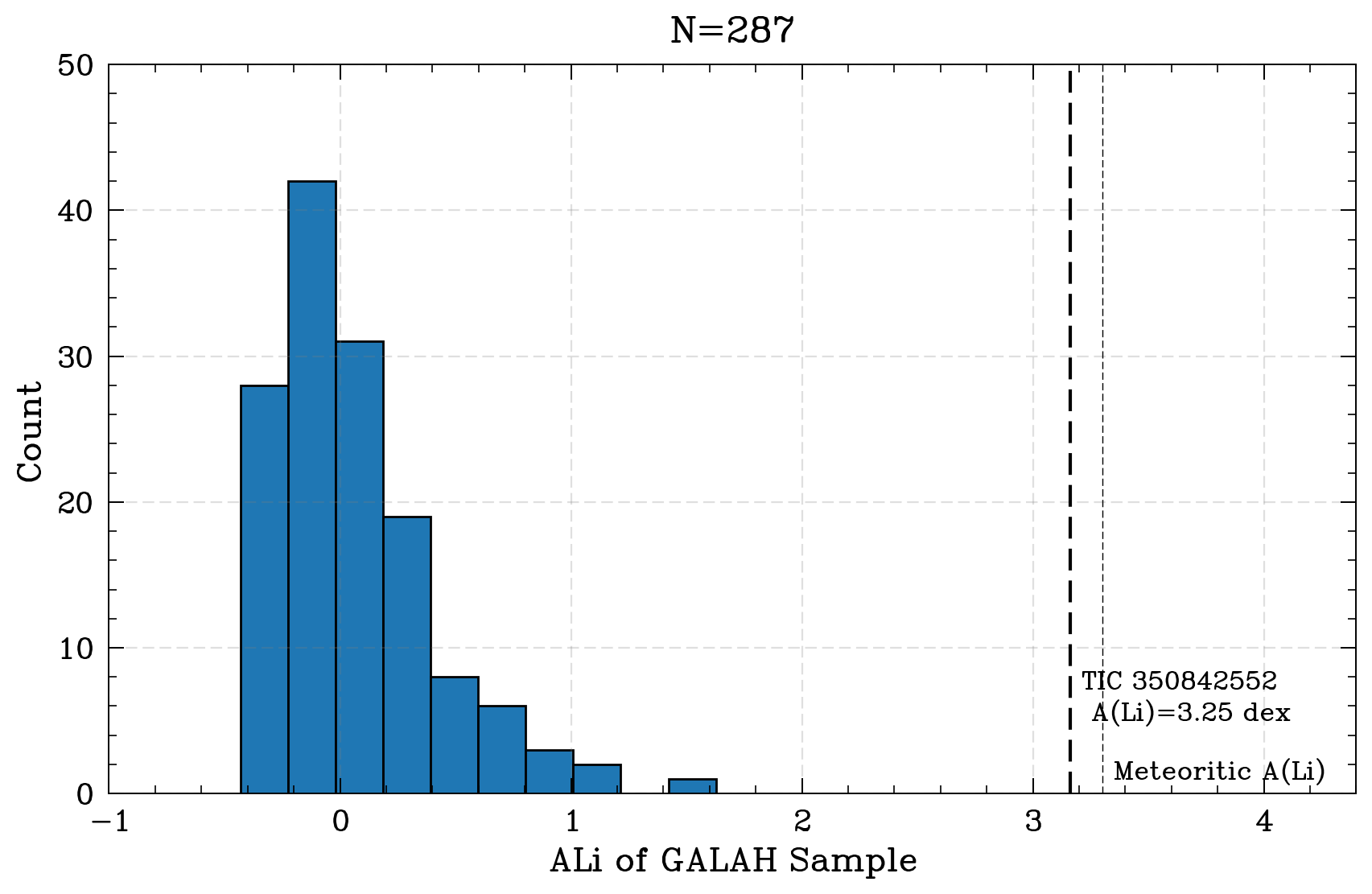}
\caption{Comparison of Zvrk's lithium abundance, shown with the dashed line, against the general population described by a control sample of GALAH stars. Specifically, we consider A(Li) measurements of stars whose reddening-corrected Gaia \(\mathrm{BP} - \mathrm{RP}\) colours were within \(1\sigma\) of Zvrk's, whose GALAH {[}Fe/H{]} values were within \(0.1\) dex of \edit1{Zvrk's value of $-0.28$}, whose Gaia absolute magnitudes were within \(1\sigma\) of Zvrk's, and whose Gaia DR3 Renormalised Unit Weight Errors (RUWE) were less than 1.4, to select single stars. This yielded a subsample of 287 stars, the histogram of whose distribution of lithium abundances is plotted with the blue bars. This control sample population has a median A(Li) of \(-0.02\) dex and a MAD of \(0.2\) dex, which is consistent with (if a little lower than) the baseline estimate provided in \citet{soaresfurtado_lithium_2021}.\label{fig:lithium1}}
\end{figure}

We recall also that our initial selection of Zvrk was on the basis of enhanced lithium enrichment, which we measure to be \(\mathrm{A(Li)} = 3.16 \pm 0.08\) dex --- almost meteoritic --- by combining \edit1{its measurements of [Fe/H] and [Li/Fe]} from GALAH DR4\footnote{\edit1{The GALAH DR4 data quality flags \texttt{flag\_fe\_h} and \texttt{flag\_li\_fe} are both 0, i.e. nominal; the quality flag \texttt{flag\_sp = 8} indicates that the Balmer lines are unusually broad, which we interpret to be a consequence of rotational broadening.}}, and the solar value of \(\mathrm{A(Li)} = 0.96 \pm 0.05\) dex of \citet{wang_3d_2021}. To assess the statistical significance of this measurement, we compare this value in \cref{fig:lithium1} to those obtained from comparable stars in the GALAH sample, chosen by way of colour, magnitude, metallicity, and Gaia RUWE cuts. Zvrk's lithium enrichment, indicated with the vertical dashed line, is far higher than of this control sample.

A high lithium abundance \edit1{suggests internal lithium production \citep[e.g.][]{charbonnel_nature_2000} combined with additional mixing, which} is customarily attributed to mass transfer or tidal interactions owing to a binary companion, or potentially its engulfment \citep[e.g.][]{aguileragomez_lithium_2016a, aguileragomez_lithium_2016b, casey_tidal_2019, soaresfurtado_lithium_2021}. Since only one APOGEE visit to Zvrk was performed, we are unable to determine the properties of any unresolved orbital companion from the time evolution of its APOGEE radial velocities. However, Zvrk has a Gaia DR3 RUWE of 1.056, which is less than the typically quoted threshold value of 1.4 for unresolved binarity \citep{rizzuto_zodiacal_2018}, so a stellar-mass companion is unlikely. Moreover, any undetected companion must produce a radial velocity (RV) signal consistent with or less than \edit1{scatter in the Gaia RVS radial-velocity measurements} of \(\sigma_v = 0.36\ \mathrm{km/s}\). We discuss such a putative companion in more detail below, in \autoref{sec:engulf}.

Zvrk is also not present in the RRRG catalogue of \citet{patton_spectroscopic_2023}, which identifies rapid rotators based primarily on their fitted temperatures being anomalously cool relative to the APOGEE DR16 catalogue. This suggests that, even if Zvrk's present rotational configuration should have been produced as a result of some kind of impulsive event causing it to puff up, such a structural perturbation should have thermally relaxed by now. Thus, our use of numerical stellar models in hydrostatic equilibrium to describe at least Zvrk's structure, although probably not its evolutionary history, appears justified.

\subsection{Rotation}\label{rotation}

\begin{figure}
\centering
\includegraphics{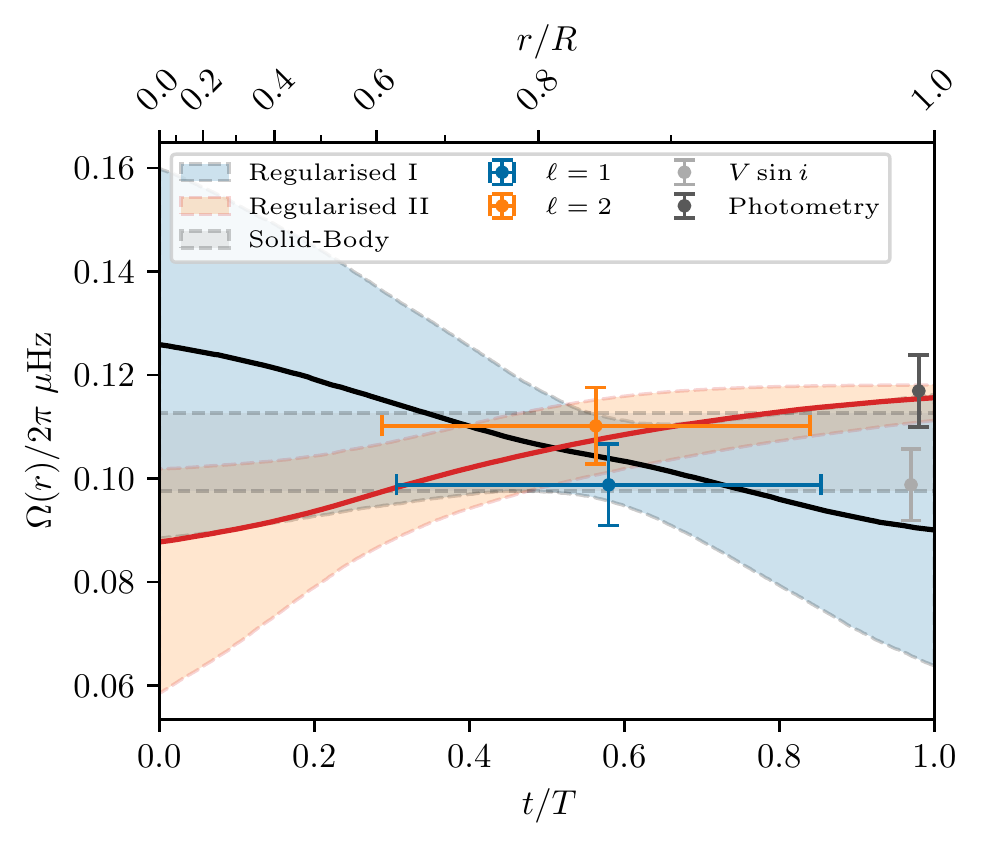}
\caption{Summary of various rotational measurements, displayed with respect to both the acoustic radial coordinate \(t\), and physical radius \(r\). Median rotational profiles from both of our regularised exercises in \cref{fig:rls} are shown with the solid lines, and their credible regions with the filled areas. The rotation rates from the pooled localisation kernels, and their localisation widths, are indicated with the two coloured data points. Our two estimates of the surface rotation rate (from photometry, and by combining spectroscopic \(V\sin i\) with asteroseismic inclination and a radius from stellar modelling) are indicated with the gray data points.\label{fig:rotsummary}}
\end{figure}

Our different values of the rotational splitting as fitted against dipole vs.~quadrupole modes (e.g.~\cref{fig:shear-obs}) suggests the existence of radial differential rotation. In less evolved stars, p-mode asteroseismology has been used to place constraints on this through so-called ``rotational inversion'' techniques \citep[e.g.][]{dimauro_internal_2016, schunker_inversion_2016a, schunker_inversion_2016b, eggenberger_rotation_2019}. Since per-multiplet rotational splittings, as ordinarily required by these methods, could not be reliably constrained for Zvrk, we instead generalise existing techniques for this problem to operate on the power spectrum directly, rather than on lists of fitted multiplet splittings. We describe our approach more fully in \autoref{sec:rotation}, and specifically report numerical results in \autoref{sec:inversion}.

We show a summary of our consolidated constraints on Zvrk's rotational configuration in \cref{fig:rotsummary}. The purely asteroseismic constraints that we derive in \autoref{sec:inversion} are shown in blue and orange, while our two measurements of the surface rotation rate --- from photometric modulations, and from combining Doppler line broadening with our asteroseismic inclination and model radius --- are shown in gray. \edit3{The paucity of independent rotationally split multiplets prevents us from fully \edit4{and unambiguously} characterising differential rotation using seismology, without the imposition of further regularisation. Both of the regularised radial-differential-rotation scenarios that we have considered appear somewhat more likely than solid-body rotation. However, they are only weakly so: while rotational shear is more likely to be present in Zvrk's convective envelope, we may not rule out a solid-body rotational configuration.}

\edit3{Moreover, both of our regularised rotational-inversion scenarios disagree on the overall sense of the bulk rotational shear. Shear in the sense of} \(\Omega\) \edit3{decreasing} outwards \edit3{would require} a secondary, very thin, shear layer near Zvrk's surface \edit3{for} consistency with the observed photometric rotation rate \citep[i.e.~the nonmonotonic configuration proposed for a different red giant in][]{tayar_spinning_2022}. \edit3{Shear in the opposite sense} may happen over longer length scales, and be situated closer to the interior.

Making envelope rotational measurements in red giants \edit1{exhibiting mixed modes} remains an open problem \edit1{outside the linear core-envelope two-zone model \cite[e.g.][]{goupil_seismic_2013}}. Our somewhat loose constraints on Zvrk's rotational stratification appear to suggest that even if the methodological barrier \edit1{of measuring pure p-mode rotational splittings} were to be overcome, one would still need both more precise constraints on the multiplet rotational splittings, and more available multiplets with precisely measurable rotational splittings, than we have at our disposal here, in order to place well-defined constraints on envelope rotational shear \edit1{from seismology alone}. Conversely, this also highlights the pressing need for far more precise measurements on rotational splittings, which will only be possible with qualitatively longer time series than currently available.

\section{Discussion}\label{discussion}

Zvrk is simultaneously more luminous, more stable in RV and photometry, and more slowly rotating, than the entire taxonomy of giant stellar-mass merger remnants discussed in \citet{phillips_seven_2023}; its rotational configuration \edit3{is thus likely} of some different origin. \edit3{In this section, we consider various aspects of the origin, the other possible observational signatures, and the astrophysical implications, of this anomalous rotation, in combination with Zvrk's other unusual observational features.}

\subsection{What happened to Zvrk?}\label{what-happened-to-zvrk}

\label{sec:engulf}
Since \edit2{we have placed observational constraints only on} Zvrk's present state (with other initial conditions being inferred from modelling), any discussion about how its rotational and chemical configuration \edit3{originated} must necessarily be speculative. \edit1{This being said, w}e identify four (mutually nonexclusive) classes of \edit1{other} explanations for it, which would roughly yield both a high rotation rate, and an enhanced lithium abundance:

\begin{enumerate}
\def\labelenumi{(\Roman{enumi})}
\tightlist
\item
  For the sake of argument, Zvrk could represent significant departures from \edit2{standard models} of single-star evolution. Its fast rotation rate would require an efficiency of angular momentum transport, and correspondingly chemical mixing, far in excess of that currently assumed of red giants.
\item
  Alternatively, interactions with an orbiting companion \edit1{could} yield enhanced lithium at Zvrk's surface. An orbiting companion overfilling its Roche lobe could \edit1{transfer mass to Zvrk}, with the accreted material directly depositing angular momentum into \edit1{its} envelope. \edit1{This} would also enrich Zvrk's envelope in lithium \edit1{were} the companion to be a sufficiently hot main-sequence star, or conversely insufficiently massive to begin breaking down its initial lithium content.
\item
  A high rotation rate, \edit1{arising from} tidal spin-up as a result of \edit1{close} binary interactions\edit1{, might} instead \edit1{induce} additional episodic mixing through rotational shear \edit2{\citep[e.g.][]{lagarde_models_2015}} or the Cameron-Fowler mechanism \citep{cameron_lithium_1971}, thereby \edit1{self-polluting} lithium into the envelope.
\item
  Finally, both enhanced lithium abundance and \edit1{fast rotation} could be attributed to Zvrk having engulfed at least one formerly orbiting companion \citep[e.g.][]{stephan_eating_2020, oconnor_giant_2023}, with both matter and angular momentum being directly deposited into its envelope, and being redistributed over the course of several mixing timescales.
\end{enumerate}

\edit2{Let us now examine each of these scenarios in turn.}

\subsubsection{Unusually Efficient Mixing}\label{unusually-efficient-mixing}

Efficient mixing \edit1{in} scenario (I) ought to cause Zvrk's envelope to become anomalously \(^{14}\mathrm{N}\)-rich relative to the APOGEE \edit1{mass-[C/N]} sequence, and hence its actual position on the opposite side of this sequence in \cref{fig:cn} renders this hypothesis unlikely. \edit2{Moreover, the kind of rotational evolution required for consistency with a primordial origin for Zvrk's present rotation --- combining a total absence of main-sequence magnetic braking, and an initial rotation rate of at least half the breakup rate, with instantaneously efficient post-main-sequence angular momentum transport from the core into the envelope --- is of extremely contrived and unphysical nature. We provide a more detailed calculation of this in \autoref{sec:braking}. This scenario is unlikely to explain Zvrk in isolation}.

\subsubsection{Roche Lobe Overflow}\label{roche-lobe-overflow}

Were accretion under scenario (II) to be ongoing, we would expect an accretion disk \edit1{to be} detectable in spectroscopy by H\(\alpha\) emission, if ionised hydrogen were to be accreted, or more generally an infrared excess. As we note in \autoref{sec:ttauri}, we do not detect the former. The latter is expected \edit1{\citep[e.g.][]{rebull_infrard_2015}} to manifest as excess colours derived from the Wide-field Infrared Survey Explorer's \citep[WISE:][]{wright_wise_2010} photometric bandpasses with central wavelengths of 3 (\(W_1\)), 4.5 (\(W_1\)), 12 (\(W_3\)), and 22 \(\mu\)m (\(W_4\)). Zvrk's measured values of \(W_1 = 7.06 \pm 0.05\), \(W_2 = 7.18 \pm 0.02\), \(W_3 = 7.06 \pm 0.02\), and \(W_4 = 6.99 \pm 0.06\), yield color excesses that either are consistent with zero (e.g., \(W_1 - W_3\)) or are far below typical detection thresholds of infrared excess \citep[e.g.][]{rizzuto_infrared_2011, nikutta_wise_2014, bromley_infrared_2021, martell_galah_2021}. Even were accretion onto Zvrk to have occurred historically, it must have stopped by now, potentially ending with the tidal disruption and engulfment of the donor. Thus, scenario \edit2{(II) is also} likely not viable in isolation.

\subsubsection{Tidal spin-up}\label{tidal-spin-up}

\begin{figure}
\centering
\includegraphics{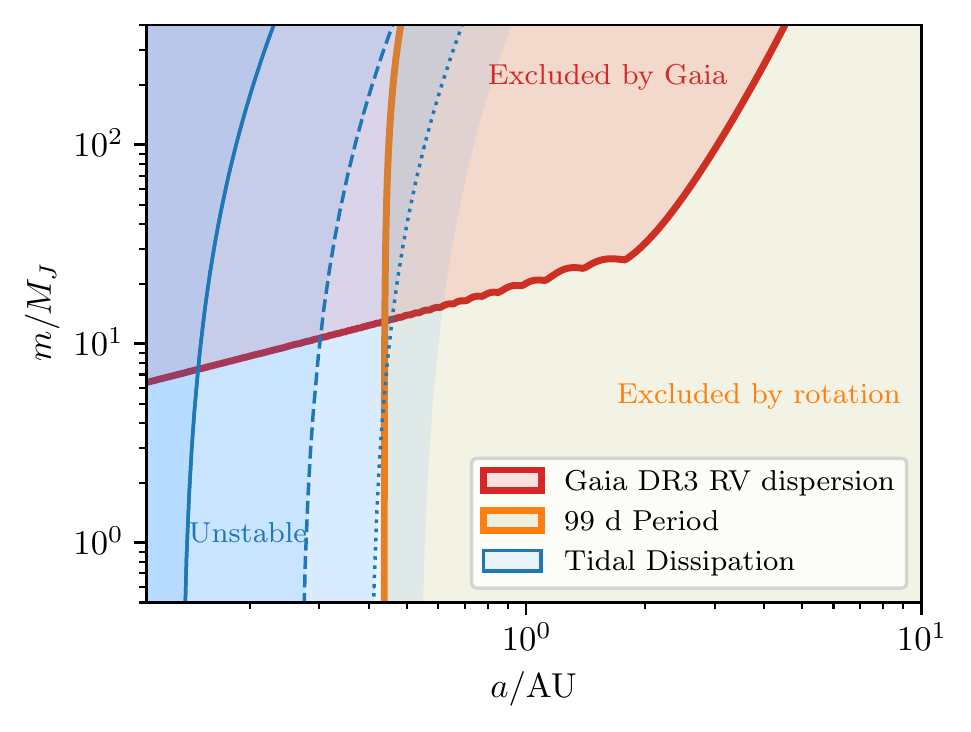}
\caption{Constraints on allowable combinations of undetected companion masses and orbital separations. Shading shows regions excluded by various constraints, with the boundaries of these regions marked out with various curves. The red solid curve, and shaded area above it, shows combinations excluded by the Gaia DR3 RV dispersion. The orange solid curve, and the shaded region to its right, shows combinations which would achieve tidal locking with Zvrk before being able to spin its envelope up to a 99-day rotational period, and are therefore excluded by the detection of such a signal. The blue curves, and shaded regions to their left, show the loci at which the effective radius of Zvrk's Roche lobe are equal to (solid curve), twice (dashed curve), and three times (dotted curve) the asteroseismic radius. The resulting large tidal bulge would lead to significant orbital decay. All combinations of companion mass and orbital separation can be seen either to be excluded by observational constraints, or to lead eventually to inspiral and engulfment. \label{fig:rv}}
\end{figure}

Let us first consider the possibility of a currently undetected companion, orbiting in a stable configuration, being the sole cause of Zvrk's unusual properties. While Zvrk's low Gaia RUWE value disfavours the presence of a stellar-mass orbital companion massive and widely-separated enough to perturb its photocentre over the existing temporal baseline, this does not exclude a lower-mass companion, which would also have to be closer in in order to induce any chemical or rotational anomalies. However, such a close-in companion would potentially induce large orbital RVs. RV constraints from Gaia in turn allow us to place limits on allowable combinations of masses and orbital separations. Specifically, at an orbit with semimajor axis \(a\), a companion of mass \(m_p\) would produce a radial velocity semiamplitude of
\begin{equation}
    K = m \sin i_p \sqrt{G \over M a(1-e^2)}, \label{eq:K}
\end{equation}
which we express in terms of the total and reduced masses
\begin{equation}
    M = M_\star + m_p; m = {M_\star m_p \over M_\star + m_p}.
\end{equation}
\edit1{If this putative companion were to have spun Zvrk up tidally, Zvrk's} final rotation would have to be aligned with the orbital angular momentum, and so our asteroseismic analysis would require \(\sin i_p \sim \sin i_\star \sim 1\). From \cref{eq:K}, one may estimate minimal RMS RV variations \(\Sigma\) given an observing duration; e.g.~\(\Sigma \to K/\sqrt{2}\) for baselines longer than the orbital period. Compared to this, the Gaia DR3 RMS RV scatter of \(\sigma_v = 0.36\ \mathrm{km/s}\), over a 2.2-year time baseline in which \(N=16\) measurements have been taken, would rule out minimal RMS variations \(\Sigma\) larger than \(\sigma_v\). This constraint then defines the boundary of those combinations of companion mass and orbital separations ruled out by Gaia, shown with the red curve in \cref{fig:rv}: the red shaded region above it is excluded by Gaia RVs.

Zvrk's 99-day rotational signal would also further exclude other regions of this parameter space: were Zvrk to have been spun up by tidal interactions alone, its final rotational frequency would be bounded from above by the companion's orbital frequency, since the \edit1{tidal torque} would disappear upon tidal locking. As such, our rotational signal demands that such an orbital companion be closely separated: it cannot have an orbital period much larger than 99 d.~Combinations of \(m\) and \(a\) with the Keplerian orbital velocity equal to 99 d are shown in \cref{fig:rv} in orange: the orange shaded region to its right is excluded by Zvrk's rotation.

What remains is a small window of the parameter space, with \(a \lesssim 0.4\ \mathrm{au}\) and \(m \lesssim 10 M_J\). However, such orbital configurations are tidally unstable. In particular, we show with the blue curves in \cref{fig:rv} the combinations of \(m\) and \(a\) such that the effective spherical radius of Zvrk's Roche lobe in such a binary system, computed using the prescription of \citet{eggleton_approximations_1983}, is equal to (solid curve), twice (dashed curve), and three times (dotted curve) our constraint on Zvrk's asteroseismic radius. Were Zvrk to take up such a substantial fraction of its Roche lobe (blue shaded region) in this binary configuration, the tidal bulge raised on it would be large, and the resulting orbital decay would cause its companion to spiral inwards, eventually to be engulfed. All these suggest that scenario (III) alone might not describe Zvrk's present configuration. Conversely, any orbital configuration in scenario (III) consistent with both the rotational and velocimetric constraints above would require us to consider scenario (IV) anyway.

\subsubsection{Engulfment}\label{engulfment}

If Zvrk's rapid rotation \edit1{derives primarily from} engulfment under scenario (IV), we can use the angular momentum of its envelope's rotation to estimate the mass and orbital properties of such a planet. For simplicity we assume that the currently observed angular momentum of the envelope stems purely from the engulfed object, as stellar expansion and magnetic braking should have significantly reduced the \edit1{envelope rotation attributable to its} primordial angular momentum (which we discuss in more detail in \autoref{sec:braking}). The spin-up of stars from planet engulfment has been investigated by several studies and is expected to occur relatively frequently \citep{qureshi_signature_2018, stephan_eating_2020, oconnor_giant_2023}.

We estimate Zvrk's moment of inertia using the best-fitting stellar structure from our optimisation exercise in \autoref{sec:opt}. To simplify our analysis further, we assume, to first order, that the whole envelope rotates with the surface rotation period of \(\sim99\) days, neglecting our weak constraints on radial differential rotation. Moreover, since the outer layers of such RRRGs carry most of their angular momentum --- even considering their much lower density than the bulk of the star --- due to their large radius, it is also safe to ignore the stellar core itself in our calculations. This yields an estimate for Zvrk's angular momentum of the form
\begin{equation}
    J_\text{env} = \alpha M_\star R_\star^2 \Omega,
\end{equation}
with the moment of inertia from Zvrk's best-fitting model corresponding to \(\alpha = 0.1055\). By assumption, this angular momentum would have mostly originated from the orbit of the engulfed planet,
\begin{equation}
    L_\text{orb} = m\sqrt{GM a (1-e^2)}
\end{equation}
where \(a\) is the semimajor axis (SMA), and \(e\) the eccentricity, of such a planetary orbit.

\begin{figure}
\centering
\includegraphics{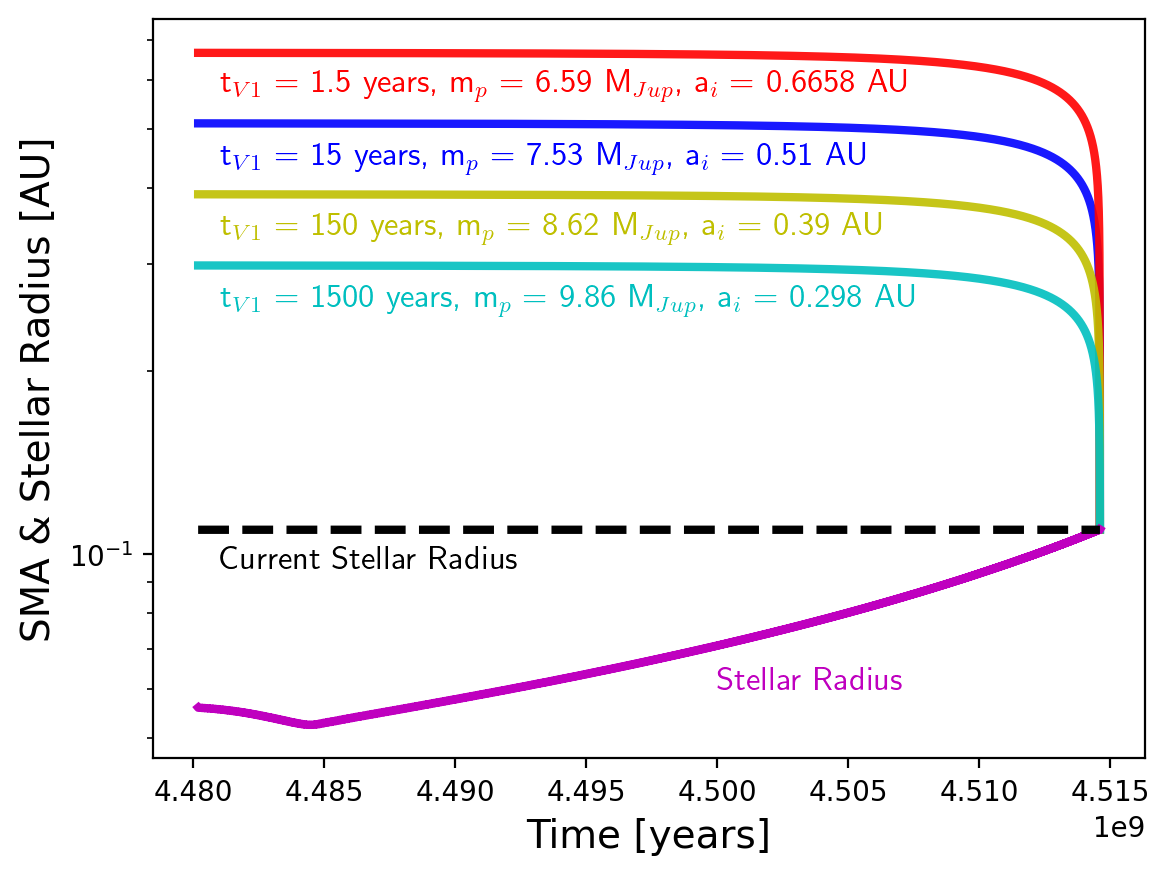}
\caption{Tidal migration tracks of pre-engulfment planetary companions for various tidal efficiencies. The angular momentum imparted on the stellar envelope depends on the orbital semimajor axis, eccentricity, and mass of the engulfed object. Here we simulate the orbital decay and eventual engulfment of various configurations of these elements (assuming, however, circular orbits), dependent on the viscous time scale \(t_{V1}\) of the star, which is uncertain \citep[see, for details, Appendix of][]{eggleton_orbital_2001}. The orbital decay is modeled such that engulfment would approximately occur at the star's current radius (shown in magenta) at its current estimated age. More efficient tidal dissipation (shown in red) would imply a planet mass of about 6.59 \(M_J\) with an initial SMA of 0.6658 au, while less efficient dissipation (shown in cyan) would imply a mass of about 9.86 \(M_J\) and an initial SMA of 0.298 au. \label{fig:tidal}}
\end{figure}

In order to supply the requisite angular momentum, a companion placed in a circular orbit at Zvrk's present radius would need to be about 16 \(M_J\), neglecting the effects of tidal dissipation entirely. This serves as an upper bound on \(m_p\), since in reality, any orbital eccentricity would permit closer periapses than semimajor axes, and in any case tidal effects at such orbital separations should drag the planet into the star long before the star could reach its current radius. Larger orbital separations, with engulfment caused by tidal dissipation and inspiral, correspond to smaller companion masses. Even following the equilibrium tide model for simplicity \citep[e.g.][]{hut_tidal_1981, eggleton_orbital_2001}, the efficiency of this tidal dissipation is rather uncertain, as the viscous dissipation timescale \(t_{V1}\) is not well-constrained for giant stars. However, candidate values for it across 4 orders of magnitude (shown in \cref{fig:tidal}) all give plausible planetary masses and orbital SMAs, ranging from about 10 \(M_J\) at 0.3 au, to about 6.5 \(M_J\) at 0.67 au.

These calculations all assume that the planet was very recently engulfed, at approximately its current size and evolutionary state, and that the planet's orbit had zero initial eccentricity. \edit1{Alternatively, if} the planet was originally at a much wider separation, a very high orbital eccentricity (\(>0.9\)) bringing periapsis within close proximity of the star (thereby also enabling strong tidal dissipation) \edit1{may also have permitted engulfment}. Such a scenario would be possible if other bodies interacted with the system and incited high eccentricity, for example through scattering or secular processes \citep[e.g.,][etc.]{weidenschilling_gravitational_1996, chatterjee_dynamical_2008, naoz_hj_2011, naoz_formation_2012, naoz_secular_2013, stephan_A_2018, stephan_white_2021}. In such a scenario the original orbit of the planet could have had an SMA of tens of au, with a minimum possible planet mass of about 6.7 \(M_J\). This scenario, however, would require the past or current presence of additional planets or stellar companions, which are so far not detected. Nonetheless, this may warrant future investigation.

\subsubsection{A Combination of (III) and (IV)?}\label{a-combination-of-iii-and-iv}

\edit2{While our discussion indicates than an engulfment is dynamically necessary to explain Zvrk's observed configuration, its} convective envelope is so massive (of order \(10^{33}\ \mathrm{g}\)) that explaining its deep lithium absorption feature proves difficult --- Zvrk would have had to ingest the lithium supply of \(\sim10^3\) gas giants to generate this chemical signature from engulfment alone. As such, this high a measured value likely may not have originated purely from the engulfment or accretion of additional material, indicating some degree of self enrichment, such as through the Cameron-Fowler process \citep{cameron_lithium_1971}. We note that this does not preclude any engulfment or accretion from having happened historically: rather, it indicates only that lithium is a degenerate tracer of history with other formation pathways.

This need for self-pollution is difficult to reconcile with Zvrk's apparent position relative to the \edit1{mass-[C/N]} sequence. A purely intrinsic origin for the observed lithium and angular momentum, with both having been brought up from the fast-rotating and nuclear-enriched core, would either require nitrogen to be somehow destroyed in the envelope, be redistributed differently from lithium, or else be incompatible with Zvrk's apparent (admittedly weakly constrained) nitrogen deficiency. Conversely, a purely extrinsic hypothesis for our derived rotational configuration, with angular momentum deposited into the envelope by either the accretion of infalling material or the engulfment of a less massive object, would produce chemical pollution compatible with the observed direction of both of these chemical anomalies, but incompatible with their observed magnitudes. Unfortunately, this tension appears only resolvable by resorting to fine-tuning some combination of scenarios (III) and (IV) --- e.g., by demanding temporally separated mixing episodes for different nuclear species, as suggested in \citet{tayar_thermohaline_2022}. We leave a more detailed examination of Zvrk's evolutionary history to future work.

\subsection{Magnetic Activity}\label{magnetic-activity}

Magnetic activity in dwarfs is known to be generated by the interposition of convective turbulence with rotational shear. This can be characterised by the Rossby number \(\mathrm{Ro} = P_\text{rot} / \tau_\mathrm{cz}\) \citep[e.g.][]{noyes_rotation_1984, brun_magnetism_2017}. \edit2{Red giants are thought to obey the same physical laws, and thereby satisfy similar scaling relations \citep[e.g.][]{charbonnel_magnetic_2017,dixon_rotationally_2020}.}
We evaluate Zvrk's Rossby number with respect to a global convective turnover timescale:
\begin{equation}
    \tau_\mathrm{cz} = \int_{r_1}^{r_2} {\mathrm d r \over V_\text{conv}},
\end{equation}
where \(r_1 = r_\text{base} + H_p\) and \(r_2 = r_\text{top} - H_p\) \edit2{are situated one pressure scale height $H_p$ above and below the base and top of the convection zone, respectively}, and \(V_\text{conv}\) is the convective velocity (which we compute from mixing-length theory). This is a similar construction to that also used for fully convective dwarfs, which likewise do not possess a well-defined tachocline. Evaluating this quantity with respect to the best-fitting model in our optimisation exercise in \autoref{sec:opt}, in conjunction with our nominal photometric rotational period, gives us \(\mathrm{Ro}_\star \sim 0.28\). By contrast, computing this quantity from a solar-calibrated MESA model with the same physics yields \(\mathrm{Ro}_\odot = 1.38\). Roughly speaking, lower values of the Rossby number indicate greater magnetic activity. The convective turnover timescale that we have obtained, \(\sim 350\) days, is comparable to the timescales over which the amplitude of the signal presented in \autoref{sec:photometry} appears to evolve, and so we cannot rule out a convective origin for the surface features producing this signal.
However, the subsolar value is consistent with an interpretation of variability due to spots.

\begin{figure}
\centering
\includegraphics{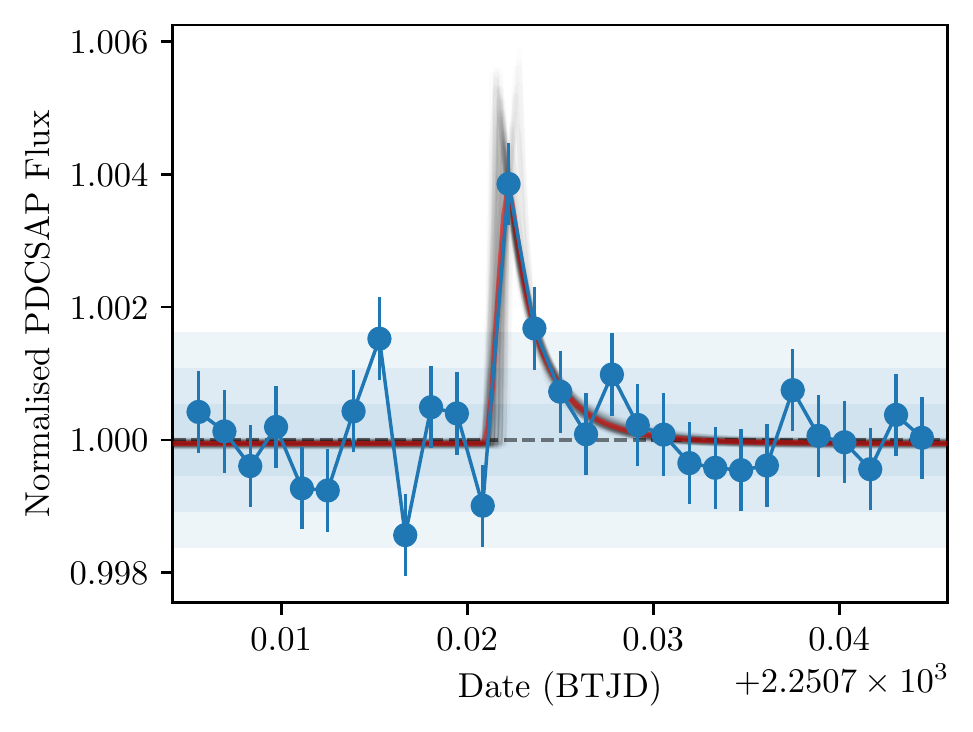}
\caption{Candidate flare identified in two-minute-cadence PDCSAP data, scaled by the mean value, so that the data points shown here are normalised to 1 (marked with the horizontal line). The horizontal bars show integer multiples of the baseline photometric scatter of data points in the 1.5 hours immediately following the peak identified here. The faint gray lines show draws from the posterior distribution over an impulse-exponential-decay model, as fitted to the main peak using nested sampling. The median of the light-curve models from the posterior distribution is shown with the thick red line. \label{fig:flare}}
\end{figure}

Aside from spotting, surface magnetism \edit1{on active dwarfs is known to manifest} in the time domain in the form of flares and eruptive outbursts. Were it to be magnetically active, it is possible that Zvrk also produces such transient events; should flares actually occur on Zvrk, we should also expect to see occasional outbursts of energetic photons. We tentatively identify at least one marginal detection of a flare in Zvrk's PDCSAP lightcurve (shown in \cref{fig:flare} at BTJD 2250.7, along with samples from the posterior distribution of fits to it of an impulse-exponential-decay model). We also find that near-UV photons have historically been detected from Zvrk's sky position in archival GALEX data \citep{gphoton}. We defer a more detailed investigation of this possible magnetic activity to future work.

\subsection{Rotational Age Constraints}\label{rotational-age-constraints}

\label{sec:braking}

\begin{figure}[htbp]
    \annotate{\includegraphics[width=.45\textwidth]{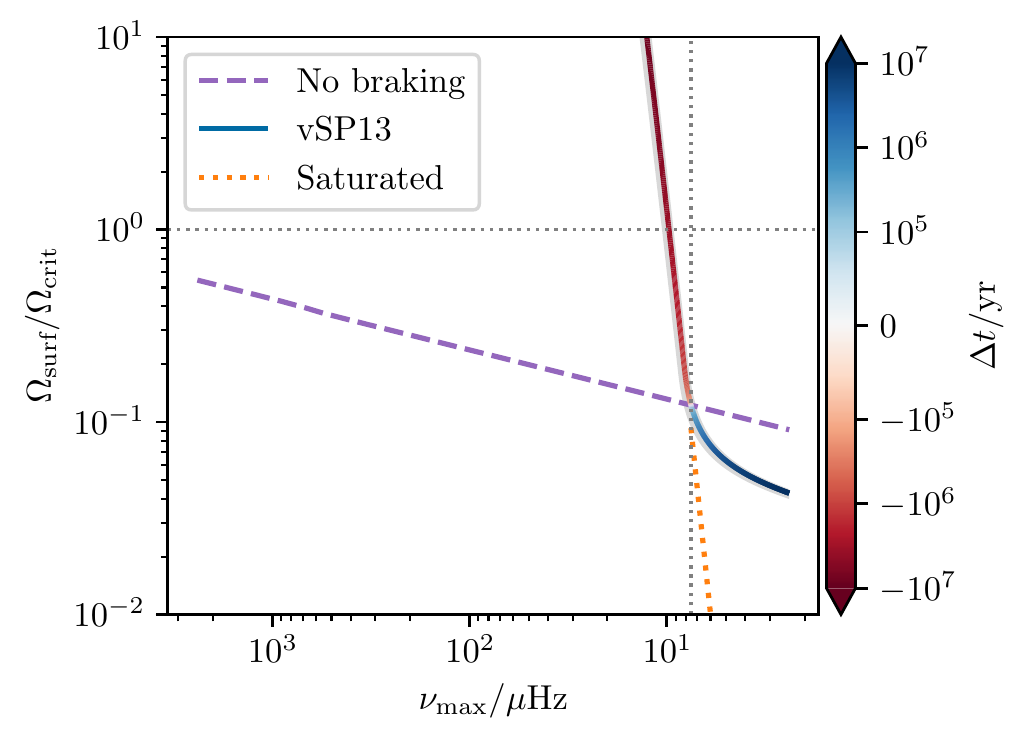}}{\node at (.22, .22){\textbf{(a)}};}
    \annotate{\includegraphics[width=.4\textwidth]{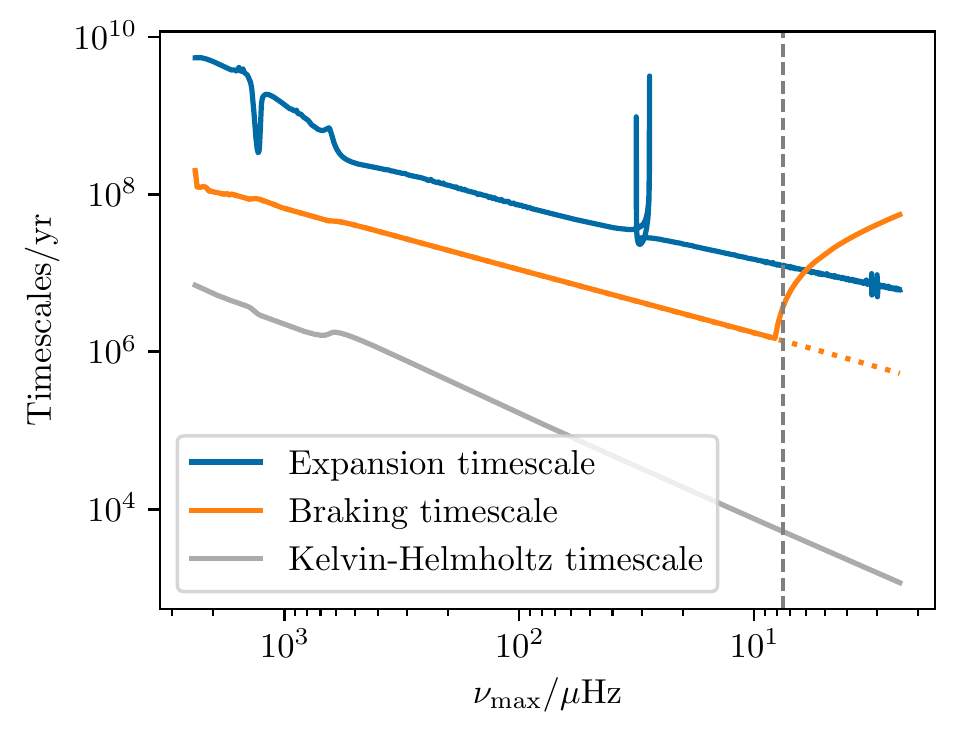}}{\node at (.9, .9){\textbf{(b)}};}
    \caption{Rotational evolution of Zvrk under different extremal scenarios. \textbf{(a)} Evolution of the surface rotational frequency $\Omega_\text{surf}$ from the main sequence up the RGB, using the frequency of maximum power $\nu_\text{max}$ as an evolutionary proxy, and shown as a fraction of the breakup rotational frequency $\Omega_\text{crit}$ (horizontal dotted line). The solid curve\edit2{, labelled vSP13,} shows magnetic braking according to the prescription of \citet{vansaders_fast_2013}, where line segments are coloured by simulation time $\Delta t$ from the present (i.e. positive represent integration into the future). The dotted curve shows rotational evolution from the present under braking parameters chosen to produce saturated braking, while the dashed curve shows the \edit2{counterfactual} scenario of no magnetic braking (i.e. angular momentum is conserved). The vertical dotted line indicates \numax\ of our best-fitting model, \edit3{at which we impose a simulated rotation period of 100 d}. The shaded region around the solid curve represents the range of solutions lying within $\pm 30\%$ of Zvrk's present rotation rate. \textbf{(b)} Comparison of several characteristic timescales. We show in particular the spindown timescale $|P_\text{rot} / \dot{P}_\text{rot}|$ associated with structural expansion in the absence of magnetic braking; the braking timescale $|J_\text{tot} / \dot{J}_\text{tot}|$ associated with angular momentum loss by the prescription of \cite{vansaders_fast_2013}, with the dotted curve showing the same saturated trajectory as in panel (a); and the Kelvin-Helmholtz timescale $t_\text{KH} = GM^2/2RL$, as a proxy for the thermal readjustment timescale.}
    \label{fig:braking}
\end{figure}

While rapid for a red giant, Zvrk's surface rotation remains far slower than its breakup rotation rate, \(\Omega_\text{crit} = \sqrt{GM/R^3}\). From our seismic constraints on Zvrk's mass and radius, we find \(P_\text{crit} = 2\pi/\Omega_\text{crit} \sim 12.5\ \mathrm{d}\). Admittedly, both Zvrk's rotational period and this breakup rotation rate would have changed significantly over the course of its evolution up the red giant branch. \edit1{In \cref{fig:braking}a, we plot} Zvrk's surface rotation rate \edit1{as a fraction of its breakup rotation rate}, integrating backwards from the present day --- \edit3{indicated with the vertical line, and at which we impose a notional rotational period of 100 d} --- along the best-fitting track from our optimisation exercise in \autoref{sec:opt}\edit1{, under a hypothetical single-star evolutionary scenario. A} counterfactual scenario with no magnetic braking \edit1{is indicated with the dashed curve,} \edit3{through which we track Zvrk's rotation accounting for structural evolution, but imposing no external torques. Under this scenario} Zvrk would have been rotating at half its breakup rotation rate on the main sequence, \edit3{to match its present rotation}.

However, if we are to interpret the apparent photometric variability as spot modulations, the same magnetic field that drives this spot activity, and the possible flaring that we discuss above, ought also to be responsible for magnetic rotational braking. In turn, the action of magnetic braking would require Zvrk to initially have been rotating still faster \edit3{than this nonbraking scenario}. To \edit1{account for} \edit3{these effects}, we \edit1{use the} magnetic braking prescription of \citet{vansaders_fast_2013}, with an angular momentum loss rate of
\begin{equation}
    \dot{J} \propto f_K \Omega^3 \mathrm{min}\left(1, {\omega_\text{crit} \tau_{\text{cz},\odot} \over \Omega \tau_\text{cz}}\right)^2,\label{eq:am-loss}
\end{equation}
\edit3{where $\omega_\text{crit}$ is a critical rotation rate such that $\tau_{\text{cz},\odot} \cdot \omega_\text{crit}$ sets the inverse Rossby number at saturation, and $f_K$ is a normalisation constant. The angular-momentum loss rate $\dot{J}$} can be seen to scale either as \(\dot{J} \propto \Omega\) for rotation rates above \edit3{this} inverse-Rossby-number saturation threshhold, \(\Omega \tau_\text{cz} > \omega_\text{crit} \tau_{\text{cz},\odot}\) (the \edit3{stronger,} ``saturated regime''), or to give \(\dot{J} \propto \Omega^3\) for lower rotation rates \edit3{below it} (the \edit3{weaker,} ``unsaturated regime''). \edit3{We show with the solid curve in \cref{fig:braking}a the surface rotation frequencies that result when the angular-momentum loss rate of \cref{eq:am-loss} is integrated both forwards and backwards in time, further demanding that all of the stellar material remains bound. We colour the segments of this curve according to simulation time $\Delta t = t - t_\text{present}$.} For this exercise, we adopt \edit3{standard values of $\omega_\text{crit} \sim 12\Omega_\odot$, and $f_K = 9.37$}, \edit3{which are} calibrated to produce the solar equatorial rotation rate of 25.4 d for our solar-calibrated MESA model with a disk rotation period of 8 d and a disk-locking timescale of 1 Myr. This setup is generally in keeping with previous uses of this angular-momentum-loss prescription. The initial conditions of our integration are again chosen to produce a rotational period of 100 d at the present observed value of \numax.

As expected given Zvrk's current Rossby number, our chosen combination of solar-calibrated \(f_K\), \(\omega_\text{crit}\), and \(\tau_\text{cz}\) are such that Zvrk leaves the saturated regime \edit3{only} shortly before the present day. Zvrk otherwise experiences \edit3{strong,} saturated braking (\(\dot{J} \propto \Omega \omega_\text{crit}^2\)) in most of its preceding evolution up the RGB. \edit3{If we were to assume that Zvrk's anomalous rotation results from engulfing a companion, this must have happened less than 5 Myr ago, when it was already a quite evolved red giant ($\numax < 20\ \mu\mathrm{Hz}$): any earlier engulfment scenario would require Zvrk to exceed its breakup rotation rate $\Omega_\text{crit}$ at the time of engulfment (horizontal line), in order to match its observed rotation at the present day upon the action of magnetic braking}. This is incompatible with Zvrk's angular momentum having been of primordial origin, ruling out a single-star hypothesis (scenario (I) in \autoref{sec:engulf}) for Zvrk's rotational configuration.

This same magnetic braking also places an upper limit on how long we might expect such a rotational signal to persist after the event. We show a comparison of \edit3{various physically relevant} timescales in \cref{fig:braking}b. The spindown timescale owing to structural expansion in the absence of magnetic braking, \(|P_\text{rot} / \dot{P}_\text{rot}|\), shown with the blue curve, can be seen to be much slower than the magnetic braking timescale \(|J_\text{tot} / \dot{J}_\text{tot}|\), shown in orange. For the latter, the solid curve shows the scenario depicted in \cref{fig:braking}a, where Zvrk's magnetic braking leaves the saturated regime shortly before the present day. We also consider \edit3{an even stronger}-braking scenario, where we extrapolate from its present rotational configuration under the assumption that the magnetic braking remains saturated, shown with the dotted curves on both \cref{fig:braking}a and b. However, even this can still be seen to yield braking (and so rotational persistence) timescales far longer than the Kelvin-Helmholtz timescale, shown in gray, which we use as a proxy for the thermal readjustment timescale. \edit3{Any} engulfment hypothesis for the deposition of angular momentum into a red giant's envelope would also necessitate heating the envelope, causing the red giant to puff up in response, and resulting in anomalously cool observed temperatures (as predicted in \citealt{oconnor_giant_2023} and \citealt{yarza_hydrodynamics_2023}, and reported in \citealt{patton_spectroscopic_2023}). \edit3{Nonetheless, since} thermal readjustment can be seen to occur over much shorter timescales than even saturated rotational braking, our observational determination of Zvrk as being both rapidly rotating, and yet not too spectroscopically unusual, remains compatible with an engulfment hypothesis.

\section{Conclusion}\label{conclusion}

We have determined that TIC~350842552, or ``Zvrk'' for short, is a rapidly-rotating first-ascent red giant. We have measured Zvrk's rotation rate with independent uses of asteroseismology, photometry (from space and from the ground), and spectroscopy (from rotational Doppler broadening). Our constraints on Zvrk's global properties and structure, derived from detailed asteroseismic modelling against constraints from individual mode frequencies, suggest that all of these rotational measurements are mutually consistent with each other. We have further developed, and then employed, new techniques in asteroseismic rotational analysis --- making use of recent theoretical developments permitting pure p-modes to be disentangled from the usual spectrum of mixed modes --- to constrain properties of Zvrk's \edit1{internal} rotation independently of the radiative core. Together, these paint a picture of Zvrk as being differentially rotating in its envelope, and possessing rotational shear in the sense of the rotation rate increasing outwards at some point: a similar configuration to that proposed for a far less evolved red giant in \citet{tayar_spinning_2022}.

Aside from this anomalous rotation, we have additionally found Zvrk's envelope to be simultaneously almost meteoritic in its lithium enrichment, and also potentially less enriched in \(^{14}\mathrm{N}\) than typical for its mass, suggestive of recent pollution. Zvrk's high surface rotation rate, combined with astrophysical constraints from magnetic braking, suggests an engulfment event as both a source of this angular momentum, and a trigger for any self-enrichment. We place limits on how long ago such an engulfment could have occurred --- no more than several million years ago. With such a recent engulfment, Zvrk's angular momentum content suggests a mass for the engulfed object of roughly 7 to 9 \(M_J\), orbiting a little closer in than 1 au. These orbital parameters are fairly consistent with existing demographic studies of giant planets \citep[e.g.][]{kennedy_planet_2008, wittenmyer_cool_2020}. However, a more detailed examination of both this engulfment, and any self-enrichment episode that it may trigger, will be necessary to reconcile Zvrk's multiple chemical anomalies. This we leave to future work.

Our characterisation of Zvrk demonstrates the diagnostic power of asteroseismology as a tool for not merely measuring rotation rates in field stars, but indeed for initially detecting rotation in the first place. The asteroseismic rotational signature remained robust against aggressive preprocessing, which had rendered direct measurement of quasiperiodic variability from the same data set otherwise nonviable. Moreover, in order to relate Zvrk's observed seismic rotational signatures to its interior structure, we have had to devise new asteroseismic techniques altogether. These new tools might also be applied to other known lithium-anomalous post-main-sequence stars, many of whose rotational properties are currently even more poorly constrained. Additionally, in the high-eccentricity engulfment scenario that we have considered, the planet's original orbit would potentially have been strongly misaligned with the pre-engulfment stellar rotational axis, which would result in rotational misalignment between Zvrk's core and envelope upon engulfment. While seismic diagnostics of any such possible misalignment would be valuable in disambiguating between engulfment scenarios, existing methods do not yet permit us to place constraints of this kind.

It is not yet well understood how efficiently such massive objects as we propose to have been engulfed would be disassociated were they actually to be deposited into a red giant envelope \citep[e.g.][]{jia_disruption_2018, yarza_hydrodynamics_2023}. While our poor constraints on the structure of Zvrk's rotational shear yield no insight on this, we anticipate firmer constraints on envelope rotational shear, which would naturally emerge with higher-quality data and/or longer time series, to better illuminate such disruption. More generally, a better understanding of such engulfment events may help to explain the origins of the current, known, population of isolated RRRGs with no detectable companions. Some of these RRRGs are also known to be magnetically active \citep[e.g.][]{gehan_surface_2022}, and the potential links between this magnetic activity and post-main-sequence rotation may also be worth further investigation. While these rotational signatures do not persist for very long (with braking timescales of \(\sim 10^{6}\ \mathrm{yr}\) this far up the RGB), they still survive for far longer than would any photometric transients (which have characteristic timescales of several days) or thermodynamic anomalies (\(\sim 10^{3}\ \mathrm{yr}\)). A systematic investigation into the origins of post-main-sequence surface rotation, to better illuminate these engulfment processes, may prove logistically easier than searching for transients as they happen.

Over broader horizons, Zvrk also serves as a demonstration of both the limitations of the presently available TESS data, and the pressing need for sustained and consistent investment in long time series for asteroseismic purposes. Between our initial asteroseismic detection and the writing of this paper, a few extra months of a third year of TESS CVZ coverage, in Cycle 5, have been collected. We have found the addition of a few months not to provide any noticeable improvements to any of our asteroseismic constraints. However, given the proposed longevity of the TESS mission (potentially decades as advertised), long-term monitoring might yield qualitative improvements in our asteroseismic constraints. Supplemental monitoring from future space missions (e.g.~PLATO) will also relieve TESS of our demands on its duty cycle. In addition to permitting multiplet splittings to be constrained without pooling as we have done, decades-long monitoring will also be a bare minimum requirement for any prospect of, e.g., measuring core rotation rates from stars as evolved as Zvrk. Were they available, these measurements would bound their total angular momentum content, instead the present situation of the core being an invisible angular momentum sink. This would in turn greatly constrain their possible histories.

\edit3{The effect of rotation on the morphology of Zvrk's echelle diagram also uncovers methodological shortcomings in the state of the art for how solar-like oscillators are analysed. While the interpretation of seismic quantities is ultimately contingent on the identification assigned to peaks in the power spectrum, mode identification for solar-like oscillations is ordinarily straightforward and unambiguous. Should this not be the case, then strictly speaking the conditional distributions from various mode-identification scenarios ought to be combined, in a manner specified by probabilities assigned to each scenario. In this case, the consistency of the rotation rates emerging from our mode identification, with those from supplemental photometric and spectroscopic analysis, suggests that our proposed scenario has a high probability of being correct. Nonetheless, the ambiguity we encountered in initially deriving a mode identification for Zvrk --- defeating automated analysis techniques --- still highlights a pressing need, in general, for the development of numerical machinery both to generate mode identifications exhaustively, and to assign probabilities to different scenarios.}

Finally, as for Zvrk itself, much remains about it to be observationally determined. Should it actually be flaring, Zvrk would make a compelling target for future space-based UV photometry missions. Moreover, although Zvrk's Gaia RUWE and RV scatter are not consistent with a binary companion with orbital periods close to the Gaia baseline, it cannot definitively rule out the presence of a very widely separated orbital companion, which may have caused a high-eccentricity engulfment scenario \citep[e.g.][]{knutson_friends_2014}, or any potential undetected close-in companions undergoing tidal decay. Zvrk's southerly location renders it amenable to contemporaneous follow-up radial-velocity observations for this purpose using cutting-edge instruments, such as ESPRESSO (for now) and G-CLEF (in the near future). Other kinds of follow-up observations, such as speckle imaging, may also prove fruitful. All of these paths lead far beyond the scope of this paper, but there is clearly much more work to be done.

\section*{Acknowledgements}

We thank S. Basu, T. Bedding, and S. Hekker for constructive feedback on preliminary versions of this work, and C. Hayes, J. Hinkle, A. Boesgaard, and C. Kochanek for productive discussions. MTYH and MSF thank the Aspen Center for Physics for their hospitality during the completion of parts
of this work. The Jupyter notebooks and MESA inlist files used for this work, and the source of this paper itself, can be found on a GitHub repository at \url{http://github.com/darthoctopus/zvrk}; the repository will be made public upon acceptance. \edit1{The \mesa\ \texttt{inlist} files used in our analysis have been placed in a Zenodo repository at \dataset[doi:10.5281/zenodo.8267322]{\doi{10.5281/zenodo.8267322}}.}


This paper includes data collected by the TESS mission. Funding for the TESS mission is provided by the NASA's Science Mission Directorate.

This work has made use of data from the European Space Agency (ESA) mission {\it Gaia} (\url{https://www.cosmos.esa.int/gaia}), processed by the {\it Gaia} Data Processing and Analysis Consortium (DPAC, \url{https://www.cosmos.esa.int/web/gaia/dpac/consortium}). Funding for the DPAC has been provided by national institutions, in particular the institutions participating in the {\it Gaia} Multilateral Agreement.

JMJO, MTYH, and MSF acknowledge support from NASA through the NASA Hubble Fellowship grants HST-HF2-51517.001-A, HST-HF2-51459.001-A, and HST-HF2-51493.001-A, respectively, awarded by STScI. STScI is operated by the Association of Universities for Research in Astronomy, Incorporated, under NASA contract NAS5-26555.
APS acknowledges partial support by the Thomas Jefferson Chair Endowment for Discovery and Space Exploration, and partial support through the Ohio Eminent Scholar Endowment. MY, Z\c{C}O, and S\"O acknowledge the Scientific and Technological Research Council of Turkey (TÜBİTAK:118F352).
SLM acknowledges funding support from the UNSW Scientia program and from the Australian Research Council through Discovery Project grant DP180101791. DS is supported by the Australian Research Council through Discovery Project grant DP190100666. Parts of this work were supported by the Australian Research Council Centre of Excellence for All Sky Astrophysics in 3 Dimensions (ASTRO 3D), through project number CE170100013.

\software{NumPy \citep{numpy}, SciPy stack \citep{scipy}, AstroPy \citep{astropy:2013,astropy:2018}, \texttt{lightkurve} \citep{lightkurve}, \texttt{Comove} \citep{kraus_comove_2022}, \texttt{dynesty} \citep{dynesty}, Pandas \citep{pandas}, \mesa\ \citep{mesa_paper_1,mesa_paper_2,mesa_paper_3,mesa_paper_4,mesa_paper_5}, \gyre\ \citep{townsend_gyre_2013}, \texttt{ADIPLS} \citep{jcd_adipls_2008}}

\facilities{TESS, ASAS-SN, APOGEE, GALAH, WISE, GALEX}

\appendix
\setcounter{table}{0}
\renewcommand{\thetable}{A\arabic{table}}

\setcounter{figure}{0}
\renewcommand{\thefigure}{A\arabic{figure}}

\section{Observational Window Function}\label{observational-window-function}

\label{sec:window}

\begin{figure}
\centering
\includegraphics{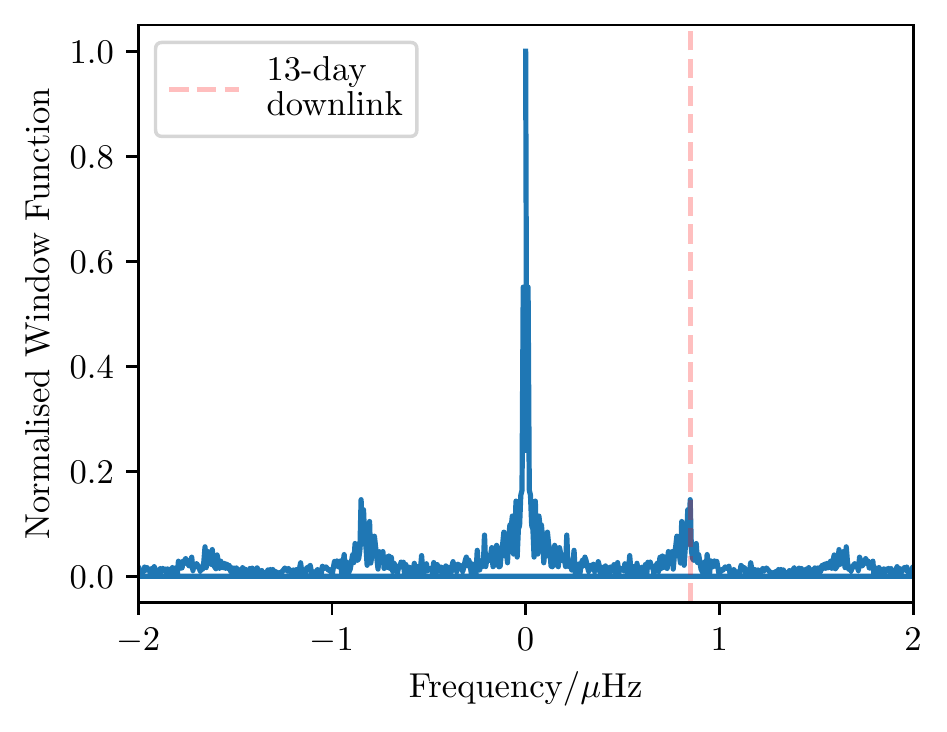}
\caption{Spectral window function of the full 2-minute-cadence time series. The primary sidelobes, whose locations are marked out with the pink dashed line, originate from TESS's 13-day downlink cycle. Other sidelobes and aliases can be seen to have significantly lower prominence. \label{fig:window}}
\end{figure}

\edit2{We plot the spectral window function of the full TESS time series in \cref{fig:window}. In addition to the nominal central peak, we see also sidelobes, the most prominent of which recur at frequencies close to the nominal recurring downlink frequency of once every 13 days (i.e. $0.89\ \mu$Hz). However, the low prominence of these sidelobes compared to the main peak is such that they cannot explain the multiple peaks of similar relative height that are visible in the power spectrum (e.g. \cref{fig:echelle-comparison}). Moreover, these sidelobes are also widely separated enough that they cannot have been misidentified as producing either the small separation between modes of alternating even degree (around $0.35\ \mu$Hz), nor our putative identified rotational splittings (around $0.1\ \mu$Hz). Other families of sidelobes and their aliases can be seen to be of far lower prominence, and are not of observational concern. As such, we believe our proposed mode identification is not the result of misinterpreting aliases in the observational window function.}

\setcounter{table}{0}
\renewcommand{\thetable}{B\arabic{table}}

\setcounter{figure}{0}
\renewcommand{\thefigure}{B\arabic{figure}}

\section{Results of Peakbagging Procedure}\label{results-of-peakbagging-procedure}

\begin{figure*}[tb]
    \centering
    \annotate{\includegraphics[width=\textwidth]{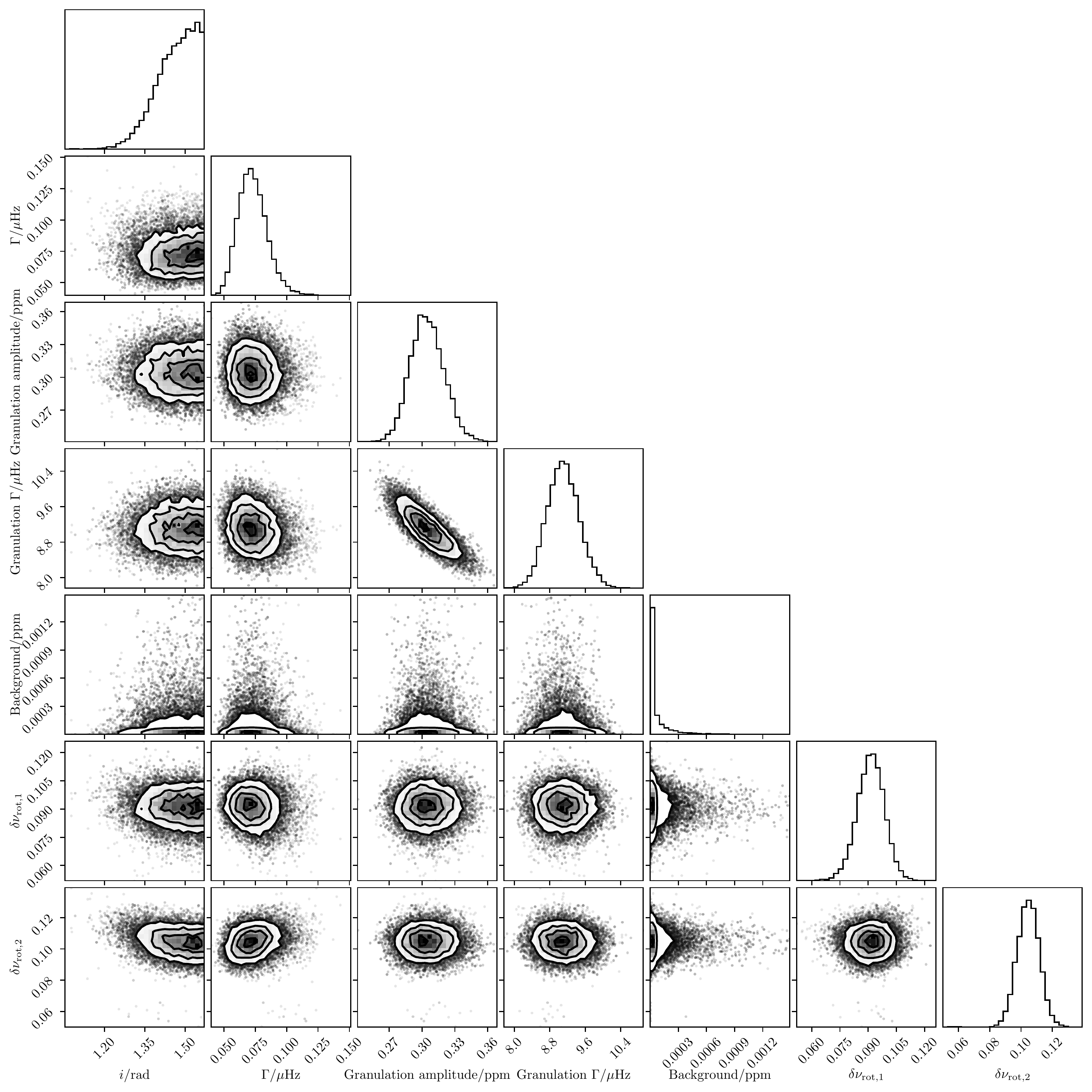}}{
    \node[below left] at (.97, .97){\includegraphics[width=.46\textwidth]{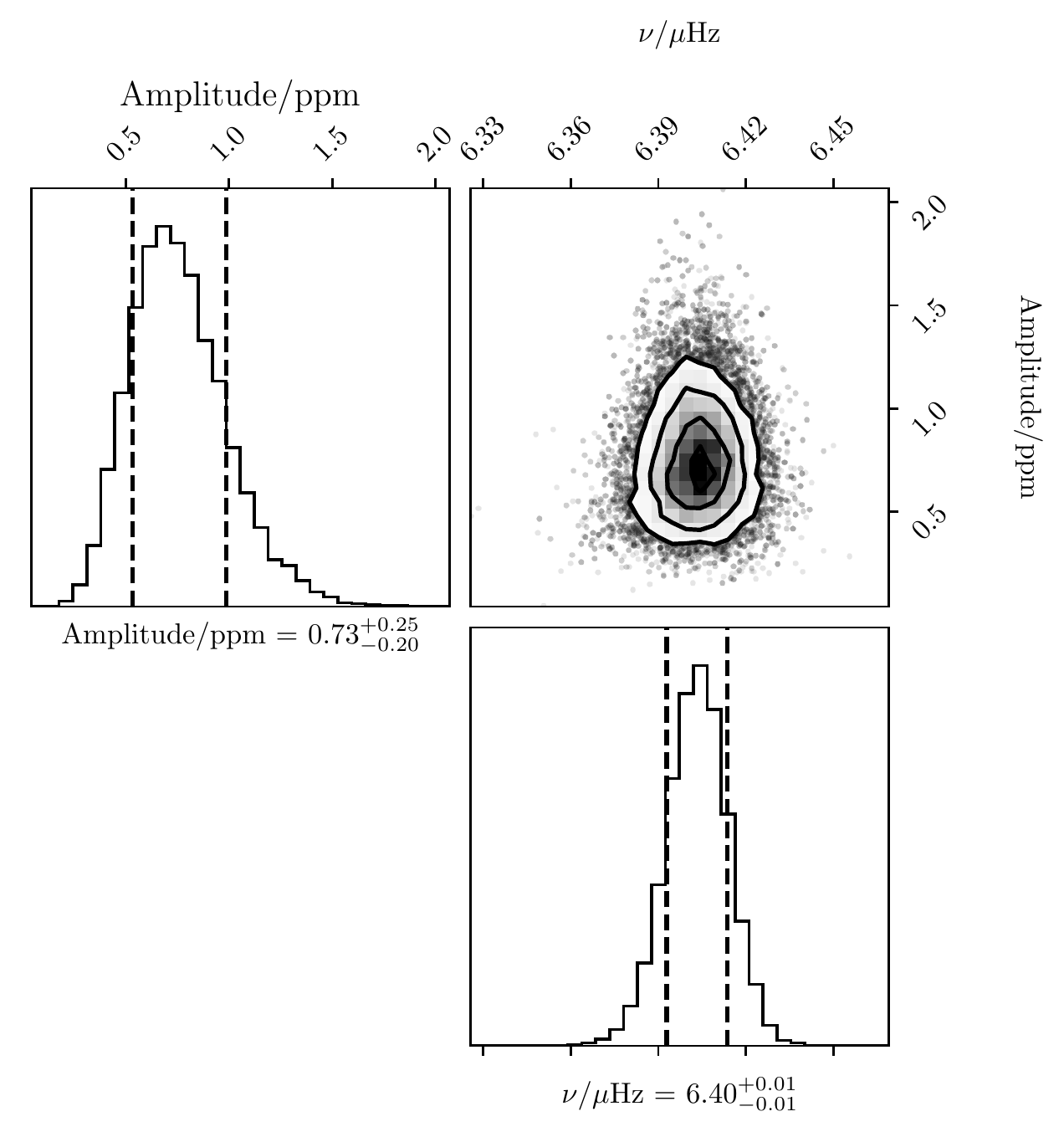}};
    }
    \caption{Posterior distributions for quantities returned from nested-sampling peakbagging procedure. The main figure, in the lower left, shows the joint posterior distribution for quantities entering into \cref{eq:model} that are pooled between modes. In particular, we see that there is mild correlation between some properties of our background model (labeled "Granulation" here), but the quantities of asteroseismic relevance are constrained to be statistically independent of each other, aside from a very mild correlation between the axial inclinations $i$, the pooled linewidths $\Gamma$, and the rotational splittings $\delta\nu_{\text{rot},\ell}$. The dipole-mode rotational splitting can be seen to be smaller than that of the quadrupole modes. The inset figure, shown in the upper right, shows the joint distribution for quantities associated with the dipole mode closest to $\numax$ (i.e. $n_p = 5, \ell = 1$), where again the notional nonrotating frequency $\nu$ is constrained independently from the mode amplitude. The joint distribution between modes also indicates that the mode frequencies for all modes are independent of each other (although that corner plot is too large to fit on a single page). The mode frequencies derived from nested sampling in this fashion, and their uncertainties, are reported in \cref{tab:peakbag}.}
    \label{fig:shared}
\end{figure*}

We illustrate the posterior distributions arising from our nested-sampling procedure in \cref{fig:shared}. These joint distributions include both the shared properties describing the power spectrum as a whole, and all of the modes collectively (main figure), as well as the properties of each individual multiplet (inset). For the purposes of stellar modelling, we assume that all rotational splittings are symmetric (i.e.~no perturbations to the mode frequencies arising from latitudinal differential rotation or magnetic fields), and use the notional \(m = 0\) mode frequencies associated with our parametric model, \cref{eq:model}, as constraints on the \(m = 0\) mode frequencies returned from MESA stellar models. We provide our fitted values of these mode frequencies in \cref{tab:peakbag}. For each mode, we report a posterior median and the size of the 1-\(\sigma\) credible region around this median, computed from the marginal posterior distribution for its nonrotating frequency. We also provide our identification of the degree \(\ell\), and of the p-mode radial order \(n_p\), supplied through the asymptotic relation, \cref{eq:asymptotic}.

\edit1{Existing significance testing techniques are intended to assess whether power excesses in selected frequency ranges can be attributed to oscillations in general \citep{basu_book_2017}, rather than to specific modes. Thus, they may not be suitable when modes overlap significantly with each other in frequency, as in this case, \edit2{since they may unfairly assess low-amplitude peaks to be significant if they should overlap with high-amplitude ones}. As such, we opt instead to perform a model-comparison likelihood-ratio test \edit2{\citep[e.g.][]{neyman_problem_1933,buse_likelihood_1982}} for each mode, in which a model power spectrum specified by \cref{eq:model}, but with that specific mode removed (taken to be the null hypothesis $H_0$), is compared against the full model specified by \cref{eq:model} (taken to be the alternative hypothesis $H_1$). Specifically, we compute
\begin{equation}
    \Lambda_i = -2 \log \left[{\max_\theta \mathcal{L}(\theta) \over \max_\theta \mathcal{L}(\theta\text{ without $i$\textsuperscript{th} mode})}\right]
\end{equation}
where maximum likelihood estimates are found for the numerator and denominator separately by \edit2{a combination of the BFGS algorithm, the simplex algorithm (to escape local minima), and the truncated Newton method, starting} from the mode of the nested-sampling posterior distribution. This likelihood-ratio statistic $\Lambda$ will be close to 0 or negative for situations favouring the null hypothesis (i.e. the associated regions of the power spectrum being well-explained without that mode, e.g. by fitted red noise), and will be large if the alternative hypothesis (i.e. our nominal model) is favoured. Since each multiplet is governed by two parameters (frequency and amplitude), as the rotational splitting widths are pooled between modes, the test statistic $\Lambda$ is asymptotically described by a $\chi^2$ distribution with two degrees of freedom, by Wilks's Theorem. This permits us to evaluate the formal probability $p$ of producing an outcome favouring $H_1$ by coincidence alone given $H_0$. We report both $\Lambda$ and $p$ from the test for each mode in \cref{tab:peakbag}. \edit2{Several modes (which we mark with $p$-values of 1)} appear not to be significant; \edit2{they} can seen on \cref{fig:asteroseismology}c not to explain much of the \'echelle power diagram in any case.}

\begin{longtable}[]{@{}
  >{\raggedleft\arraybackslash}p{(\columnwidth - 10\tabcolsep) * \real{0.1143}}
  >{\centering\arraybackslash}p{(\columnwidth - 10\tabcolsep) * \real{0.1429}}
  >{\centering\arraybackslash}p{(\columnwidth - 10\tabcolsep) * \real{0.1429}}
  >{\centering\arraybackslash}p{(\columnwidth - 10\tabcolsep) * \real{0.1429}}
  >{\centering\arraybackslash}p{(\columnwidth - 10\tabcolsep) * \real{0.1714}}
  >{\raggedright\arraybackslash}p{(\columnwidth - 10\tabcolsep) * \real{0.2857}}@{}}
\caption{Notional \(m=0\) p-mode frequencies from peakbagging procedure with multiplet model\edit1{; we also report likelihood-ratio test statistics $\Lambda$, and associated $p$-values.} \label{tab:peakbag}}\tabularnewline
\toprule\noalign{}
\begin{minipage}[b]{\linewidth}\raggedleft
\(\ell\)
\end{minipage} & \begin{minipage}[b]{\linewidth}\centering
\(\nu/\mu\text{Hz}\)
\end{minipage} & \begin{minipage}[b]{\linewidth}\centering
\(e_\nu/\mu\text{Hz}\)
\end{minipage} & \begin{minipage}[b]{\linewidth}\centering
\(n_p\)
\end{minipage} & \begin{minipage}[b]{\linewidth}\centering
\(\Lambda\)
\end{minipage} & \begin{minipage}[b]{\linewidth}\raggedright
\(p\)
\end{minipage} \\
\midrule\noalign{}
\endfirsthead
\toprule\noalign{}
\begin{minipage}[b]{\linewidth}\raggedleft
\(\ell\)
\end{minipage} & \begin{minipage}[b]{\linewidth}\centering
\(\nu/\mu\text{Hz}\)
\end{minipage} & \begin{minipage}[b]{\linewidth}\centering
\(e_\nu/\mu\text{Hz}\)
\end{minipage} & \begin{minipage}[b]{\linewidth}\centering
\(n_p\)
\end{minipage} & \begin{minipage}[b]{\linewidth}\centering
\(\Lambda\)
\end{minipage} & \begin{minipage}[b]{\linewidth}\raggedright
\(p\)
\end{minipage} \\
\midrule\noalign{}
\endhead
\bottomrule\noalign{}
\endlastfoot
0 & 5.81 & 0.01 & 4 & \(14.605\) & \(6.738\times10^{-4}\) \\
0 & 6.91 & 0.02 & 5 & \(6.509\) & \(0.039\) \\
0 & 8.20 & 0.02 & 6 & \(5.112\) & \(0.078\) \\
0 & 9.41 & 0.05 & 7 & \(-6.031\) & \(1\) \\
0 & 10.66 & 0.03 & 8 & \(12.428\) & \(0.002\) \\
1 & 5.34 & 0.03 & 3 & \(2.254\) & \(0.324\) \\
1 & 6.40 & 0.01 & 4 & \(17.626\) & \(1.488\times10^{-4}\) \\
1 & 7.64 & 0.01 & 5 & \(19.316\) & \(6.390\times10^{-5}\) \\
1 & 8.78 & 0.02 & 6 & \(23.218\) & \(9.086\times10^{-6}\) \\
1 & 10.09 & 0.03 & 7 & \(-4.060\) & \(1\) \\
2 & 5.63 & 0.05 & 3 & \(-2.937\) & \(1\) \\
2 & 6.70 & 0.02 & 4 & \(-0.410\) & \(1\) \\
2 & 7.90 & 0.03 & 5 & \(11.731\) & \(0.003\) \\
2 & 9.17 & 0.02 & 6 & \(10.920\) & \(0.004\) \\
2 & 10.43 & 0.04 & 7 & \(8.857\) & \(0.012\) \\
\end{longtable}

\section{Generalised Rotational Inversions}\label{generalised-rotational-inversions}

\label{sec:rotation}

\setcounter{table}{0}
\renewcommand{\thetable}{C\arabic{table}}

\setcounter{figure}{0}
\renewcommand{\thefigure}{C\arabic{figure}}

In principle, each observed rotational multiplet permits us to probe Zvrk's internal rotational profile \(\Omega(r)\) with sensitivity to different parts of the stellar interior. In the perturbative regime of slow rotation, where \(\Omega \ll 2\pi \Delta\nu\), the spacing between modes in a rotational multiplet is related to \(\Omega(r)\) through an integral expression of the form \citep[e.g.][]{gough_1981, gough_rotation_1990}
\begin{equation}
    \omega_{n\ell m} - \omega_{n\ell 0}\equiv m \ \delta \omega_{n\ell}\sim m\beta_{n\ell} \int K_{n\ell}(r) \Omega(r) \mathrm d r ,\label{eq:rotintegral}
\end{equation}
where \(\beta\) is an overall sensitivity constant, and \(K\) is a normalised integral kernel, associated with a mode indexed by \(n\) and \(\ell\). For low-degree p-modes, \(\beta \to 1\) in the limit of large \(n_p\) and/or large \(\ell\), while for g modes, \(\beta \to 1 - 1/\ell(\ell + 1)\) in the same limit. If a large enough number of rotational multiplet splittings \(\delta\omega_{n\ell}\) have been measured precisely, \cref{eq:rotintegral} may be inverted to yield inferences about the nature of the rotational profile. There are two main classes of methods by which such ``rotational inversions'' have been performed in the asteroseismic literature, both inheriting in large part from similar efforts in helioseismology \citep[e.g.][]{schunker_inversion_2016a, schunker_inversion_2016b, eggenberger_rotation_2019}. In the method of optimally localised averages \citep[OLA: e.g.][]{backus_resolving_1968}, one chooses judiciously some linear combination of \(\beta_i K_i\) such that the coadded kernels approximate an averaging kernel of finite width localised at some preselected position \(r_0\), with the shape of this kernel, and the quality of this approximation, being determined by some optimal compromise between spatial resolution and effective measurement uncertainty (e.g.~MOLA: \citealt{gough_inverting_1985}; SOLA: \citealt{pijpers_sola_1994}); that same linear combination of \(\delta\omega_i\) then approximates \(\Omega(r_0)\). A full picture of \(\Omega(r)\) is built up by repeating this procedure for various choices of \(r_0\). By contrast, in the regularised least-squares (RLS) method \citep[e.g.][]{jcd_comparison_1990}, one first parameterises a subset of permissible rotational profiles, and then uses \cref{eq:rotintegral} to generate predictions for the multiplet rotational splittings; an optimal description of the rotational profile from this subset is then found by minimising the discrepancy between the predicted and measured rotational splittings, in a least-squares sense, with the potential inclusion of additional regularisation to penalise e.g.~discontinuities and other pathological features.

Rotational inversions in evolved stars have mostly been applied to dipole modes of mixed gravitoacoustic character. Morphologically, the rotational kernels of these mixed modes are combinations of g-mode components, which abruptly truncate at the radiative core boundary, and p-mode components, which sample the envelope and are insensitive to the radiative core. Such a combination has historically been described with a mixing fraction \(\zeta\) in the form \(\beta_{\text{mixed},n\ell}K_{\text{mixed},n\ell} = \zeta\beta_{\text{g},n\ell}K_{\text{g},n\ell}+(1-\zeta)\beta_{\text{p},n\ell}K_{\text{p},n\ell}\) \citep{goupil_seismic_2013}. However, since \(\zeta\) is determined by the proximity of mixed modes to the underlying p- and g-modes as they undergo avoided crossings, small changes to the frequencies of these underlying modes may result in large changes to \(\zeta\) --- particularly where the coupling strength between the mode cavities is weak --- for the most p-dominated mixed modes which are closest to these avoided crossings, and which are most amenable to observation. Equivalently, the radius of convergence for perturbative expansions for the rotating mode frequencies, of which linear expressions like \cref{eq:rotintegral} are but the first-order term, is significantly reduced for near-degeneracy configurations like those seen in red giants \citep{deheuvels_near_2017, ong_rotation_1}. Any use of these mixed-mode kernels to perform inversions (not merely rotational) would accordingly be highly sensitive to small differences in the structure --- and thus, on the red giant branch, evolution --- of the reference models used to generate these kernels numerically, compared to that of the actual star. Obversely, the stellar structure would have to be constrained exquisitely well, both precisely and accurately, in order for inversions to be feasible with mixed modes.

For these reasons, rotational inversions have never been attempted on red giants as evolved as Zvrk, since the coupling between its p- and g-mode cavities is exceedingly weak, which, ordinarily, would render its mixed-mode kernels pathologically sensitive to \(\zeta\). However, since this coupling is so weak as to be observationally irrelevant, we now apply new techniques to generate notional pure p-mode kernels, and perform inversions with respect to them instead. Since per-multiplet rotational splittings are not available, we also generalise existing rotational-inversions techniques to operate with respect to the power spectrum directly, rather than with respect to preprocessed mode frequencies.

\subsection{Rotational Kernels}\label{rotational-kernels}

Having isolated Zvrk's p-mode cavity through the \(\pi\)-mode construction of \citet{ong_semianalytic_2020} in our fitting and optimisation exercises, we now compute its rotational kernels using \(\pi\)-mode eigenfunctions, as described in \citet{ong_rotation_1}. In order for such pure p-mode kernels to be used in inversion problems for subgiants and less evolved red giants, the notional ``pure'' p-mode rotational splittings must be determined from a set of mixed modes. In general, this remains an open problem. Fortunately, the observational irrelevance of any coupling to the g-mode cavity works in our favour here, as we already have effectively measured the p-mode rotational splittings directly, rendering the problem moot.

\begin{figure*}[htbp]
    \centering
    \annotate{\includegraphics[width=.435\textwidth]{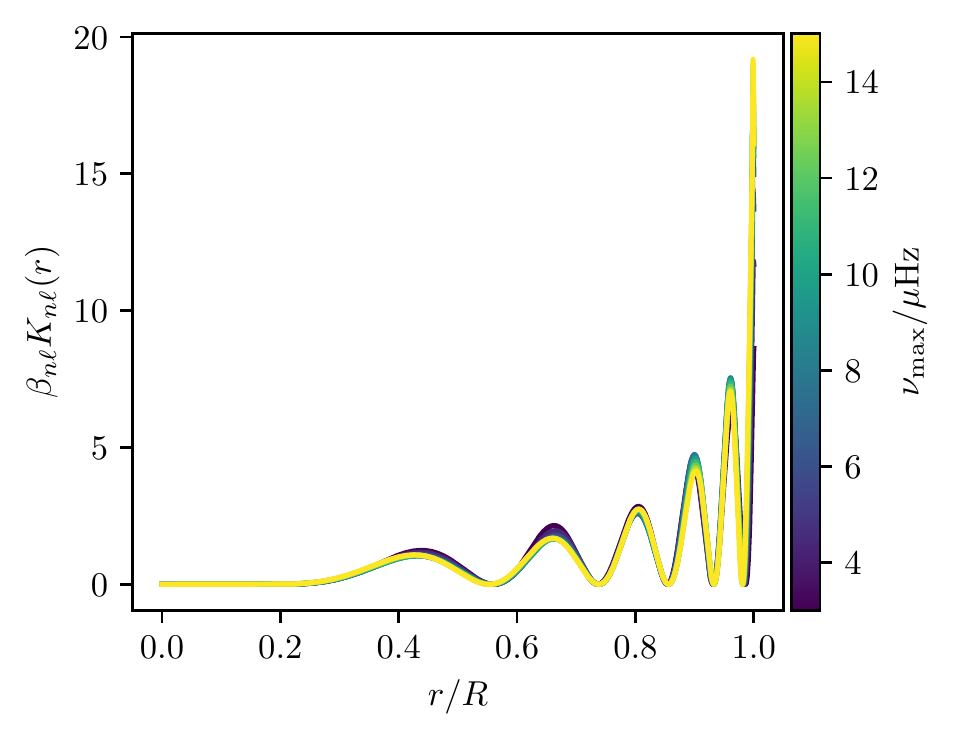}}{\node[right] at (.15, .9){\textbf{(a)}: $n_p = 6, \ell = 1$};}
    \annotate{\includegraphics[width=.435\textwidth]{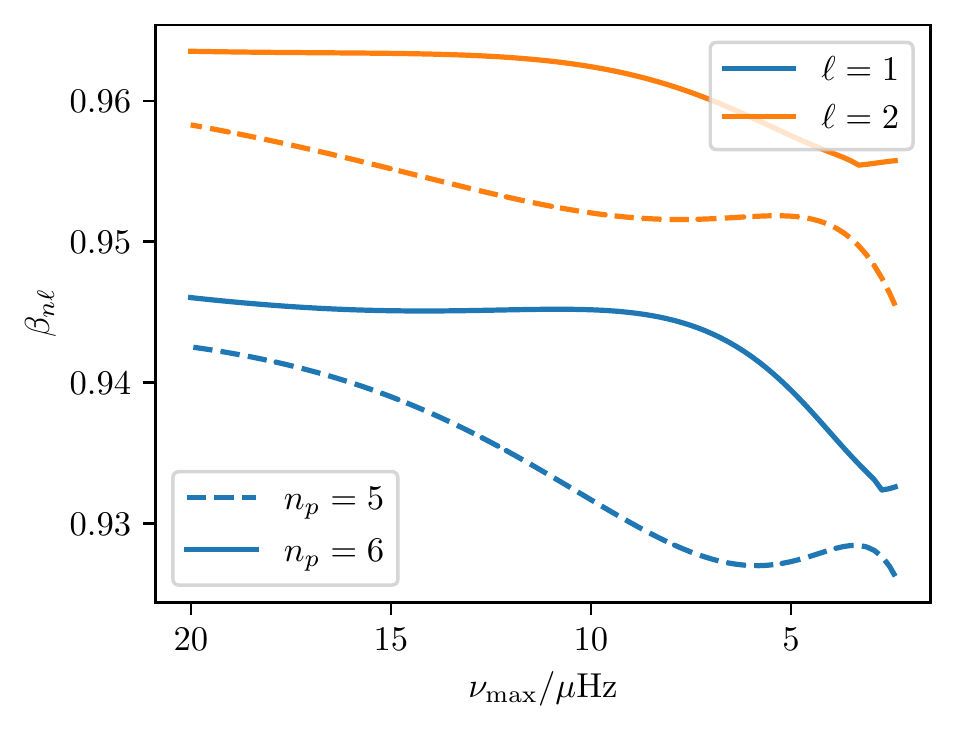}}{\node[right] at (.19, .9){\textbf{(b)}};}
    \annotate{\includegraphics[width=.435\textwidth]{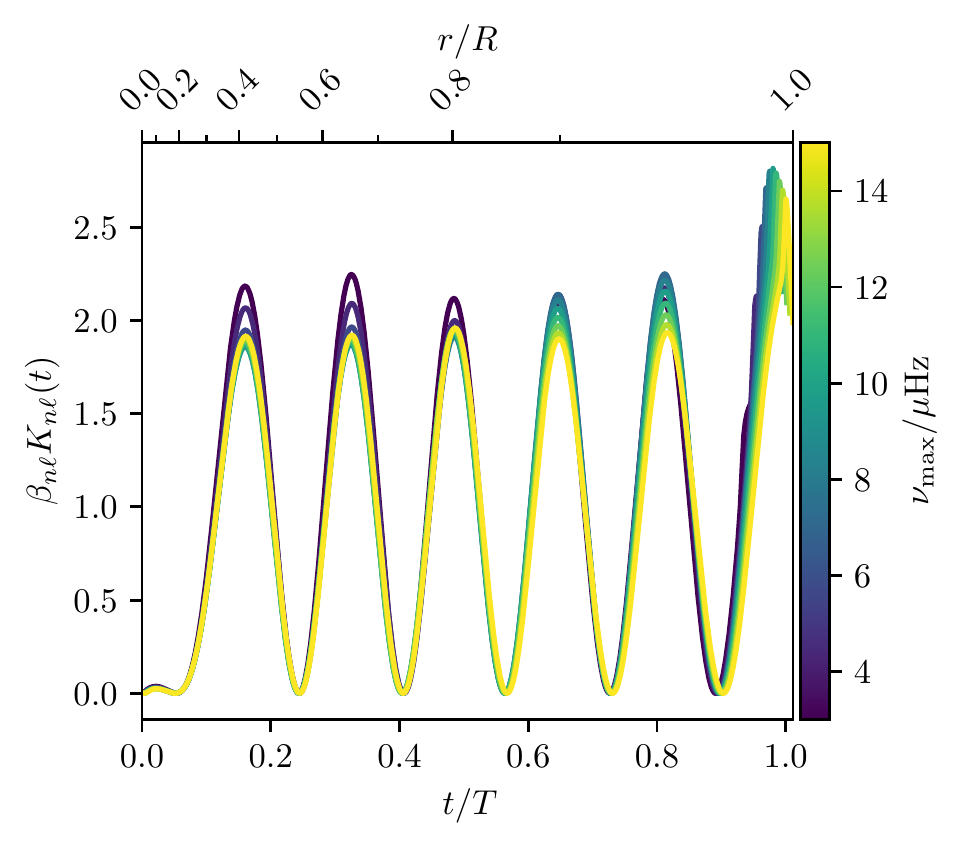}}{\node[right] at (.15, .78){\textbf{(c)}: $n_p = 6, \ell = 1$};}
    \annotate{\includegraphics[width=.435\textwidth]{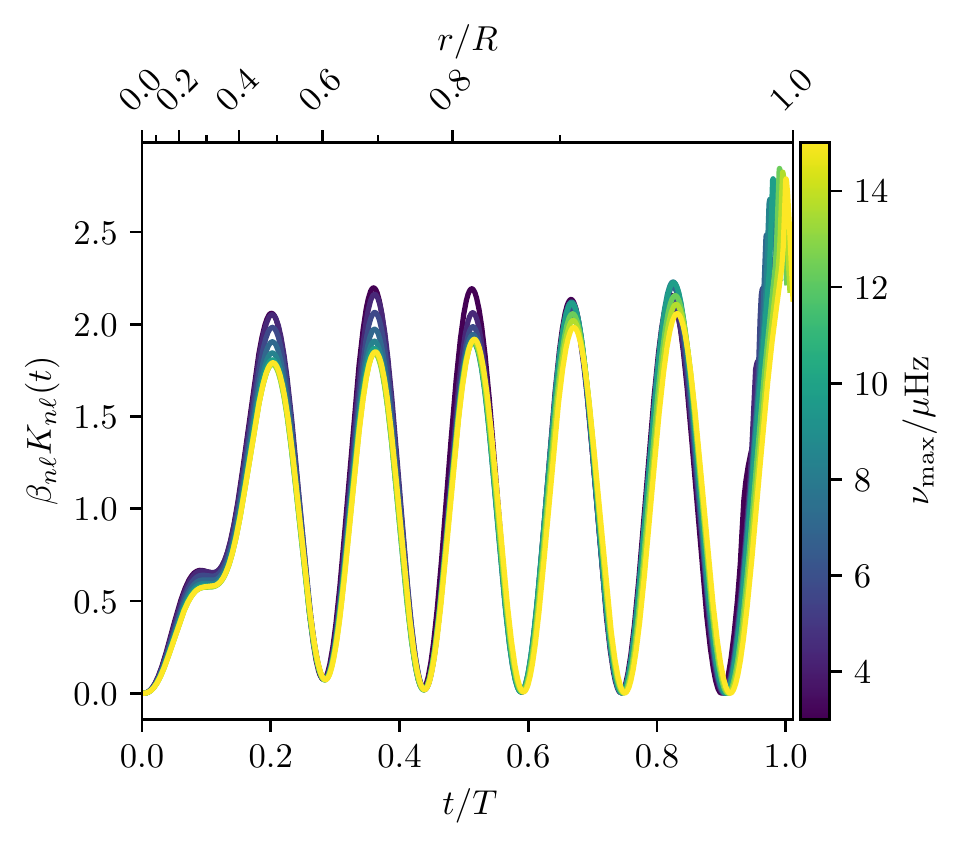}}{\node[right] at (.15, .78){\textbf{(d)}: $n_p = 6, \ell = 2$};}
    \caption{Stability of pure p-mode rotational kernels to stellar evolution. \textbf{(a)} Rotational kernel of the $n = 6, \ell = 1$ p-mode, shown as a function of fractional physical radius for $\numax$ between half and double the observed value. Despite the large range of $\numax$ considered, the shape of the kernel can be seen to remain very stable, unlike those of mixed modes. \textbf{(b)} Variations in the overall sensitivity $\beta$ of several p-mode kernels over the course of evolution (using $\numax$ as its proxy). Rather than rapidly switching between 1 and $1-1/\ell(\ell + 1)$ as those of mixed modes would, the sensitivity of p-mode kernels remains very stable over time. \textbf{(c)} The same kernel as (a), but reparamaterised to permit the use of the acoustic radius $t$ rather than the physical radius $r$ as the integration variable. \textbf{(d)} The same transformation applied to the $n_p = 6, \ell = 2$ kernel. The shape of the kernel (particularly close to the centre of the star) can be seen to be slightly different. It can also be seen to be likewise insensitive to evolution.}
    \label{fig:kernels}
\end{figure*}

Freed of any dependence on the g-mode cavity, the pure p-mode rotational kernels turn out to be extremely robust to even large variations in structure and evolution. We show in \cref{fig:kernels} how they evolve over a very large range of \(\numax\) --- from double to half of our nominal observed value --- over the course of the best-fitting evolutionary track from our optimisation exercise. Specifically, we show in \cref{fig:kernels}a the evolution of a representative rotational kernel --- that associated with the \(n_p = 6, \ell = 1\) mode --- plotted as a function of fractional radius. The overall morphology and normalisation of this kernel, as well as the locations of null sensitivity, can be seen to remain extremely stable despite the very large range of \numax~considered here. This is as opposed to the behaviour of mixed-mode kernels, where relative fractional sensitivity to the core, which is determined by \(\zeta\), would vary rapidly over time, as p-modes evolve onto and off resonance with the underlying g-modes. Likewise, the overall normalisations \(\beta\) are also very stable (\cref{fig:kernels}b), changing by no more than 1\% over a large range of \(\numax\). Again, this is stark contrast with mixed modes, whose overall normalisations \(\beta \sim \zeta \beta_g + (1-\zeta) \beta_p\) would switch between 1 and \(1-1/\ell(\ell + 1)\) as modes evolve through avoided crossings. Thus, despite the TESS data having constrained Zvrk's structure in \autoref{sec:opt} with somewhat less precision than has traditionally been possible with \emph{Kepler}, we can be reasonably confident that our subsequent attempts at inspecting its rotational stratification using p-mode kernels will remain robust against \edit2{small} misestimates of its interior \edit2{or surface} structure, unlike the conventional situation with mixed-mode kernels \edit2{(e.g. \citealt{ong_redgiant_2023})}.

The morphology of the kernel shown in \cref{fig:kernels}a is typical of the pure p-mode rotational kernels encountered in helioseismology, and in the asteroseismology of main-sequence stars. Its sensitivity can be seen to be concentrated towards the surface, as are its zero points. Thus, if we were to place uninformative priors on the localisation of rotational shear, such priors cannot be uniform in the physical radial coordinate, but must rather be proportional to the apparent envelope, and density of zeros, of these kernels. To ease our subsequent analysis, we will find it beneficial to reparameterise these kernels into a radial coordinate system such that uninformative priors are also uniform in our preferred radial coordinate. For this purpose, we choose to use the acoustic radius, \(t(r) = \int_0^r (1/c_s)\mathrm{d}r\), as our preferred radial coordinate \citep[as originally proposed in][]{pijpers_sola_1994}. The p-mode displacement eigenfunctions are known to admit an approximate description in terms of Riccati-Bessel functions of the first kind\footnote{These are related to the usual spherical Bessel functions of the first kind as \(s_\ell(x) = x j_\ell(x)\). Here we use lowercase \(s\) rather than the more conventional uppercase to avoid confusion with the Lamb frequency, which is also written as \(S_\ell\) in the context of asteroseismology.}, \(s_\ell\) \citep[see e.g.][]{roxburgh_asteroseismology_2010, lindsay_near_2023}:
\begin{equation}
     \psi_r \equiv r\sqrt{\rho c_s} \xi_r \sim s_\ell(\omega t - \delta_\ell),\label{eq:bessel}
\end{equation}
where \(c_s\) is the sound speed, and \(\delta_\ell\) is an inner phase function with only weak dependence on frequency and position away from \(t = 0\) and \(t = T\). Since the p-mode eigenfunctions are orthogonal with respect to the integral measure \(r^2 \rho\) when integrated with respect to the physical radial coordinate, the scaled wavefunctions \(\psi\) form an orthogonal basis with respect to the unit measure when integrated with respect to the acoustic radial coordinate. Thus, we may treat a uniform prior on the acoustic radius as being uninformative. Under the action of such reparameterisation, the kernels themselves transform as \(K_t(t) = c_s K_r(r(t))\). We show the \(n_p = 6, \ell = 1\) kernel under the action of this reparameterisation in \cref{fig:kernels}c, plotting the kernels now as functions of the fractional acoustic radius. Again, we see that any variations to the shape and the loci of sensitivity arising from structural evolution are slight. Away from the boundaries set by the centre and the surface of the star, the shape of the kernels is roughly sinusoidal; this is because the Riccati-Bessel functions appearing in \cref{eq:bessel} may be approximated as sinusoids at large argument \citep{arfken_weber_2005}, so this is a property of p-modes generally, and not just this mode in particular. For illustration, we also show a similar diagram for the \(n_p = 6, \ell = 2\) p-mode in \cref{fig:kernels}d, which exhibits much the same qualitative insensitivity to evolution.

A further consequence of our use of pure p-mode rotational kernels is that, as can be seen in \cref{fig:kernels}, the nonradial p-mode wavefunctions decay to zero at small radius, and so these kernels are almost entirely insensitive to the radiative core (whose boundary is situated in our best-fitting stellar structure at \(r/R \sim 0.03\)). For rotational inferences made using mixed modes, it is common to parameterise radial differential rotation in terms of a phenomenological two-zone model, with the core and envelope treated as separately rotating solid bodies \citep[e.g.][]{klion_diagnostic_2017, triana_internal_2017, mosser_period_2018, gehan_core_2018, eggenberger_asteroseismology_2019, fellay_asteroseismology_2021, ahlborn_improved_2022}. By contrast, any constraints on radial differential rotation that we might obtain using our observed p-modes, through these p-mode rotational kernels, will be localised entirely within the convective envelope.

\subsection{Radial Differential Rotation}\label{radial-differential-rotation}

\label{sec:inversion}

Our marginal posterior distribution for the difference between the dipole and quadrupole rotational splittings, \cref{fig:shear-obs}, suggests that Zvrk's envelope is not rotating as a solid body, and possibly exhibits radial differential rotation. Given the preceding discussion, we might wish to investigate this differential rotation using existing rotational inversion techniques. However, our attempts to do so are hindered by the limitations of the existing TESS data set --- including its poorer photometric stability, redder bandpass, and presently shorter duration compared to the nominal \emph{Kepler} mission --- which resulted in us having to pool measurements of the rotational splitting width between different modes in \autoref{sec:nested}. Even then, each of our pooled rotational splittings is constrained with less precision than that of individual multiplets in \emph{Kepler} data \citep[e.g.~as in][]{tayar_spinning_2022}. Were we to have constrained these splittings for each multiplet individually, the resulting uncertainties would be larger still --- too large for the rotational inversion techniques that we have described above to be feasibly prosecuted. To proceed further, we must adapt existing methods to this present situation.

\subsubsection{Optimally-Localised Averages}\label{optimally-localised-averages}

\begin{figure*}[htbp]
    \centering
    \annotate{\includegraphics[width=3in]{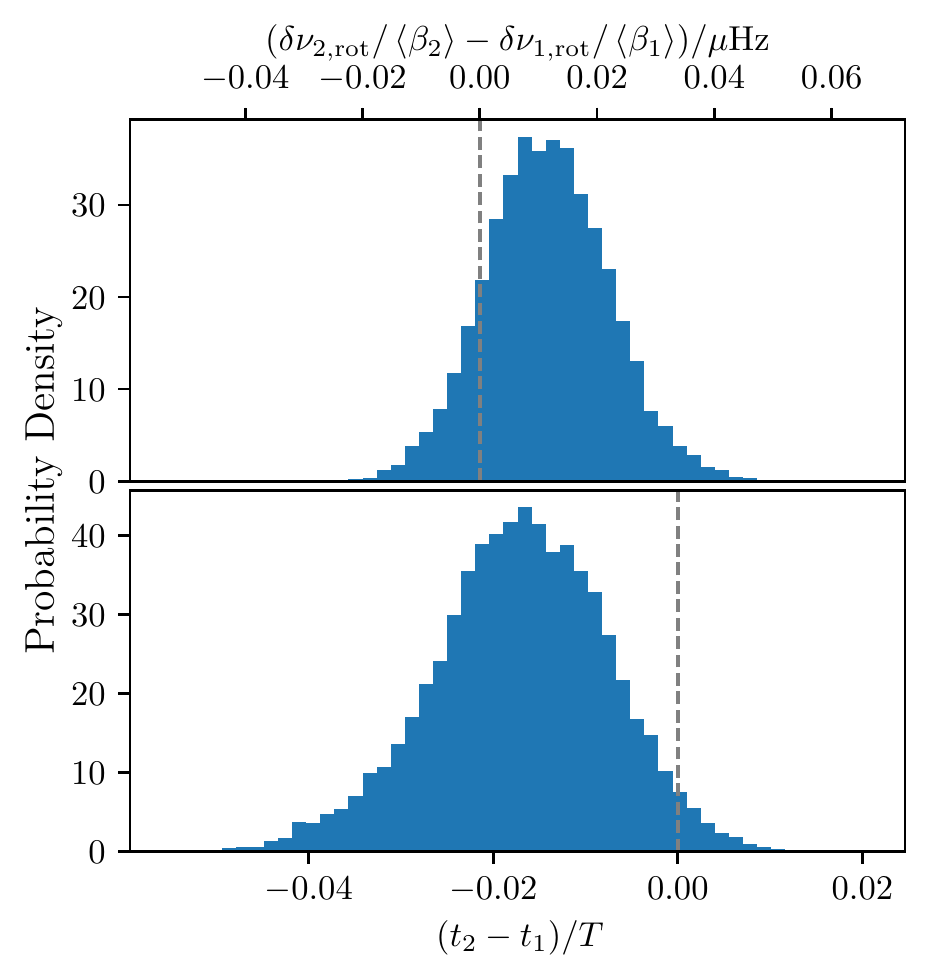}}{\node at (.9,.2){\textbf{(a)}};}
    \annotate{\includegraphics[width=3in]{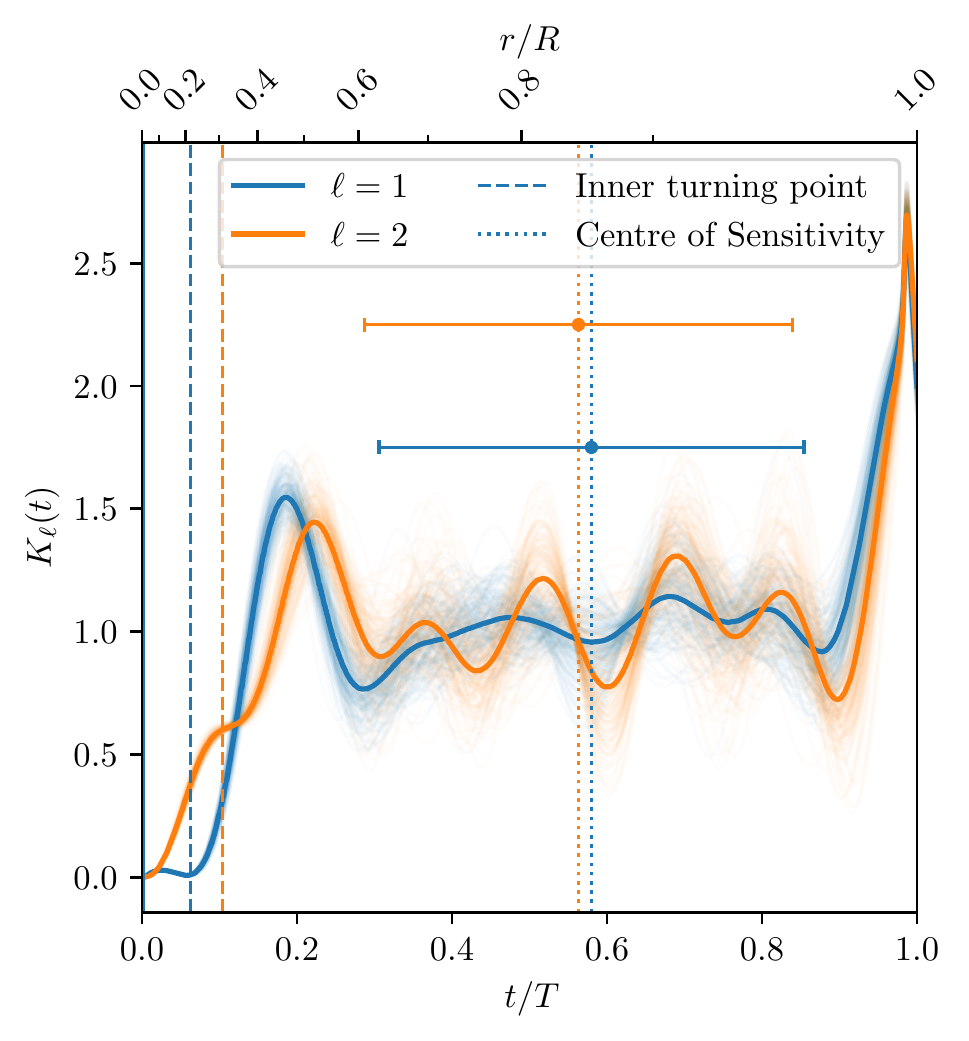}}{\node at (.9,.15){\textbf{(b)}};}
    \caption{Raw asteroseismic constraints on sensitivity to rotational shear. \textbf{(a)} Posterior distributions for differences between the pooled quadrupole and dipole rotational splitting. The upper panel shows the posterior distribution of the differences in estimated rotation rates, which are in turn the rotational splittings divided by effective sensitivity constants $\left<\beta_\ell\right> \sim 1$. The lower panel shows the posterior distribution of the differences in acoustic radii of the centres of sensitivity associated with the underlying averaged rotational kernels of each degree, considered separately. The two quantities are statistically independent. \textbf{(b)} Differential sensitivity kernels associated with distributions in (a). Associated with each pooled rotational splitting $\delta\nu_\ell$ is an effective rotational kernel $K_\ell$, which is a weighted average of the rotational kernels associated with each mode of that degree. Faint curves show samples from the posterior distribution, while the solid coloured curves show the posterior median effective kernel for each degree. The inner turning points at $\numax$ of the best fitting stellar model from our optimisation procedure are shown with the vertical dashed lines, while the centres of sensitivity for the median kernels are indicated with the vertical dotted lines. The ordering of their positions can be seen to reversed compared to the theoretical inner turning points. The localisation widths of the median kernels are indicated by the horizontal intervals placed around the centres of sensitivity.}
    \label{fig:shear}
\end{figure*}

We first consider how the OLA method may be adapted to operate with respect to our pooled rotational measurements. We observe that the pooled rotational splitting for dipole modes must be related to those of the underlying individual dipole multiplets through some kind of visibility-weighted averaging procedure, not necessarily linear. If we approximate this with a linear combination of mode-wise rotational splittings as \(\delta\omega_{\ell = 1}\sim \sum_{n,\ell = 1} w_n \delta\omega_{n,\ell}\), then the corresponding linear combination of dipole-mode kernels effectively provides us with a ``pooled'' sensitivity kernel, \(K_{\text{eff}, 1}\), for the observed dipole rotational splitting. We may also divide the averaged splitting by the associated linear combination of sensitivity constants, \(\left<\beta_1\right> = \sum_{n,\ell=1} w_n \beta_{n\ell}\), to yield an estimate of \(\Omega(t_1)\) as localised by this pooled kernel. Such an estimate is localised, roughly speaking, at the centre of sensitivity for this pooled kernel, \(t_1 = \int t \cdot K_{\text{eff}, 1}(t)\ \mathrm{d}t\). Since we have two averaged rotational splittings at our disposal --- one for dipole modes, and one for quadrupole modes --- we in principle then should be able to constrain rotational shear in terms of the differences in localised rotation rates, \(\Delta\Omega \sim \Omega(t_2) - \Omega(t_1)\), as well as differences in the locations at which these estimates are made, \(\Delta t = t_2 - t_1\).

It remains for us to determine how these coefficients \(w_i\) are to be related to the mode visibilities. By ansatz, we set \(w_i \propto H_i\), the observed mode heights appearing in \cref{eq:model}. Through the above discussion, we may then evaluate \(\Delta\Omega\), \(\Delta t\), and the per-degree pooled kernels in closed form, given values for all of the nonradial mode heights and averaged rotational splittings. Since our nested-sampling procedure in \autoref{sec:nested} provides us with a large number of samples from the joint distributions between all of these quantities, applying this procedure to all of these samples permits us to estimate the entire posterior distributions for \(\Delta\Omega\), \(\Delta t\), and the pooled kernels themselves, constrained directly by the power spectrum.

We examine these posterior distributions in \cref{fig:shear}. In \cref{fig:shear}a, we show the marginal distributions for \(\Delta\Omega\) (top) and \(\Delta t\) (bottom), in both cases marking out 0 with a dashed line, which is the value demanded by a null hypothesis of solid-body rotation. We find that \(p(\Delta\Omega > 0) \gtrsim 0.85\), weakly suggesting that rotational shear is indeed present, although unfortunately we are not able to rule out the null hypothesis altogether. Since we have defined \(\Delta t\) in the sense \(t_2 - t_1\), rudimentary JWKB analysis would suggest that it be positive, as the inner turning points of the quadrupole p-modes, set by where the Lamb frequency \(S_\ell = \sqrt{\ell(\ell+1)} c_s/r\) matches the mode frequency \(\omega\), lies outside of that of the dipole modes. Contrary to this expectation, the posterior distribution for \(\Delta t\) is more consistent instead with negative values. This would indicate that the inner portions of the convective envelope are rotating more quickly than the outer portions.

We examine these kernels in more detail in \cref{fig:shear}b, where we plot with the faint thin curves 100 samples from the posterior distribution over both pooled kernels, and with the thick solid curves the posterior median pooled kernels, using different colours for each degree. Additionally, we mark out the location of the JWKB inner turning points (evaluated at \(\numax\)) from our best-fitting model with vertical dashed lines, in matching colours. The quadrupole inner turning point can indeed be seen to lie exterior to the dipole inner turning point. However, the shape of the quadrupole-mode pooled kernel can also be seen to be more spread out than the dipole-mode kernel. Ultimately, this is because the expressions for the rotational kernels contain explicit dependence on \(\ell\) through contributions from a term \([\ell(\ell + 1) - 1]\xi_h^2\), in addition to the implicit dependences of the eigenfunctions themselves on \(\ell\). For low-degree p-modes, the displacement eigenfunctions become approximately radial in the outer parts of the star, where the approximation of plane-parallel stratification holds well. Thus, this degree-dependent enhancement to the rotational kernel from the horizontal component of the eigenfunction is most significant near the interior of the star. Accordingly, the quadrupole-mode kernels feature a ``leg'' protruding inwards from their first local maximum after the inner turning point, which is not present in the dipole-mode kernels. This has the effect of rendering the quadrupole-mode kernels more spread out compared to the dipole mode kernels, and of moving their centres of sensitivity inwards (shown with the dotted vertical lines). For Zvrk in particular, this adjustment to the centre of sensitivity is comparable to, and slightly larger than, the differences in the inner turning points. In any case, any differences in the locations of the centres of sensitivities are minute compared to both the spread implied by the posterior distribution on \(\Delta t\), and the localisation widths of the pooled kernels themselves (indicated with the horizontal intervals). Thus, without access to the individual rotational splittings, our constrained modification to the OLA techique is unfortunately unable to pin down the location, or even direction, of \edit3{any} shear in the envelope.

\begin{figure*}[htbp]
    \centering
    \annotate{\includegraphics[width=.85\textwidth]{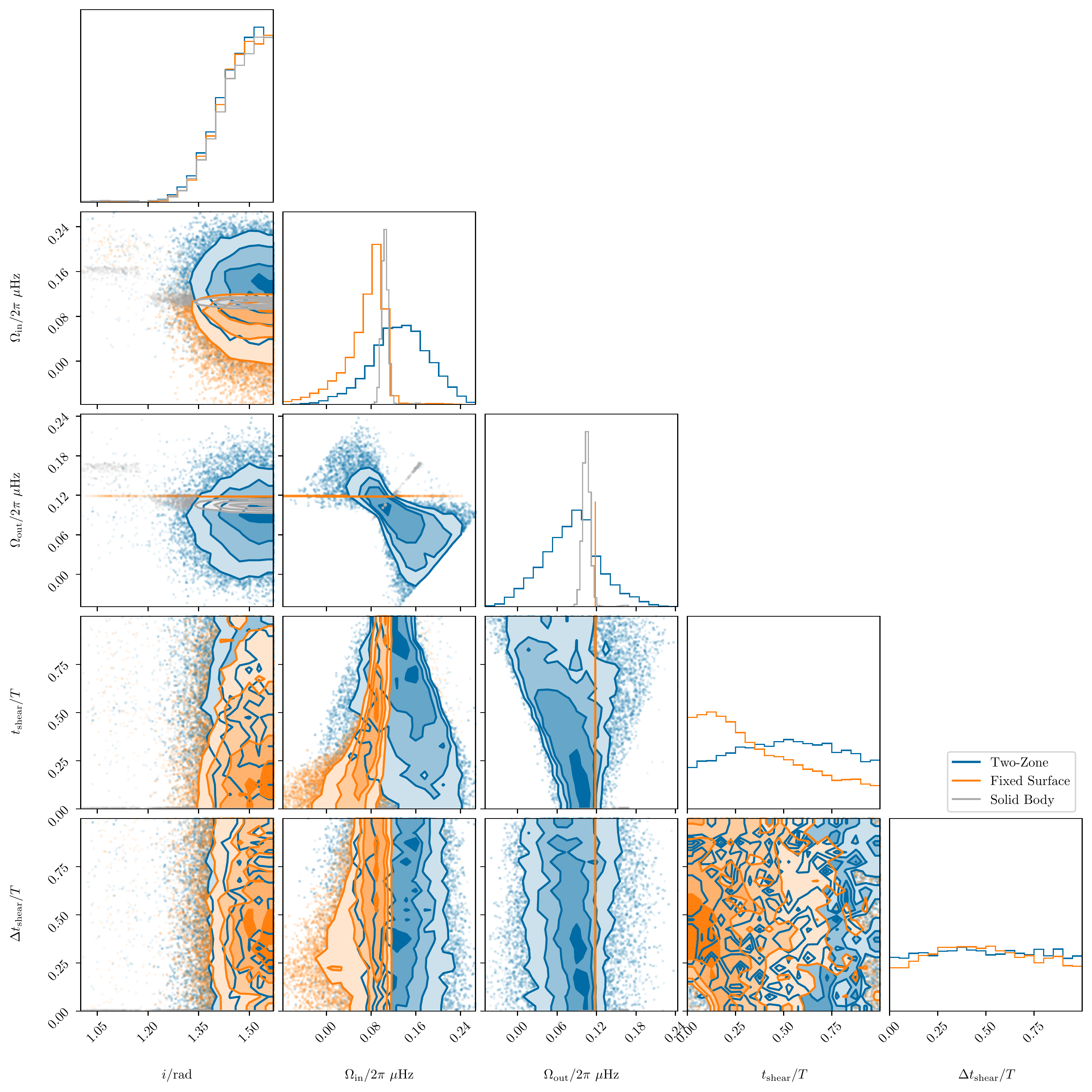}}{
    \node[below,left] at (.65, .95){\includegraphics[width=.33\textwidth]{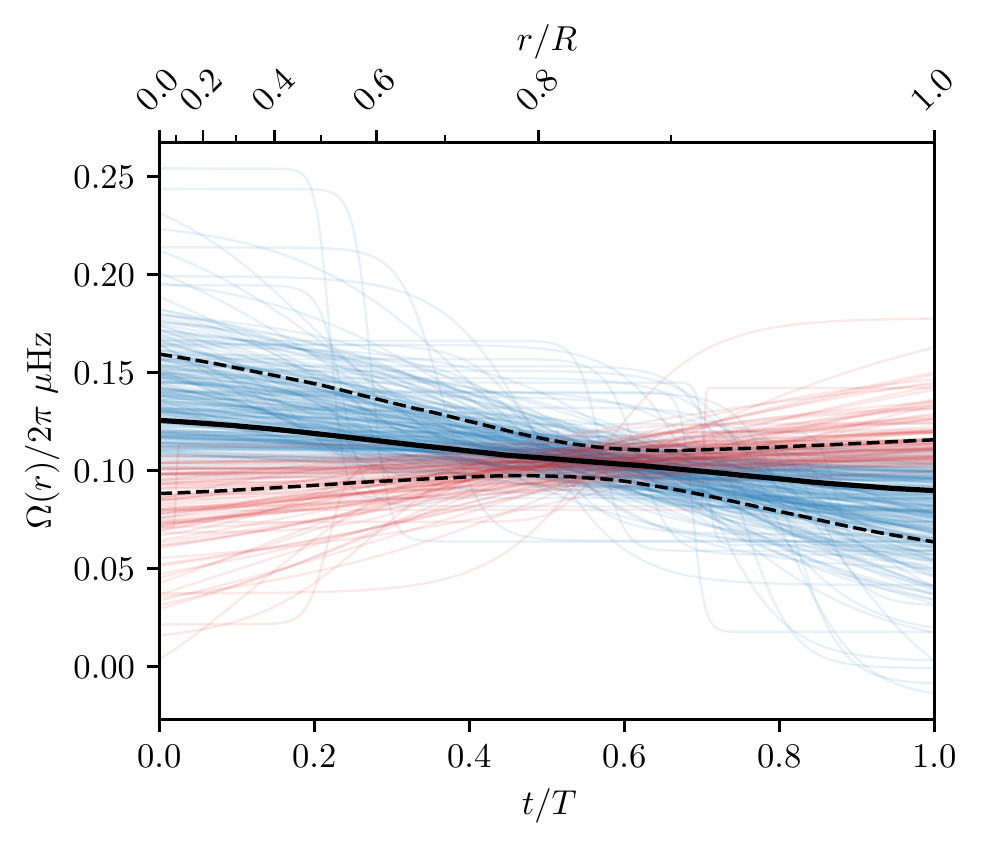}};
    \node[below,left] at (1.03, .95){\includegraphics[width=.33\textwidth]{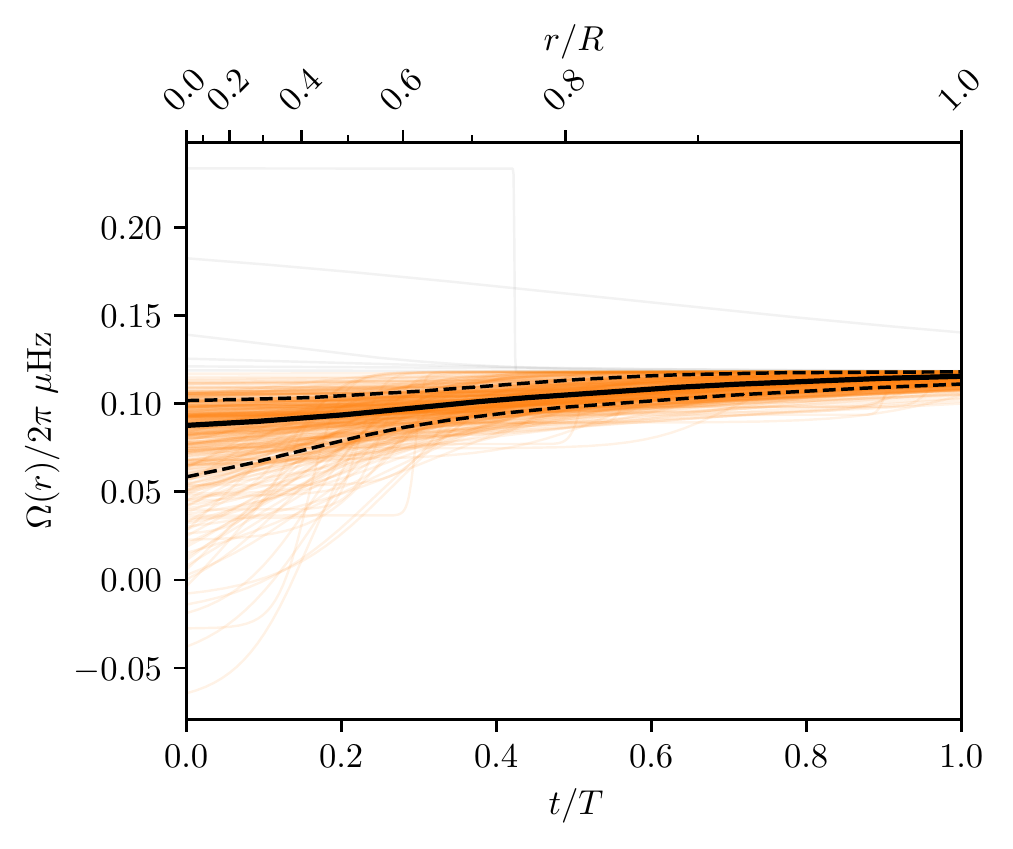}};
    \node[below,left] at (1.03, .62){\includegraphics[width=.33\textwidth]{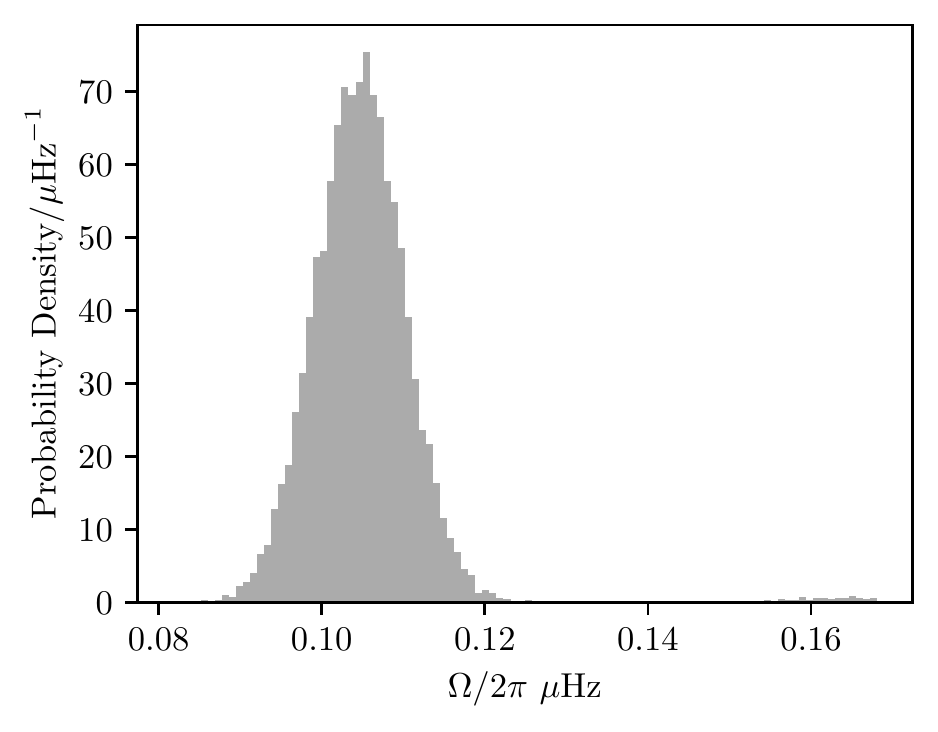}};
    \node at (.1, .97){\textbf{(a)}};
    \node at (.6, 1.04){\textbf{(b)}};
    \node at (.98, 1.04){\textbf{(c)}};
    \node at (.99, .74){\textbf{(d)}};}
    \caption{Regularised asteroseismic constraints on rotational shear. \textbf{(a)} Joint posterior distributions for the stellar inclination, inner and outer rotation rates, location, and thickness of the rotational shear layer, which parameterise our regularised two-zone description of the rotational profile, \cref{eq:shear}. We show in blue the posterior distributions obtained from the power spectrum with the flat priors described in the main text; with orange a conditional distribution obtained by demanding that the outer rotation rate matches the value provided by photometry; \edit3{and with gray a conditional distribution from our null-hypothesis scenario in which solid-body rotation is enforced.} \textbf{(b)} Samples from the posterior distribution over rotational profiles from the main exercise are shown with the thin faint lines. These are coloured blue if they are consistent with differential rotation in the sense of the rotation rate decreasing towards the surface, and red for the converse sense. The median rotational profile is shown with the thick black curve, and the $\pm 1\sigma$ posterior credible region around it is demarcated with the dashed curves. \textbf{(c)} The same as (b), but for our restricted scenario where the outer rotation rate is matched to the photometric value. Draws from the posterior distribution with the rotation rate increasing outwards are shown in orange, while those with the converse sense of differential rotation are shown in gray. \edit3{\textbf{(d)} A closer look at the conditional posterior distribution for the rotation rate under a solid-body rotation scenario.}}
    \label{fig:rls}
\end{figure*}

\subsubsection{Regularised Least-Squares}\label{regularised-least-squares}

The preceding discussion has implicitly relied on a two-zone model of radial differential rotation, where we have used the separation between pooled dipole and quadrupole modes to provide us both the averaged rotation rate in these two zones, as well as the locations of the zones themselves. Since the underlying rotational kernels have proven uncooperative in allowing us to distinguish the locations of these two zones, we may instead make this assumption explicit by parameterising \(\Omega(t)\) to contain two physically separated regions of solid-body rotation, with a shear layer between the two. More concretely, we describe the rotational profile as
\begin{equation}
    \Omega(t; \Omega_\text{in}, \Omega_\text{out},t_\text{shear},\Delta t) = \Omega_\text{in} + \left(\Omega_\text{out} - \Omega_\text{in} \over 2\right)\left(\tanh\left(t - t_\text{shear}\over \Delta t\right) + 1\right),\label{eq:shear}
\end{equation}
which we use to generate predictions for the rotational splittings through \cref{eq:rotintegral}, as done in the RLS technique. Rather than comparing these predictions against measured rotational splittings (which we do not have in hand), we may instead constrain the parameters \((\Omega_\text{in}, \Omega_\text{out},t_\text{shear},\Delta t)\) of this description directly against the power spectrum, by inserting these predictions into our parametric model of the power spectrum, \cref{eq:model}, in lieu of the pooled rotational splittings, and evaluating likelihoods using the same \(\chi^2\)-2-degree-of-freedom likelihood function (\cref{eq:lnlike2dof}) as we had earlier used for the pooled rotational splittings. This allows us to derive posterior distributions for these parameters of \cref{eq:shear} (and the other existing parameters of the model) through the same nested sampling procedure as we had used in \autoref{sec:nested}. For this purpose, we place the same uniform priors on \(\Omega_\text{bulk} = (\Omega_\text{in} + \Omega_\text{out})/2\) as we did on the pooled rotational splittings in \autoref{sec:nested} (up to factor of \(2\pi\)), and we also place very wide uniform priors on \(\Delta\Omega = \Omega_\text{out} - \Omega_\text{in}\), permitting it to vary between \(-0.016\ \mu\)Hz and \(+0.016\ \mu\)Hz --- i.e.~our prior is chosen explicitly so as not to favour either sense of the rotational shear.

We show the resulting joint posterior distribution for these parameters, together with the inclination, using blue points, contours, and histograms, in \cref{fig:rls}a. As we encountered in our earlier discussion of the rotational kernels, even this regularised construction is unable to constrain the location, or the sharpness, of the rotational shear. \edit3{It appears only to} weakly favour a sense of the rotational shear where the interior of the convective envelope is rotating faster than the exterior. To better illustrate this, we show in \cref{fig:rls}b several random samples for the posterior distribution over possible shapes of \(\Omega(r)\), using the thin faint curves. These curves are coloured blue if \(\Omega_\text{in}>\Omega_\text{out}\), and red otherwise. The overall median rotational profile, which we show with the solid black curve, is such that the rotation rate decreases outwards. \edit3{Rotational shear in this sense is also implied by the maximum-likelihood estimator for this parameterisation of radial differential rotation.} \edit3{However, this technique does not} rule out the opposite sense of differential rotation, as some rotational profiles in the converse sense (i.e.~increasing outwards, shown in red) are encompassed within the \(1\sigma\)-credible region, shown with the dashed black curves.

\edit3{To assess the statistical support behind this result, we perform a likelihood-ratio model-comparison test against a solid-body rotation scenario, in which $\Omega_\text{out}$ is constrained to be equal to $\Omega_\text{in}$, such that $t_\text{shear}$ and $\Delta t$ are entirely unconstrained. For illustrative purposes, we also estimate, again using nested sampling, the full posterior distribution for all parameters under this solid-body scenario. We show the joint distributions of the inclination $i$ and rotation rate $\Omega$ with the gray contours in \cref{fig:rls}a, and show the marginal posterior distribution for $\Omega$ in more detail in \cref{fig:rls}d. The above two-zone differential rotation scenario yields a likelihood-ratio test statistic of $\Lambda = 3.832$ against this solid-body rotation scenario: it is more consistent with the power spectrum. However, this comes at the expense of introducing 3 additional free parameters; accordingly, there is a fairly high probability of $p = 0.280$ that this improvement in likelihood can be explained purely by coincidence.} Differential rotation in the sense of \(\Omega\) decreasing outwards \edit3{is thus more likely than solid-body rotation, but not unambiguously so}.

\edit3{Differential rotation in this direction} would possibly require a second shear layer to exist near the surface, not described by \cref{eq:shear}, for consistency with the photometric rotation rate. This shear layer would also have to be located very near the surface, in order not to overly affect values returned from the integral expression in \cref{eq:rotintegral} from the low-degree kernels that we have access to. Such an asteroseismically inaccessible (at least at low degree) near-surface shear layer is not strictly speaking impossible. Indeed, it is the preferred explanation adopted by \citet{tayar_spinning_2022} to explain the rotational configuration of the red giant considered in that work, whose surface rotation rate, obtained from spectroscopy, is similarly faster than a pooled estimate of bulk rotation in the envelope recovered from quadrupole and octupole p-modes. A near-surface shear layer (albeit with rotational shear in the opposite sense, and at different optical depth than demanded here) is also known to exist in the Sun, where its presence is confirmed by the observational availability of intermediate and high-degree modes, whose rotational kernels, being confined almost entirely to the near-surface layers, would not find it similarly invisible \citep[e.g.][]{schou_helioseismic_1998, howe_rotation_2009}.

However, without measurements of modes of intermediate or high degree available, we cannot claim that this explanation is unique. For one, although we have so far discussed differential rotation in only the radial direction, it is possible that it also exists latitudinally. While the spherical harmonics of the sectoral modes through which we observe the rotational splitting are sensitive mostly to an equatorial band \citep[i.e.~\(P_\ell^\ell(\cos \theta)\) can be expressed in terms of only \(\sin \theta\):][]{arfken_weber_2005}, the latitudes of the surface features producing our photometric rotational signal are entirely unconstrained. Should they be located at high latitude, and should Zvrk additionally exhibit latitudinal differential rotation in an anti-solar sense, then we would expect its equator to be rotating more slowly than the photometric modulations would imply; this would resolve our discrepancy from purely geometrical considerations. However, such an explanation would be fairly ad-hoc, as it would require Zvrk's surface features to persist at much higher latitudes than the spots on magnetically active stars. In any case, latitudinal differential rotation in general, and on stars that are not Sun-like in particular, is very poorly understood.

We can also explore alternative solutions to this issue by further regularising our asteroseismic inferences. To demonstrate this, we repeat our nested-sampling exercise, but further demand that \(\Omega_\text{out}\) match our photometric surface rotational period of 99 days (i.e.~we place a \(\delta\)-function prior on \(\Omega_\text{out}\)). We place a uniform prior on \(\Omega_\text{in}\) centered on this value, extending \(0.016\ \mu\)Hz in both directions --- again, so as not to favour either sense of the rotational shear a priori. In this fashion, we recover a conditional posterior distribution for the scenario where the outer rotation rate of the convective envelope matches the surface. We show the results of this procedure in \cref{fig:rls}a using the orange points, contours, and histograms. \edit3{Again, the maximum-likelihood point estimate under this scenario is more likely than that of solid-body rotation, with a likelihood-ratio test statistic of $\Lambda = 3.213$. While this is a lesser improvement than the unconstrained two-zone model, such an improvement in likelihood comes with fewer additional free parameters (only 2), yielding a lower test probability of $p=0.201$.}

Under this more restrictive scenario, the inferred conditional posterior distribution on \(\Omega_\text{in}\) can now be seen to favour differential rotation in the sense of \(\Omega\) increasing outwards. This can also be seen when we examine the implied conditional distribution over the rotational profiles themselves, shown in \cref{fig:rls}c, where sampled profiles which increase outwards are shown in orange, and those which do not are shown in gray: the latter can be seen to be exceedingly rare. However, we see also that the imposition of this hard prior on \(\Omega_\text{out}\) also requires \(t_\text{shear}\) to be situated much farther away from the surface, as can be seen from the strong peak in its marginal distribution in \cref{fig:rls}a, at roughly \(t/T \sim 0.1\) (i.e.~\(r/R \sim 0.3\) in our best-fitting stellar structure), as opposed to naturally producing a near-surface shear layer as we originally posited. This does not, however, contraindicate the existence of such a near-surface shear layer, as we have little a priori justification for adopting such a \(\delta\)-function prior on \(\Omega_\text{out}\) in the first place.

We compare these rotational constraints against those from photometry and spectroscopy in \cref{fig:rotsummary}. While both of our direct measurements for the surface rotation rate sit slightly higher than our unconstrained regularised predictions for the surface rotation rate from asteroseismology (blue shaded region), we note that the discrepancy between the two direct measurements is greater than between either of the two, and our asteroseismic prediction. As we discuss above, an ad-hoc combination of anti-solar latitudinal differential rotation and high-latitude spotting could bring the equatorial surface rotation rate implied by photometry into agreement with our asteroseismic measurements of the bulk rotation rate in the envelope. Conversely, in the more likely scenario of low-latitude features and solar-like latitudinal differential rotation, our photometric value would be more representative of an equatorial rotation rate than the value produced by Doppler line-broadening, which is integrated over the visible disk.

  \bibliography{biblio.bib}

\end{document}

%% file: macros.tex
\usepackage[capitalize]{cleveref}
\usepackage{CJKutf8}
\usepackage{newunicodechar}

\newcommand{\Dnu}{\ensuremath{\Delta\nu}}
\newcommand{\numax}{\ensuremath{{\nu_\text{max}}}}
\newcommand{\amlt}{\ensuremath{{\alpha_\text{MLT}}}}
\newcommand{\teff}{\ensuremath{{T_\text{eff}}}}

\newcommand\mesa{{\textsc{mesa}}}
\newcommand\gyre{{\textsc{gyre}}}

\newcommand{\chinesename}{{\begin{CJK}{UTF8}{gbsn}(王加冕)\end{CJK}}}

\DeclareRobustCommand{\okina}{%
  \raisebox{\dimexpr\fontcharht\font`A-\height}{%
    \scalebox{0.8}{`}%
  }%
}
\newunicodechar{ʻ}{\okina}

\newcommand{\annotate}[2]{\begin{tikzpicture}
    \node[anchor=south west,inner sep=0,align=center] (image) at (0,0) {
    #1
    };
    \begin{scope}[x={(image.south east)},y={(image.north west)}]
    #2
    \end{scope}
\end{tikzpicture}}

\graphicspath{{./},{./figures/}}

\renewcommand{\edit}[2]{{\ifnum#1<1000%
#2%
\else%
\textbf{#2}%
\fi}}


%% file: preamble.tex


\correspondingauthor{Joel Ong}
\email{joelong@hawaii.edu}
\author[0000-0001-7664-648X]{J. M. Joel Ong \chinesename}
\altaffiliation{The \textit{Gasing Pangkah} Collaboration}
\affiliation{Institute for Astronomy, University of Hawaiʻi, 2680 Woodlawn Drive, Honolulu, HI 96822, USA}
\affiliation{NASA Hubble Fellow}

\author[0000-0003-2400-6960]{Marc Teng Yen Hon}
\altaffiliation{The \textit{Gasing Pangkah} Collaboration}
\affiliation{Institute for Astronomy, University of Hawaiʻi, 2680 Woodlawn Drive, Honolulu, HI 96822, USA}
\affiliation{NASA Hubble Fellow}


\author[0000-0001-7493-7419]{Melinda Soares-Furtado}
\altaffiliation{The \textit{Gasing Pangkah} Collaboration}
\affiliation{Department of Astronomy, University of Wisconsin-Madison, 475 N. Charter St., Madison, WI 53703, USA}
\affiliation{MIT Kavli Institute for Astrophysics and Space Research, 77 Massachusetts Ave., Cambridge, MA 02139, USA}
\affiliation{NASA Hubble Fellow}


\author[0000-0001-8220-0548]{Alexander P. Stephan}
\altaffiliation{The \textit{Gasing Pangkah} Collaboration}
\affiliation{Department of Astronomy, The Ohio State University, Columbus, OH 43210, USA}
\affiliation{Center for Cosmology and Astroparticle Physics, The Ohio State University, Columbus, OH 43210, USA}


\author[0000-0002-4284-8638]{Jennifer van Saders}
\affiliation{Institute for Astronomy, University of Hawaiʻi, 2680 Woodlawn Drive, Honolulu, HI 96822, USA}


\author[0000-0002-4818-7885]{Jamie Tayar}
\affiliation{Department of Astronomy, University of Florida, Bryant Space Science Center, Stadium Road, Gainesville, FL 32611, USA}


\author[0000-0003-4631-1149]{Benjamin Shappee}
\affiliation{Institute for Astronomy, University of Hawaiʻi, 2680 Woodlawn Drive, Honolulu, HI 96822, USA}

\author[0000-0003-3244-5357]{Daniel R. Hey}
\affiliation{Institute for Astronomy, University of Hawaiʻi, 2680 Woodlawn Drive, Honolulu, HI 96822, USA}


\author[0000-0002-8849-9816]{Lyra Cao}
\affiliation{Department of Astronomy, The Ohio State University, Columbus, OH 43210, USA}


\author[0000-0002-7772-7641]{Mutlu  Y{\i}ld{\i}z}
\affiliation{Department of Astronomy and Space Sciences, Science Faculty, Ege University, 35100, Bornova, \.Izmir, Turkey}

\author[0000-0002-9424-2339]{Zeynep \c{C}el{\normalshape \.i}k Orhan}
\affiliation{Department of Astronomy and Space Sciences, Science Faculty, Ege University, 35100, Bornova, \.Izmir, Turkey}

\author[0000-0001-5759-7790]{S{\normalshape \.i}bel  \"Ortel}
\affiliation{Department of Astronomy and Space Sciences, Science Faculty, Ege University, 35100, Bornova, \.Izmir, Turkey}






\author[0000-0001-7516-8308]{Benjamin Montet}
\affiliation{School of Physics, University of New South Wales, Sydney, NSW 2052, Australia}
\affiliation{UNSW Data Science Hub, University of New South Wales, Sydney, NSW 2052, Australia}







\author[0000-0001-9206-3460]{Thomas W.-S. Holoien}
\affiliation{Carnegie Observatories, 813 Santa Barbara Street, Pasadena, CA 91101 USA}


\author[0000-0001-7516-4016]{Joss Bland-Hawthorn}
\affiliation{Sydney Institute for Astronomy, School of Physics, A28, The University of Sydney, NSW 2006, Australia}
\affiliation{ARC Centre of Excellence for All Sky Astrophysics in 3 Dimensions (ASTRO 3D), Australia}

\author[0000-0002-4031-8553]{Sven Buder}
\affiliation{Research School of Astronomy and Astrophysics, Australian National University, Canberra, ACT 2611, Australia}
\affiliation{ARC Centre of Excellence for All Sky Astrophysics in 3 Dimensions (ASTRO 3D), Australia}

\author[0000-0001-7362-1682]{Gayandhi M. De Silva}
\affiliation{School of Mathematical and Physical Sciences, Macquarie University, Sydney, NSW 2109, Australia}
\affiliation{Australian Astronomical Optics, Macquarie University, 105 Delhi Rd, North Ryde, NSW 2113, Australia}
\affiliation{ARC Centre of Excellence for All Sky Astrophysics in 3 Dimensions (ASTRO 3D), Australia}

\author[0000-0001-6280-1207]{Ken C. Freeman}
\affiliation{Research School of Astronomy and Astrophysics, Australian National University, Canberra, ACT 2611, Australia}
\affiliation{ARC Centre of Excellence for All Sky Astrophysics in 3 Dimensions (ASTRO 3D), Australia}



\author[0000-0002-3430-4163]{Sarah L. Martell}
\affiliation{School of Physics, University of New South Wales, Sydney, NSW 2052, Australia}
\affiliation{ARC Centre of Excellence for All Sky Astrophysics in 3 Dimensions (ASTRO 3D), Australia}

\author[0000-0003-3081-9319]{Geraint F. Lewis}
\affiliation{Sydney Institute for Astronomy, School of Physics, A28, The University of Sydney, NSW 2006, Australia}


\author[0000-0002-0920-809X]{Sanjib Sharma}
\affiliation{Space Telescope Science Institute, 3700 San Martin Drive, Baltimore, MD, 21218, USA}


\author[0000-0002-4879-3519]{Dennis Stello}
\affiliation{School of Physics, University of New South Wales, Sydney, NSW 2052, Australia}
\affiliation{Sydney Institute for Astronomy, School of Physics, A28, The University of Sydney, NSW 2006, Australia}
\affiliation{ARC Centre of Excellence for All Sky Astrophysics in 3 Dimensions (ASTRO 3D), Australia}
\affiliation{Stellar Astrophysics Centre, Aarhus University, Ny Munkegade 120, DK-8000 Aarhus C, Denmark}



\received{June 30, 2023}
\revised{February 23, 2024}
\accepted{February 26, 2024}
\submitjournal{\apj}
\shortauthors{Ong et al.}
\def\sectionautorefname{Section}
\def\subsectionautorefname{Section}
\def\subsubsectionautorefname{Section}